\documentclass{aa}
\usepackage{graphicx}
\usepackage{grffile}
\usepackage{array}

\usepackage{color}

\newcommand{\be}{\begin{equation}}
\newcommand{\ee}{\end{equation}}

\newcommand{\ifm}[1]{\relax\ifmmode#1\else$\mathsurround=0pt #1$\fi}
\newcommand{\kms}{\ifmmode\,{\rm km}\,{\rm s}^{-1}\else km$\,$s$^{-1}$\fi}

\newcommand{\ltsima}{$\; \buildrel < \over \sim \;$}
\newcommand{\lsim}{\lower.5ex\hbox{\ltsima}}
\newcommand{\gtsima}{$\; \buildrel > \over \sim \;$}
\newcommand{\gsim}{\lower.5ex\hbox{\gtsima}}

\newcolumntype{C}{ >{\centering\arraybackslash} m{2cm} }
\newcolumntype{D}{ >{\centering\arraybackslash} m{3.5cm} }

\definecolor{green}{rgb}{0,0.5,0}
\definecolor{grey}{rgb}{0.4,0.5,0.7}

\def\M11{M_{11}}
\def\V100{V_{100}}
\def\R1{R_{Mpc}}
\def\T6{T_6}

\begin{document}

\title{Bulge formation through disc instability - I. Stellar discs}
\author{T.~Devergne\inst{1}
\and 
A.~Cattaneo\inst{2,3}
\and 
F.~Bournaud\inst{4}
\and
I.~Koutsouridou\inst{2}
\and
A.~Winter\inst{5}
\and
P.~Dimauro\inst{6}
\and
G.~A.~Mamon\inst{3}
\and
W.~Vacher\inst{7}
\and
M.~Varin\inst{7}}
\institute{
Universit{\'e} Paris XI, 91405 Orsay, France \and
Observatoire de Paris/LERMA, PSL University, 61 av. de l'Observatoire, 75014 Paris, France \and
Institut d'Astrophysique de Paris, CNRS, 98bis Boulevard Arago, 75014 Paris, France \and
CEA/IRFU/SAp, 91191 Gif-sur-Yvette Cedex, France \and
Universit{\'e} de Cergy-Pontoise, 33 Boulevard du Port, 95011, Cergy, France \and
Observatório Nacional do Rio de Janeiro (ON), Rua Gal. José Cristino
77, São Cristóvão, 20921-400 Rio de Janeiro, RJ, Brazil\and
Observatoire de Paris/GEPI, PSL University, 61 av. de l'Observatoire, 75014 Paris, France 
}

\date{Received \emph{---} / Accepted \emph{---}}


\abstract{
We use simulations to study the growth of a pseudobulge in an isolated thin
exponential stellar disc embedded in a static spherical halo. 
We observe a transition from later to earlier morphological types and an increase in bar prominence  for higher disc-to-halo mass ratios, for lower disc-to-halo size ratios, and for lower halo concentrations.
{We compute bulge-to-total stellar mass ratios $B/T$ by fitting a two-component S{\'e}rsic-exponential surface-density distribution.}
The final $B/T$ is strongly related to the disc's fractional contribution $f_{\rm d}$
to the total gravitational acceleration at the optical radius. The
formula $B/T=0.5f_{\rm d}^{1.8}$ fits the simulations to an accuracy of $30\%$,
is consistent with observational measurements of $B/T$ and $f_{\rm d}$
as a function of luminosity,
and reproduces the observed relation between $B/T$ and stellar mass
when incorporated into the {\sc GalICS~2.0} semi-analytic model of
galaxy formation.
}


\titlerunning{Bulge formation through disc instability - I.}
\authorrunning{Bulge formation through disc instability - I.Devergne et al.}
\maketitle

\label{firstpage}

 
\section{Introduction}

According to the standard theory of galaxy formation, the dissipative infall of gas in the gravitational potential wells of  dark-matter (DM) haloes forms discs \citep{fall_efstathiou80};
elliptical galaxies are formed by mergers
\citep{toomre_toomre72}. 
Semianalytic models (SAMs) of galaxy formation build on this theory
and describe a galaxy as the sum
of two components: a disc and a bulge. 
Observers perform
a similar decomposition when they fit galaxies with the sum of an
exponential and a \citet{sersic63} profile to compute quantitative morphologies
\citep{simard_etal11,meert_etal15,meert_etal16,dimauro_etal18}.

This simplification brushes over the complexity and diversity of
galactic morphologies, for example, the distinction between normal and barred
spirals \citep{hubble26}.
{\citet{gadotti09}, \citet{weinzirl_etal09}, \citet{salo_etal15}, and \citet{erwin18,erwin19} have considered more detailed models that
decompose galaxies into three components:
a disc, a bulge, and a bar if present.
Despite the clear merit of such decompositions, this approach is possible only for relatively small samples of a few thousand nearby
galaxies, such as the Spitzer Survey of Stellar Structure in Galaxies (S$^4$G; \citealp{sheth_etal10}). 
It would be much more difficult to perform the same analysis on large samples, such as the Sloan Digital Sky Survey (SDSS), or on high-redshift data with even poorer spatial resolution\footnote{\citet{erwin18} compared the frequency of bars in the S$^4$G and the SDSS. He finds that SDSS-based studies underestimate the fraction of barred galaxies at low masses because of poor spatial resolution and the correlation between bar size and stellar mass.}.

Explaining the  broad statistical properties of galaxies in large surveys is the main purpose of SAMs.
If the goal is a detailed study of the morphological structure of galaxies, then
hydrodynamic simulations are a much better tool.}
If the observations that we aim to explain cannot distinguish between
different types of bulges, then it makes sense to compute the bulge-to-total
mass ratio $B/T$ in such a way that any stellar surface-density
excess above an
exponential fit is assigned to the bulge component, independently of its
origin, structure, and kinematics. That is not to say that all bulges
are the same.

 Many spiral bulges ressemble miniature ellipticals, especially those
in galaxies with stellar mass $M_\star\gsim 10^{11}\,M_\odot$
 \citep{fisher_drory11}. They are called classical bulges.
This similarity suggests a common formation mechanism.
 In the earliest SAMs (for example,
 \citealp{kauffmann_etal93}), all bulges were formed through mergers.
 Some of them never accreted any gas. We call them elliptical galaxies.
 Others regrew a disc and became spiral galaxies \citep{baugh_etal96}.

In this picture, most spirals should be bulgeless because
mergers make a negligible contribution to the mass growth of galaxies
with $M_\star< 10^{11}\,M_\odot$ \citep{cattaneo_etal11}.
Bulgeless galaxies are observed \citep{kormendy_etal10}, but they do not
constitute the majority of the spiral population unless we include
dwarf galaxies. \citet{fisher_drory11} find that the fraction of galaxies in which a bulge is detected increases smoothly
from $\sim 20\%$ at  $M_\star=10^{9}\,M_\odot$ to $\sim 100\%$ at
$M_\star=10^{10.7}\,M_\odot$.

Most of the bulges in spirals
with $M_\star< 10^{11}\,M_\odot$ are pseudobulges with different
kinematics than elliptical galaxies and do not follow the
fundamental plane \citep{kormendy82,kormendy93,kormendy_kennicutt04}.
To explain these systems, SAMs began to incorporate a second formation mechanism:
disc instabilities
\citep{cole_etal00,hatton_etal03,shen_etal03}.

Self-gravitating thin discs are dynamically unstable \citep{hohl71,kalnajs72}.
The observation of disc galaxies despite such instability has been one of the historical arguments for DM  \citep{ostriker_peebles73}.
Morphological features, such as spiral arms, bars, peanut-shaped boxy
pseudobulges, rings, and ovals, show that haloes do not completely stabilise discs, however.

\citet{combes_sanders81}  used N-body simulations to study the
stability of a truncated \citet{toomre63} disc and demonstrated that, if the
mass of the disc was larger than the mass of the DM within its
maximum radius, a persistent bar developed quickly and, after some time, took a more or
less pronounced peanut shape  when seen edge-on.
Observations of peanut-shaped pseudobulges confirm that they are
connected with bars and owe their origin to them \citep{kormendy_kennicutt04}.
The stronger conclusion that peanut-shaped pseudobulges  are nothing more nor less than bars seen edge-on
\citep{combes_sanders81,combes_etal90,pfenniger_friedli91,berentzen_etal98,athanassoula_misiriotis02,athanassoula03}
is less straightforward from an observational standpoint, but observations of galaxies such as
NGC~7582, where the bar is very flat and three times longer than the
pseudobulge \citep{quillen_etal97}, are consistent with a picture
in which the peanut is the vertical extension of a longer, flatter bar
\citep{athanassoula05,wegg_etal15}.

The literature above demonstrates that classical bulges and
pseudobulges are different entities. {\sc GalICS~2.0}
\citep{cattaneo_etal17} has been the
first (and to date only) SAM 
to treat them as
separate components. More detailed investigations aimed at separating pseudobulges from
bars or at distinguishing different types of pseudobulges\footnote{Not all pseudobulges are peanut-shaped. Some are nuclear
spirals and they are as flat as discs \citep{carollo_etal98,kormendy_kennicutt04}. However, the
distribution of S{\'e}rsic index, ellipticity and $B/T$ is
similar for pseudobulges in barred, oval, and normal galaxies,
suggesting that the difference between classical bulges and pseudobulges is more
important than the one between different types of pseudobulges
\citep{fisher_drory08}.
Moreover, \citet{athanassoula05} has shown that disc-like bulges
result from the dissipational inflow of gas into the central regions of spiral
galaxies. Hence, it is reasonable for us to neglect them 
in an article that is about stellar discs.}
are beyond the scope of SAMs\footnote{We acknowledge that a bar and a bulge are different dynamical entities, and that it is possible to separate in a morphological decomposition \citep{gadotti09,weinzirl_etal09,salo_etal15,erwin18,erwin19}. We merely state that SAMs are not an appropriate tool for such decomposition.}.

\citet*[ELN]{efstathiou_etal82} extended the analysis by \citet{combes_sanders81}
to the more realistic case of an exponential disc and found a condition for the
circular velocity $V_{\rm c}$ at $2.2$ exponential scale-lengths, where the rotation curve of a self-gravitating exponential disc peaks
\citep{freeman70}. A thin exponential stellar disc embedded in a
static spherical DM halo becomes unstable and develops a bar if
\begin{equation}
V_{\rm c}(2.2R_{\rm d})<\epsilon\sqrt{{\rm G}M_{\rm d}\over R_{\rm d}},
\label{efstathiou}
\end{equation}
where $M_{\rm d}$ is the disc mass, $R_{\rm d}$ is the exponential
scale-length and $\epsilon=1.1$.
\citet{christodoulou_etal95} used analytic arguments to conclude that a similar criterion with $\epsilon=0.9$ should apply to gaseous discs.

Since \citet{mo_etal98} and \citet{vandenbosch98,vandenbosch00},
the ELN criterion (Eq.~\ref{efstathiou}) has provided the standard description of disc
instabilities that all current SAMs adopt
({\sc Galacticus}: \citealp{benson12}; {\sc
  GalICS~2.0}: \citealp{cattaneo_etal17}; {\sc Morgana}:
\citealp{lofaro_etal09}; {\sc Sag}: \citealp{gargiulo_etal15}; {\sc
  SantaCruz}: \citealp{porter_etal14}; {\sc ySAM}: \citealp{lee_yi13};
{\sc Galform}: \citealp{gonzalez_etal14}; {\sc Lgalaxies}:
\citealp{henriques_etal15}; {\sc Sage}: \citealp{croton_etal16}).

More realistic simulations
\citep{athanassoula08} found that,
even in cases where the criterion predicts stability, a bar can still
form if resonances destabilise the disc by transferring angular
momentum to the halo (ELN's assumption of a static halo prevents this
possibility in their simulations). Moreover,
in cases where the ELN criterion predicts instability, the disc can
still be stabilised by random motions, which ELN did not consider
because of the assumption of thin discs.

Several reasons explain why SAMs have kept using the ELN criterion
despite these criticisms:
\begin{itemize}
\item[$\bullet$]{The goal of SAMs is to explain the global properties of galaxies (such
as the trend of $B/T$ with $M_\star$) 
in a cosmological context. 
A detailed description of
galactic dynamics is beyond their scope.}
\item[$\bullet$]{SAMs separate the formation of haloes from the evolution of baryons
    within haloes.  In
    reality, baryons affect the radial distribution of subhaloes
    within groups \citep{libeskind_etal10} as well as the density profiles
    \citep{pontzen_governato12,maccio_etal12,teyssier_etal13,dicintio_etal14,tollet_etal16},
    spins, and shapes \citep{bryan_etal13} of DM haloes,
    and the clustering of galaxies on scales as large as $\sim 1\,$Mpc
    \citep{vandaalen_etal14}.
    Assuming a static halo is a simplification,
    but it is consistent with the semi-analytic approximation.}
 \item[$\bullet$]{From the perspective of SAMs, the ELN criterion is first
     and foremost a criterion for the ratio of the disc mass $M_{\rm d}$
     to the halo mass $M_{\rm vir}$. This ratio is largely determined
     by feedback processes
     \citep{dekel_silk86,silk_rees98,brook_etal12,angles_etal17,tollet_etal19} that
     cannot be modelled accurately. Disc sizes are based on the crude
assumption \citep{kimm_etal11,stewart_etal13,jiang_etal19} that  angular
momentum is conserved \citep{mo_etal98} and they, too, affect the
circular velocity in Eq.~(\ref{efstathiou}).
The errors from these uncertainties are likely to be more
significant than those from the 
modelling of disc instabilities themselves.}
\item[$\bullet$]{A thick-disc component with higher velocity dispersion makes
    discs more stable \citep{toomre64} and there is observational
    evidence that thick discs are ubiquitous
    \citep{yoachim_dalcanton06,comeron_etal11}, but what is their origin?
    If it is secular heating of the thin disc
    \citep{villumsen85,villalobos_helmi08,schoenrich_binney09,steinmetz12},
    then it is logical that simulators should not put in their
    inititial conditions the effects of processes they want to simulate.
    If it is interactions with satellites
    \citep{quinn_etal93,abadi_etal03}, then it is legitimate to
    neglect them in a model for the evolution of isolated galaxies that
    have not experienced any merger.
    A third possibility is that thick discs are relics from gas-rich,
    turbulent, clumpy discs at high redshift and that thin discs
    formed later \citep{brook_etal04,bournaud_etal09}.
    The problem is that there is no SAM to compute disc scale-heights
    (but see \citealp{efstathiou00} for an attempt in this direction).
    In absence of physical arguments for one scale-height or another,
    the simplest assumption (thin discs) is the most reasonable.}
\end{itemize}

The main problem is another.
The ELN criterion tells us whether a disc is likely to become unstable
but not the mass of the bar or pseudobulge that is likely to form because
of that instability. To compute this mass, SAMs must incorporate
additional assumptions.
Cole et al. (2000; also see \citealp{gonzalez_etal14} and
\citealp{gargiulo_etal15}) considered an extreme model in which discs evolve into bulges whenever the
instability criterion in Eq.~(\ref{efstathiou}) is satisfied.
\citet{hatton_etal03} and \citet{shen_etal03} proposed a more conservative model in which matter is transferred from the disc
to the bulge so that $M_{\rm d}$ decreases until the disc becomes
stable again.
In discs where $V_{\rm c}(2.2R_{\rm d})$ falls just slightly short of
the critical value required for stability, the results obtained with
the two methods vary wildly. A galaxy for which $B/T=1$ in the first
model may have $B/T=0.1$ in the second one.
Therefore, even if the ELN criterion were fully reliable, 
its implications for galactic morphologies would still be considerably uncertain.

In this article, we present a series of simulations in which we follow
the evolution of a thin exponential stellar disc embedded in  a static
spherical \citet*[NFW]{navarro_etal97} halo.
Our computational set-up is very similar to the one by ELN, although
we explore a larger space of parameters and have better resolution.
Unsurprisingly, our simulations confirm ELN's previous findings, but that is not the purpose of our research.
Our question is: as different SAMs make very
different predictions for $B/T$ even if they are all based on the ELN
criterion, which approach (if any) agrees better with the values of $B/T$ measured
in the simulations used to establish the ELN criterion?

As any approximations in the simulations are passed on to SAMs through the ELN criterion, the agreement of a SAM with the simulations is no guarantee of its
being a correct description of disc instabilities, but, at least, it proves self-consistency (the SAM faithfully reproduces any biases
of the simulations).
On the contrary,  lack of agreement indicates that the SAM is adding biases of its own on top
of those already present in the simulations. This is why it is high time for a sanity check and, possibly, new
prescriptions that may improve those used to calculate $B/T$ in SAMs.

The structure of the article is as follows.
In Section~2, we describe our initial conditions, which are entirely specified by three parameters 
(the ratio $r_{\rm d}=R_{\rm d}/R_{\rm vir}$ of the disc scale-length $R_{\rm d}$ to the virial radius $R_{\rm vir}$,
the ratio $m_{\rm d}=M_{\rm d}/M_{\rm vir}$ of the disc mass $M_{\rm d}$ to the total mass $M_{\rm vir}$ within the virial radius, 
and the concentration $c$ of the DM halo), the explored parameter
space and the computational strategy
(we use the adaptive-mesh-refinement [AMR] code {\sc ramses};
\citealp{teyssier02}).
In Section~3, we present our findings for the dependence of $B/T$ on
$r_{\rm d}$, $m_{\rm d}$, and $c$.
{In Sections~4 and 5, we compare our results with previous models and}
the observed
morphology--luminosity relation, {respectively.}
In Section~6 we explore the effects of incorporating our findings
into the {\sc GalICS~2.0} SAM and their implications for galactic morphologies.
Section~7 discusses our results and summarises the conclusions of the article.

\section{Computational set-up}

\subsection{Initial conditions}

We make our problem dimensionless by expressing all lengths in units of the virial radius $R_{\rm vir}$,
all masses in units of the virial mass $M_{\rm vir}$ and all speeds in units of the virial velocity $V_{\rm vir}=\sqrt{{\rm G}M_{\rm vir}/R_{\rm vir}}$.
Owing to the axial symmetry of our initial configuration, we adopt cylindrical coordinates $(r,z,\phi)$, where $z$ is the direction of the disc's rotation
axis. Here and throughout this article, upper and lower-case letters refer to dimensional and dimensionless quantities, respectively.
As $M_{\rm vir}$ is the total mass within $R_{\rm vir}$, $m_{\rm d}=M_{\rm d}/M_{\rm vir}$ and $1-m_{\rm d}$ are the dimensionless masses of the disc and the
DM halo, respectively. 

The dimensionless mass of the DM within a sphere of radius $r$ is given by:
\begin{equation}
m_{\rm DM}(r)=(1-m_{\rm d}){f(cr)\over f(c)},
\end{equation}
where $f(x) = \ln(1+x)-x/(1+x)$ and $c$ is the concentration of the NFW profile.

\begin{figure}
\includegraphics[width=1.05\hsize]{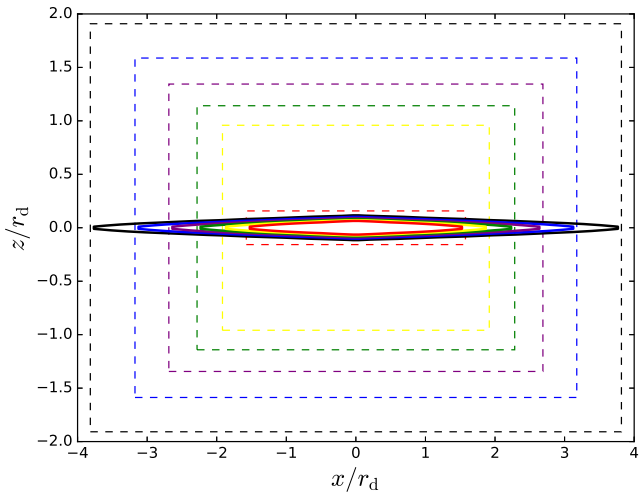} 
\caption{Initial conditions and refinement regions.
The black, blue, violet, green, yellow, and red curves show the isodensity contours that contain 90$\%$, 80$\%$, 70$\,\%$, 60$\%$, 50$\%$, and 40$\%$ of the disc mass at $t=0$, respectively.
The black, blue, violet, green, yellow, and red dashed lines show the cylinders within which the cell size equals 1/8, 1/16, 1/32, 1/64, 1/128, and 1/256 of the disc exponential scale length, respectively.}
\label{in_cond}
\end{figure}

We assume that the disc is exponential and isothermal in the vertical
direction (for example, \citealp{villumsen85}; \citealp{efstathiou00}).
These assumptions give the density distribution:
\begin{equation}
\rho(r,z)={1\over 2h}{\rm sech}^2\left({z\over h}\right){m_{\rm d}\over 2\pi r_{\rm d}^2}e^{-{r\over r_{\rm d}}},
\label{rho}
\end{equation}
where $h$ is disc's vertical scale-length (all quantities in Eq.~\ref{rho} are adimensional).
As our goal is to study the stability of thin discs (discussion in
Section~4), we run all our simulations for $h/r_{\rm d}=0.044$. {
This value  is small but not unrealistic. \citet{bland_ortwin16} find that the Milky
  Way has a thin-disc vertical scale-height of 220 -- 450{\rm\,pc}
  and an exponential scale-length of $\sim 2.5\,$kpc.
  Hence, for the Milky Way, $h/r_{\rm d}$ is in the range  0.088 -- 0.18. }
Fig.~\ref{in_cond} shows the isodensity contours that contain 40$\%$, 50$\%$, 60$\,\%$, 70$\%$, 80$\%$, and 90$\%$ of the disc mass for our initial configuration.
 
Eq.~(\ref{rho}) is used to generate random coordinates for a million stellar particles within the optical radius $r_{\rm opt}=3.2r_{\rm d}$ that contains $83\%$ of the mass of the disc
(\citealp{fujii_etal11} demonstrated that numerical heating through close encounters becomes negligible when the disc is resolved with at least a million particles).
All stellar particles have equal mass. This article is on stellar
discs, but our discs contain a small amount of gas to pave the way for a second article
on the formation of bulges in discs with gas.
{We use a gas fraction of 
  2$\%$ because in massive low-redshift galaxies the gas fraction
  varies from close to zero to a few percent  and is rarely larger
  than 5--10$\%$ (for example, \citealp{combes_etal13}),
but we also ran four simulations without gas to check the
  impact that even a small gas fraction could have on our results.}
The gas in our simulations is isothermal at $10^4\,$K and is not allowed to form stars.
The total mass of the stellar particles is $0.98\times 0.83\,m_{\rm d}$ in dimensionless virial units.

The circular velocity $v_{\rm c}(r)$ is the speed that a star must have to be on a circular orbit of radius $r$. Its value is the sum in quadrature of the contributions from
the disc and the halo:
\begin{equation}
v_{\rm c}^2=v_{\rm d}^2+v_{\rm h}^2, 
\label{vc}
\end{equation}
where $v_{\rm d}(r)$ is computed as in \citet{freeman70} and
\begin{equation}
v_{\rm h}^2={m_{\rm DM}(r)\over r}.
\end{equation}

Stars have velocity dispersion:
\begin{equation}
\sigma^2(r)=\pi h{m_{\rm d}\over 2\pi r_{\rm d}^2}e^{-{r\over r_{\rm d}}}
\label{sigma}
\end{equation}
determined from Eq.~(\ref{rho}) through the requirement that our initial condition should be in equilibrium (albeit unstable).
Hence, their velocities:
\begin{equation}
{\bf v}={\bf v}_{\rm rot}+\Delta{\bf v}
\label{vp}
\end{equation}
will be the sum of an ordered rotational component (oriented as $\hat{\bf e}_\phi$)
and a random deviate from a Maxwellian distribution with velocity
dispersion $\sigma$.
{The assumption of an isotropic velocity dispersion is
motivated by simplicity, but it is not unreasonable, since the radial
and vertical velocity dispersions in the Solar Neighbourhood are
$(35\pm 5){\rm\,km\,s}^{-1}$ and $(25\pm 5){\rm\,km\,s}^{-1}$,
respectively \citep{bland_ortwin16}.}

The rotation speed $v_{\rm rot}(r)=\langle v_\phi(r)\rangle$ equals the circular velocity $v_{\rm c}(r)$
only for $\sigma=0$. For $\sigma>0$, $v_{\rm rot}$ and $v_{\rm c}$ are linked by the condition:
\begin{equation}
\langle(v_{\rm rot}+\Delta{v}_\phi)^2\rangle=v_{\rm c}^2,
\label{vrot_vc}
\end{equation}
which gives:
\begin{equation}
v_{\rm rot}^2=v_{\rm c}^2-\sigma^2, 
\label{vrot}
\end{equation}
since $\langle\Delta v_\phi\rangle = 0$ and $\langle\Delta v_\phi^2\rangle = \sigma^2$.

\begin{table}
\begin{center}
\caption{Simulations with relaxed initial conditions: parameter values and difference in $B/T$ with respect to the unrelaxed simulations.}
\begin{tabular}{ ccrr}
\hline
\hline 
$\lambda$         & $m_{\rm d}$ &  $c$ & $\Delta{B\over T}\,(\%)$        \\
 \hline
0.025	& 	0.01 	&5 & $-7$ \\
0.025	&	0.04	&5 & $-17$\\
0.050	&	0.02	&5& $+2$\\
0.025	&	0.01	&15& $-12$\\
0.025	&	0.04	&15&$+7$\\
0.050	&	0.02	&15&$-8$\\
\hline
\hline
\end{tabular}
\end{center}
\label{model_parameters1}
\end{table}

\begin{table}
\begin{center}
\caption{Simulations with a live halo: parameter values and difference in $B/T$ with respect to those with a static halo.}
\begin{tabular}{ ccrr}
\hline
\hline 
$\lambda$         & $m_{\rm d}$ &  $c$ & $\Delta{B\over T}\,(\%)$        \\
 \hline
0.025	& 	0.01 	&5 & $+9$ \\
0.1	        &	0.04	&5 & $+14$\\
0.025	&	0.01	&10& $+7$\\
0.1	       &	0.04	&10&$+18$\\
\hline
\hline
\end{tabular}
\end{center}
\label{model_parameters2}
\end{table}
\begin{table}
\begin{center}
\caption{Simulations without gas: parameter values and difference in $B/T$ with respect to those with $f_{\rm gas}=0.02$.}
\begin{tabular}{ rrrr}
\hline
\hline 
$\lambda$         & $m_{\rm d}$ &  $c$   & $\Delta{B\over T}\,(\%)$        \\
  \hline
0.011& 0.01 & 5 &-4\\
0.011& 0.01 & 10&-9\\
0.025 & 0.04 & 10&-1\\
0.025& 0.04 & 15&+3\\
\hline
\hline
\end{tabular}
\end{center}
\label{model_parameters3}
\end{table}
Our rotation speeds and thus our particle velocities are computed using Eq.~(\ref{vrot})
everywhere except in a small central region where we set $v_{\rm rot}=0$ because $v_{\rm c}<\sigma$.
This central region has $r\ll r_{\rm d}$ by construction, since the normalisation of $\sigma(r)$ is determined by the disc scale-height (Eq.~\ref{sigma}) and we have assumed that $h\ll r_{\rm d}$.
{However, the fact that $\sigma$ is computed considering the vertical equilibrium only and that
$3\sigma^2>v_{\rm c}^2$ at $r\lsim 0.07\,r_{\rm d}$ implies that the central region will expand a little bit when the initial conditions are allowed to relax.}

{The global stability of our initial conditions is explored through the simulations presented in this article (Section 3).
Their local stability is a different problem. The discussion of how local disc instabilities can affect the final $B/T$ of globally unstable discs is postponed to Section 4.}

\subsection{Parameter space}

Having fixed the disc scale height $h=0.044\,r_{\rm d}$ and the gas fraction $f_{\rm gas}=0.02$, our initial conditions are entirely determined by three parameters: $r_{\rm d}$,
$m_{\rm d}$, and $c$.

\begin{figure*}
\begin{center}
\begin{tabular}{ C D D D D }
&$\lambda=0.011$&$\lambda=0.025$&$\lambda=0.05$&$\lambda=0.1$\\
$m_{\rm d}=0.005$&
 \includegraphics[width=1.122\hsize]{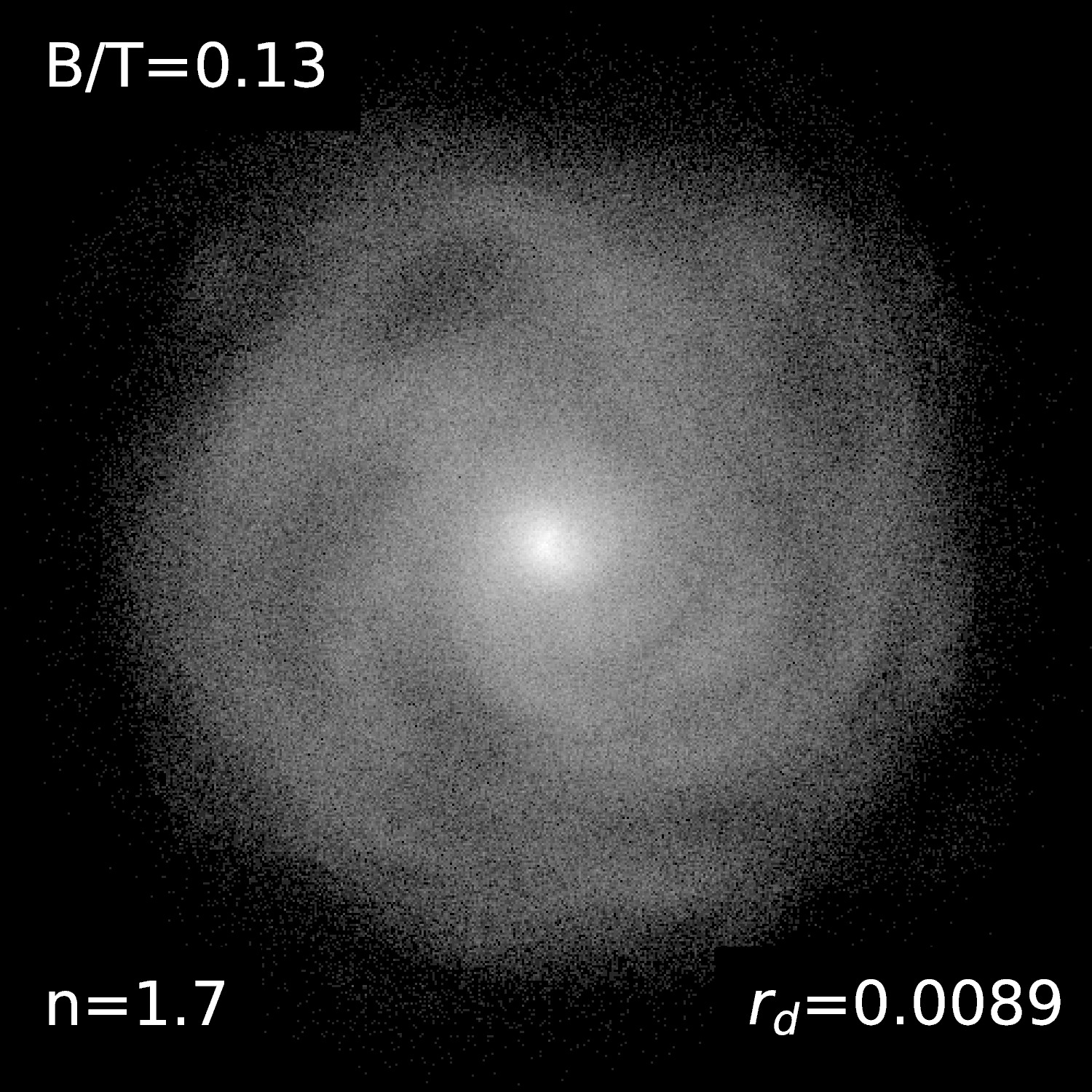}&&&\\
$m_{\rm d}=0.01$&
\includegraphics[width=1.122\hsize]{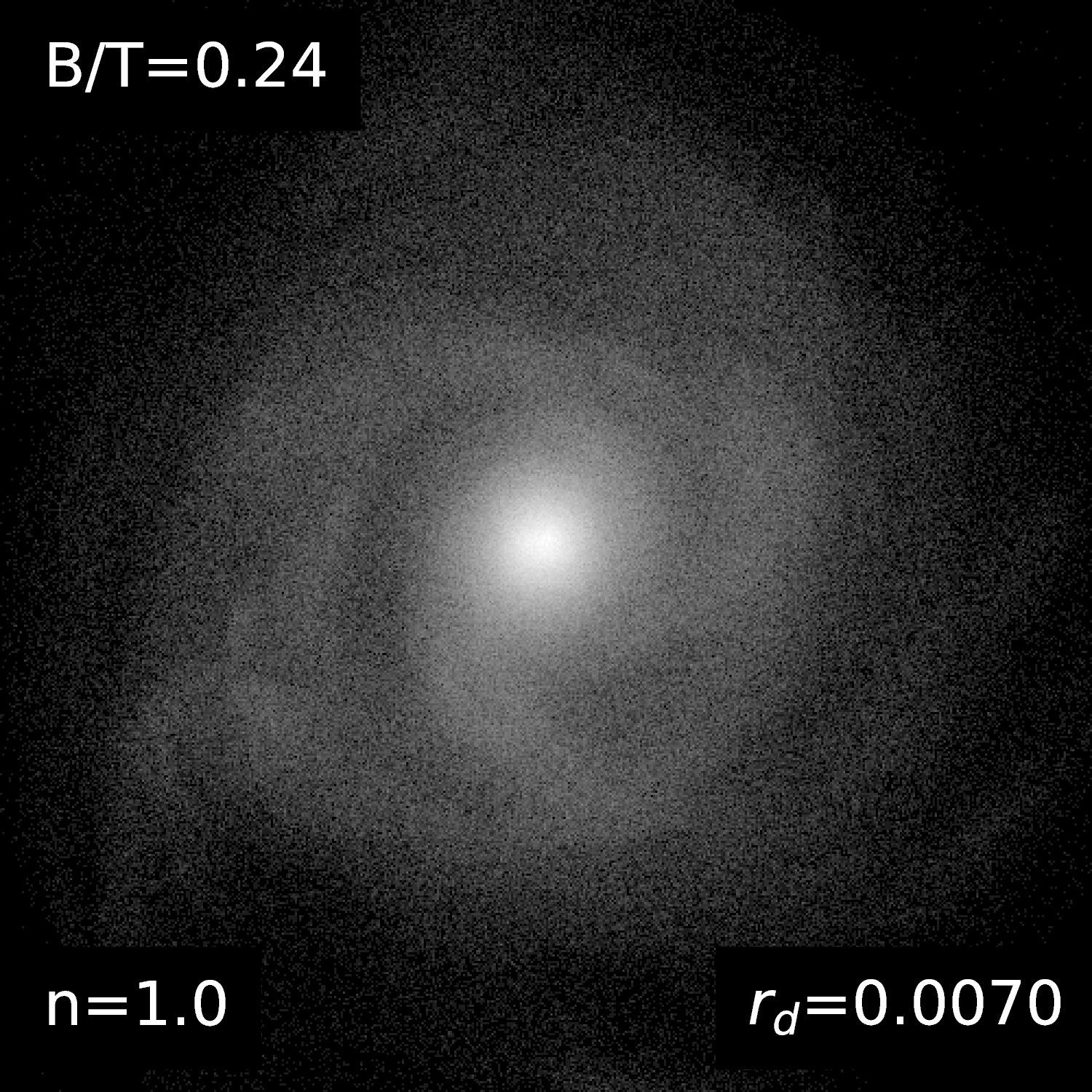}&
\includegraphics[width=1.122\hsize]{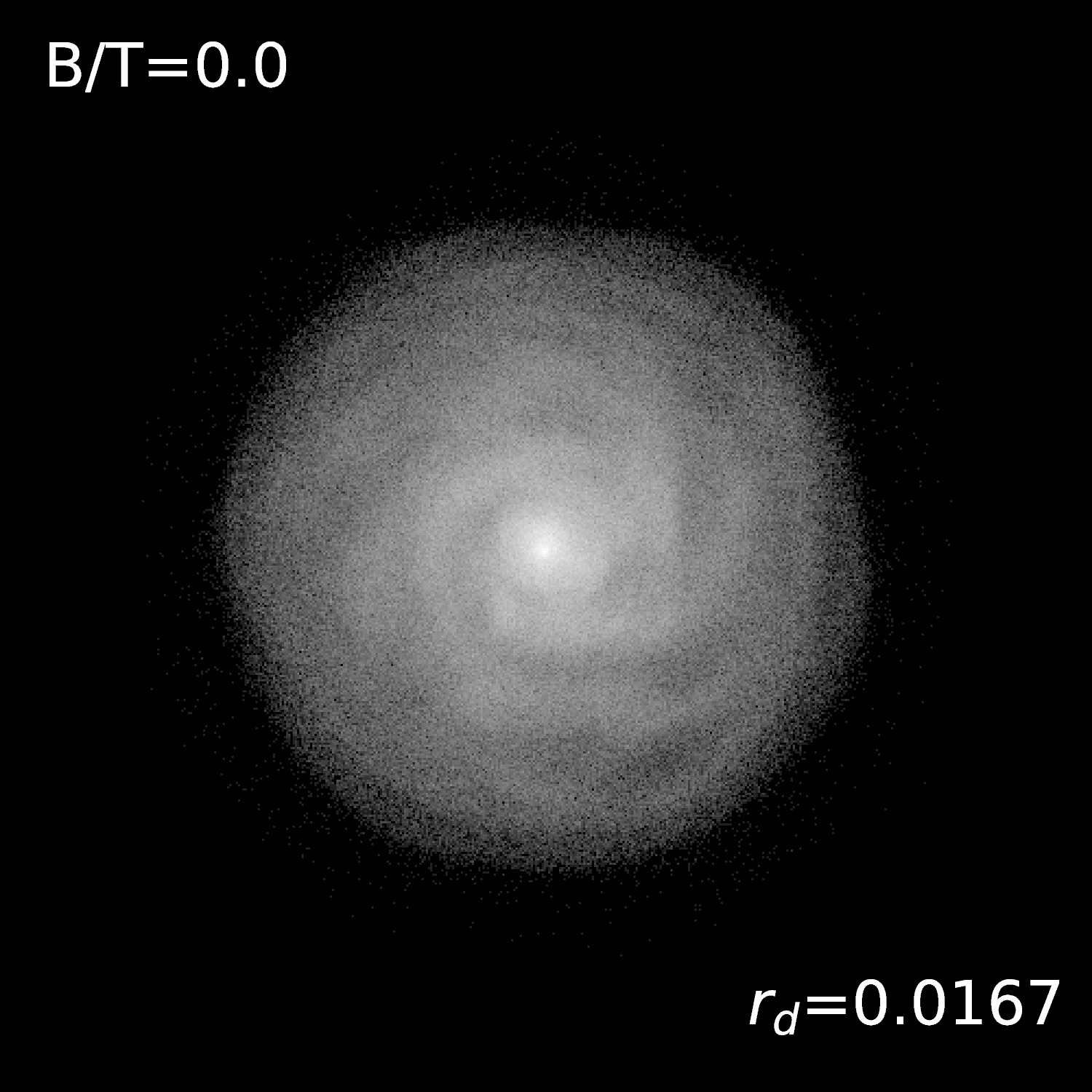}&&\\
$m_{\rm d}=0.02$& 
\includegraphics[width=1.122\hsize]{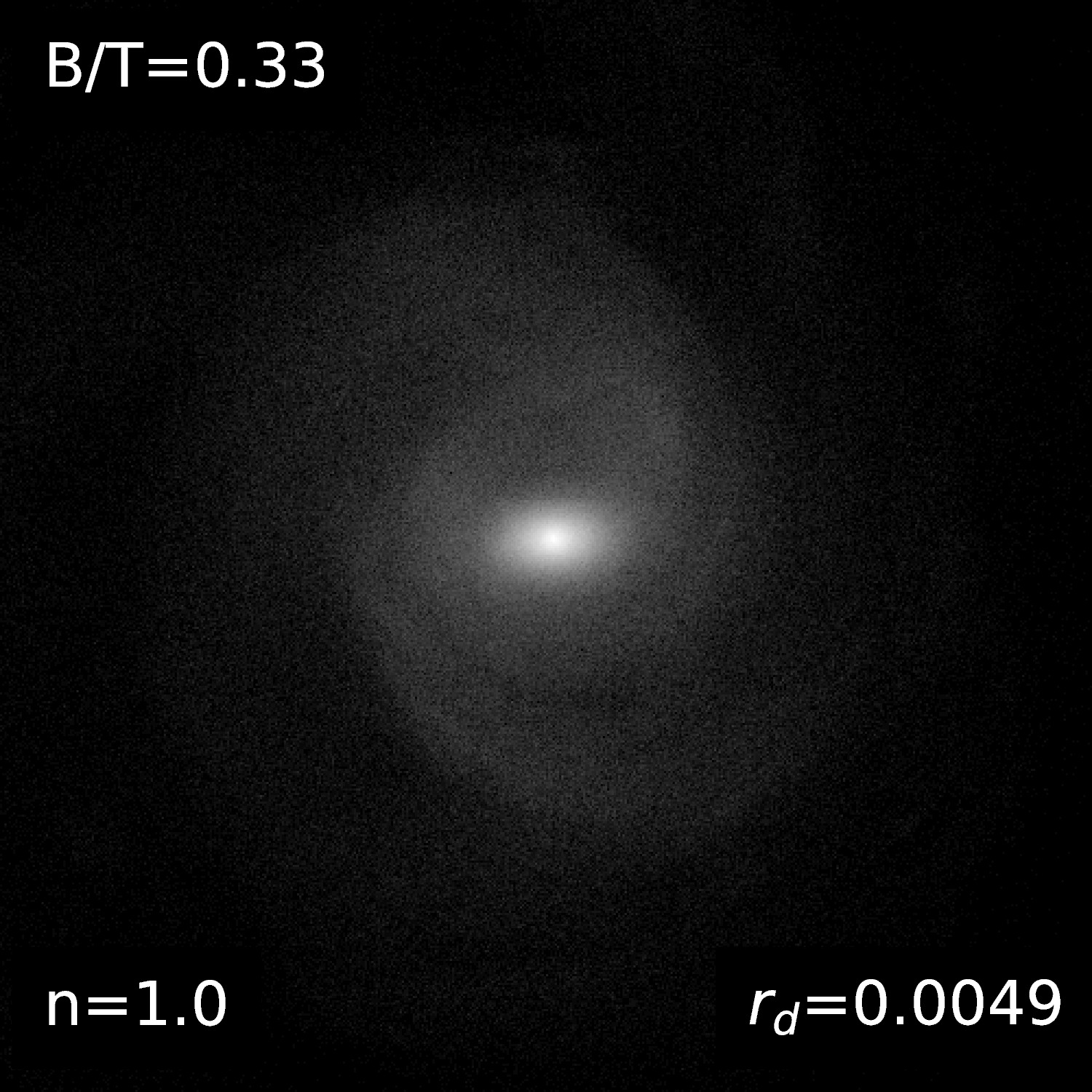}&
\includegraphics[width=1.122\hsize]{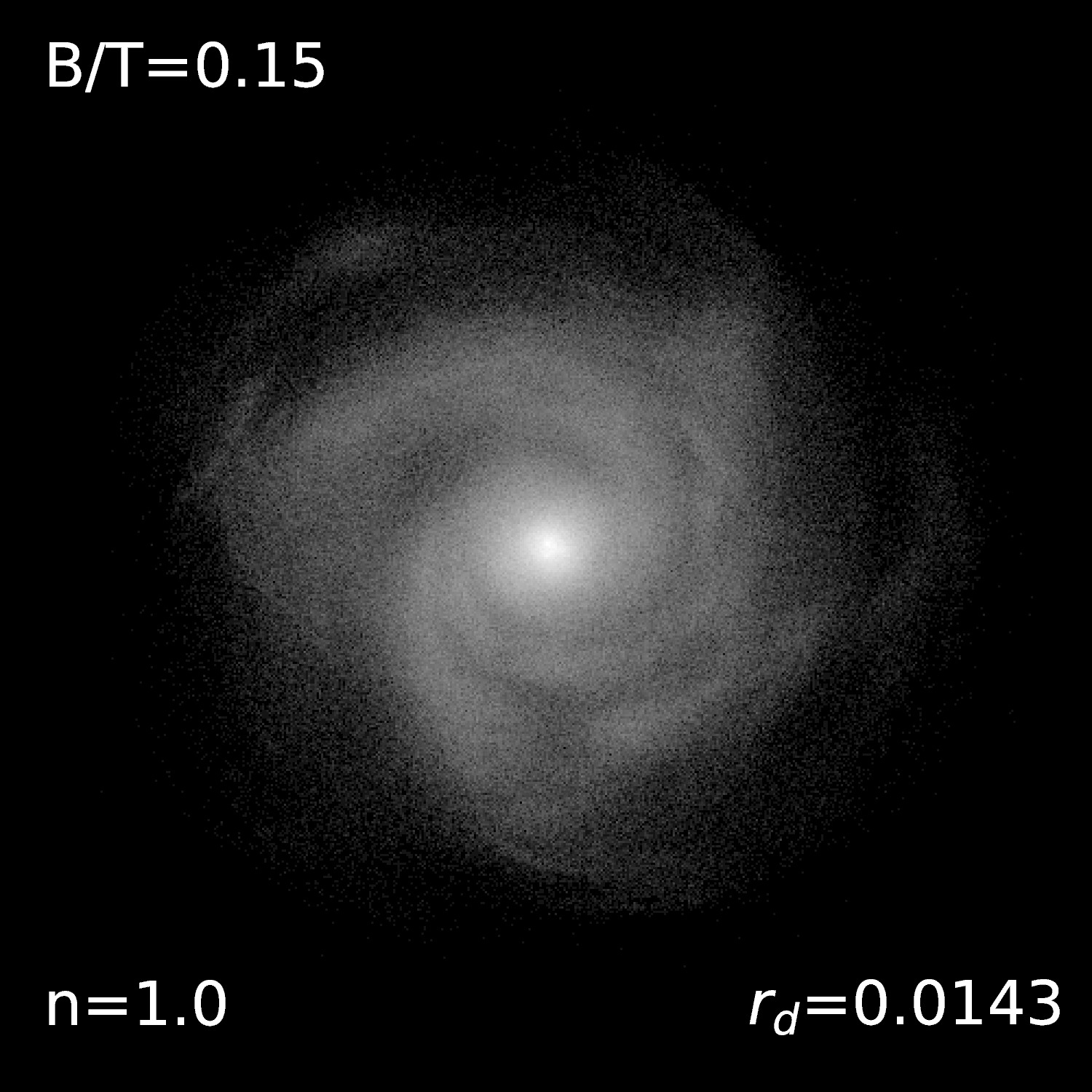}& 
\includegraphics[width=1.122\hsize]{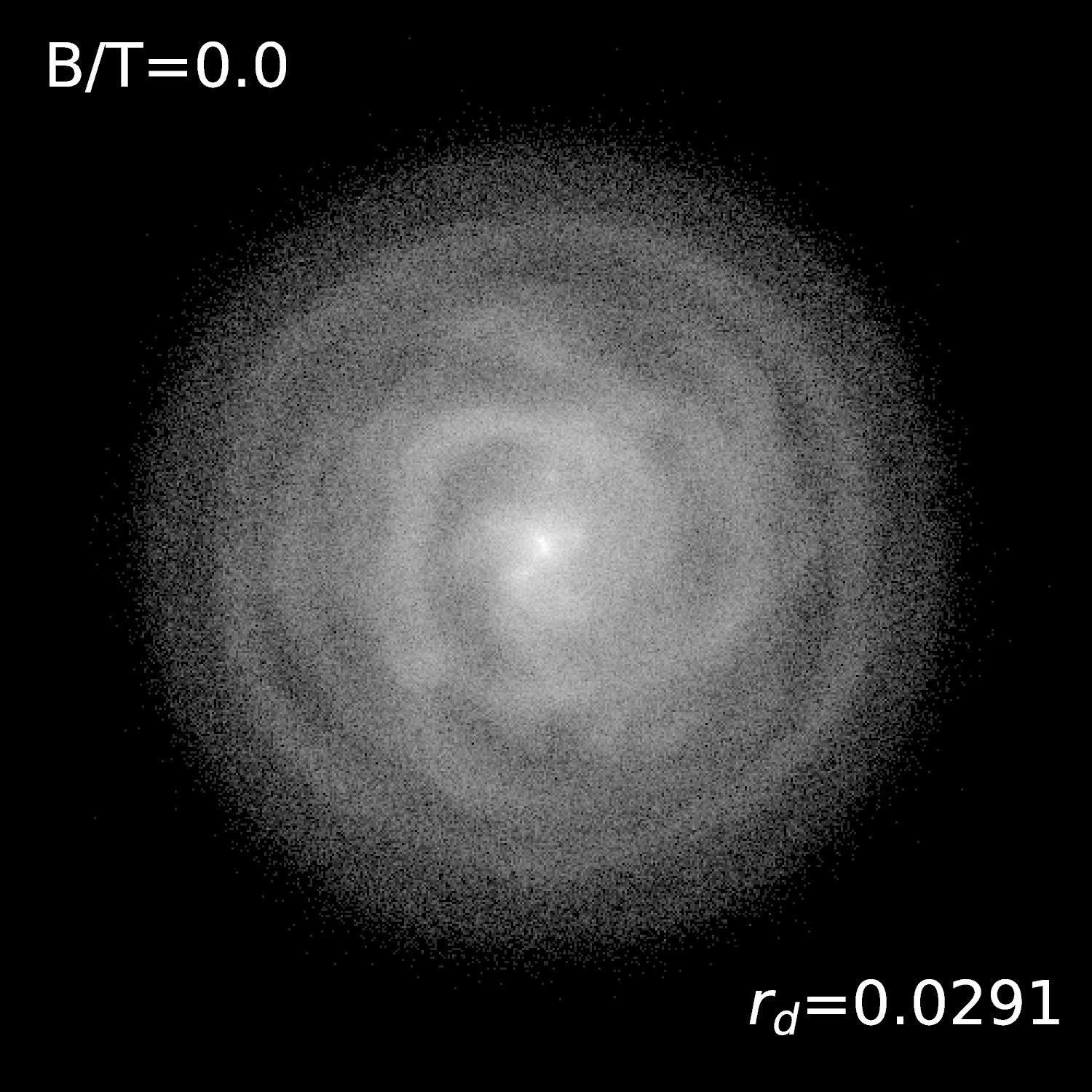}&\\ 
$m_{\rm d}=0.04$&
\includegraphics[width=1.122\hsize]{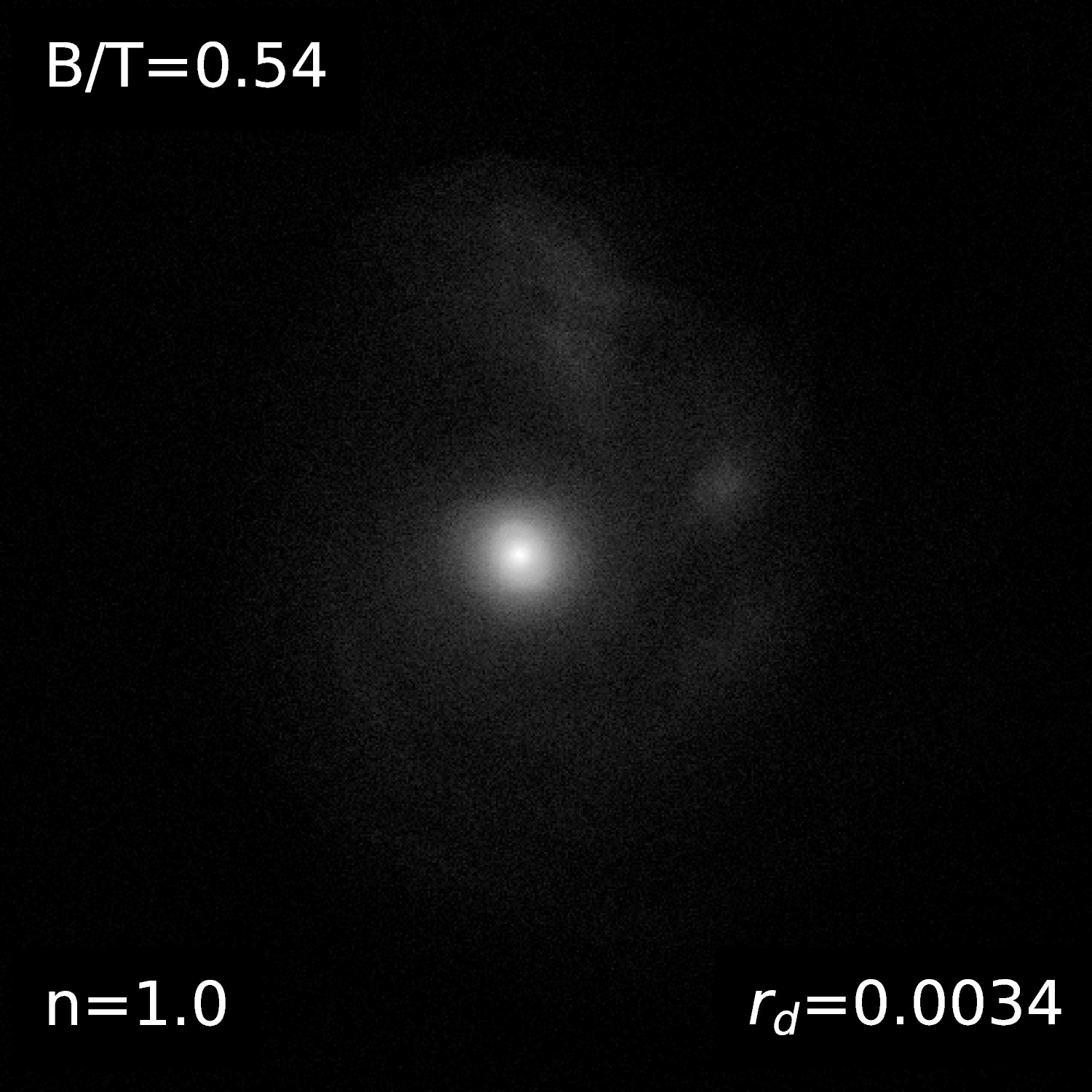}&
\includegraphics[width=1.122\hsize]{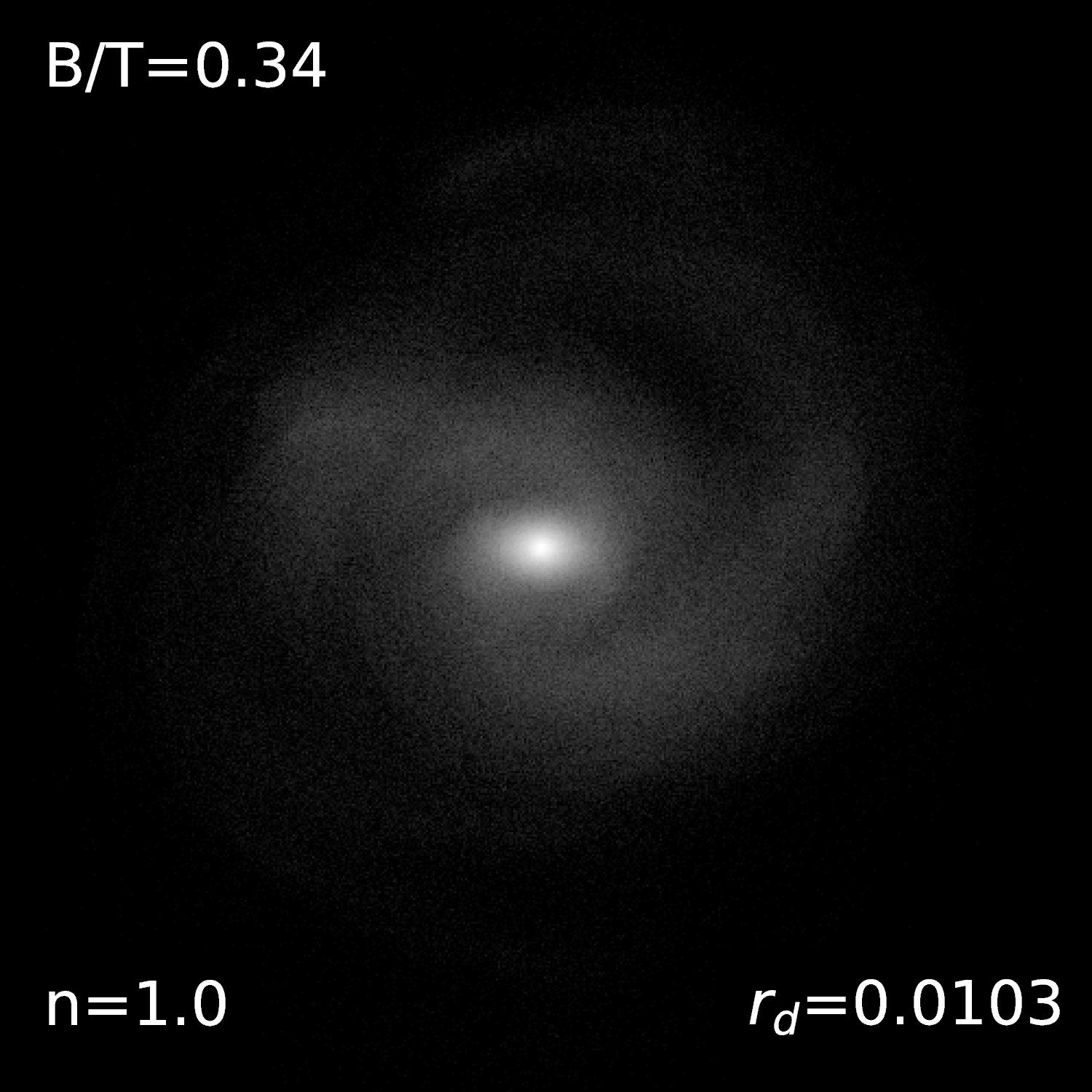}& 
\includegraphics[width=1.122\hsize]{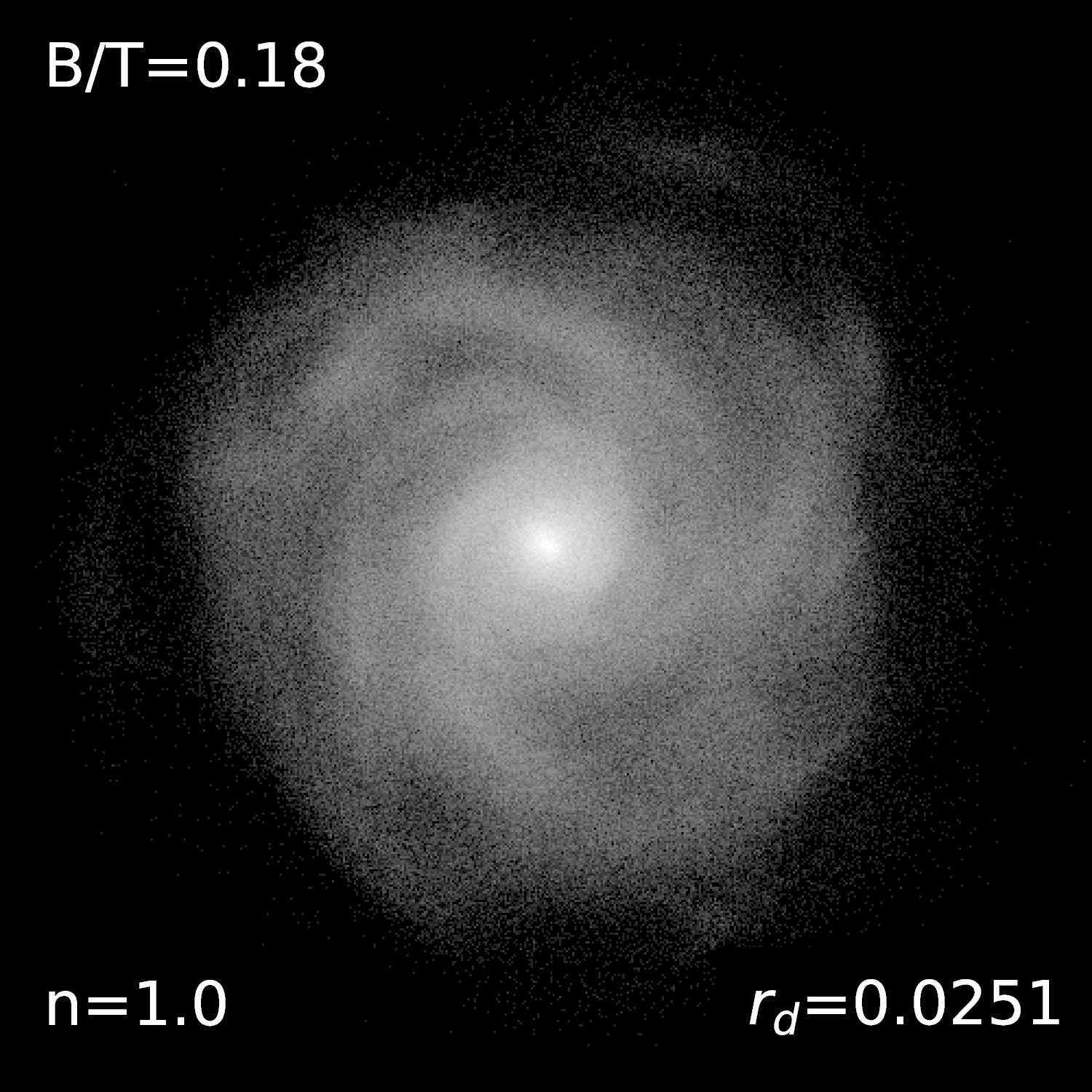}&
\includegraphics[width=1.122\hsize]{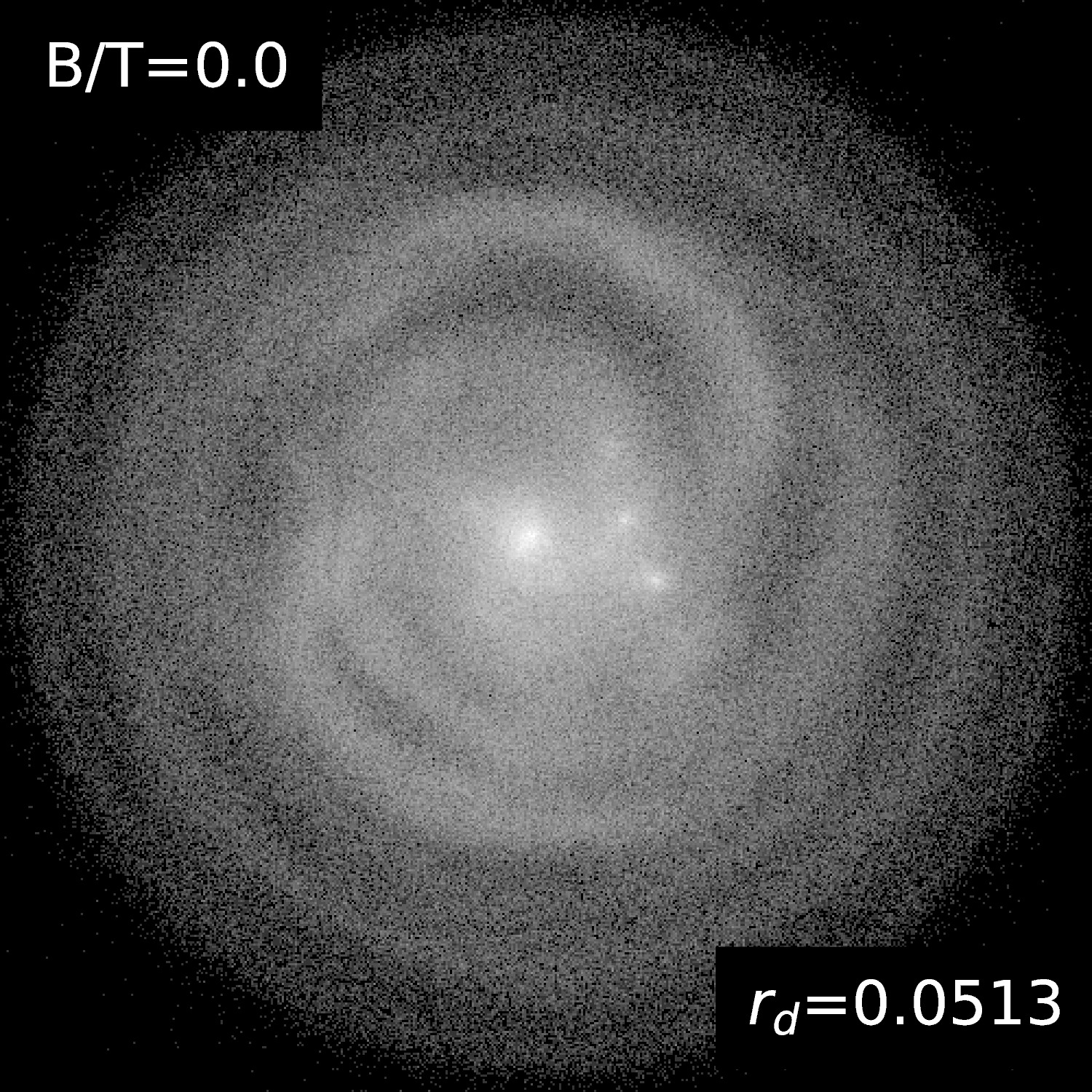}\\
\end{tabular}
\end{center}
\caption{Face-on view of the simulated galaxies at the first time $t$
  when $B/T$ has stopped growing ($2{\rm\,Gyr}<t<3{\rm\,Gyr}$). The image brightness is proportional to the logarithm of stellar surface density.
This figure show the simulations with $c=5$, sorted by $m_{\rm
  d}=M_{\rm d}/M_{\rm vir}$ (lines) and halo spin $\lambda$ (columns).
{On each image, we have shown the corresponding value of $r_{\rm
    d}=R_{\rm d}/R_{\rm vir}$.}
Bulge-to-total stellar mass ratios $B/T$ have been computed by decomposing the stellar surface-density distribution into an exponential and a \citet{sersic63} profile.
Galaxies that could be fitted with a single exponential profile have $B/T=0$.
When the fit required a bulge component, its S{\'e}rsic index $n$ has
been shown.
\label{conc5}
}
\end{figure*}

\begin{figure*}
\begin{center}
\begin{tabular}{ C D D D D }
&$\lambda=0.011$&$\lambda=0.025$&$\lambda=0.05$&$\lambda=0.1$\\
$m_{\rm d}=0.005$&
\includegraphics[width=1.122\hsize]{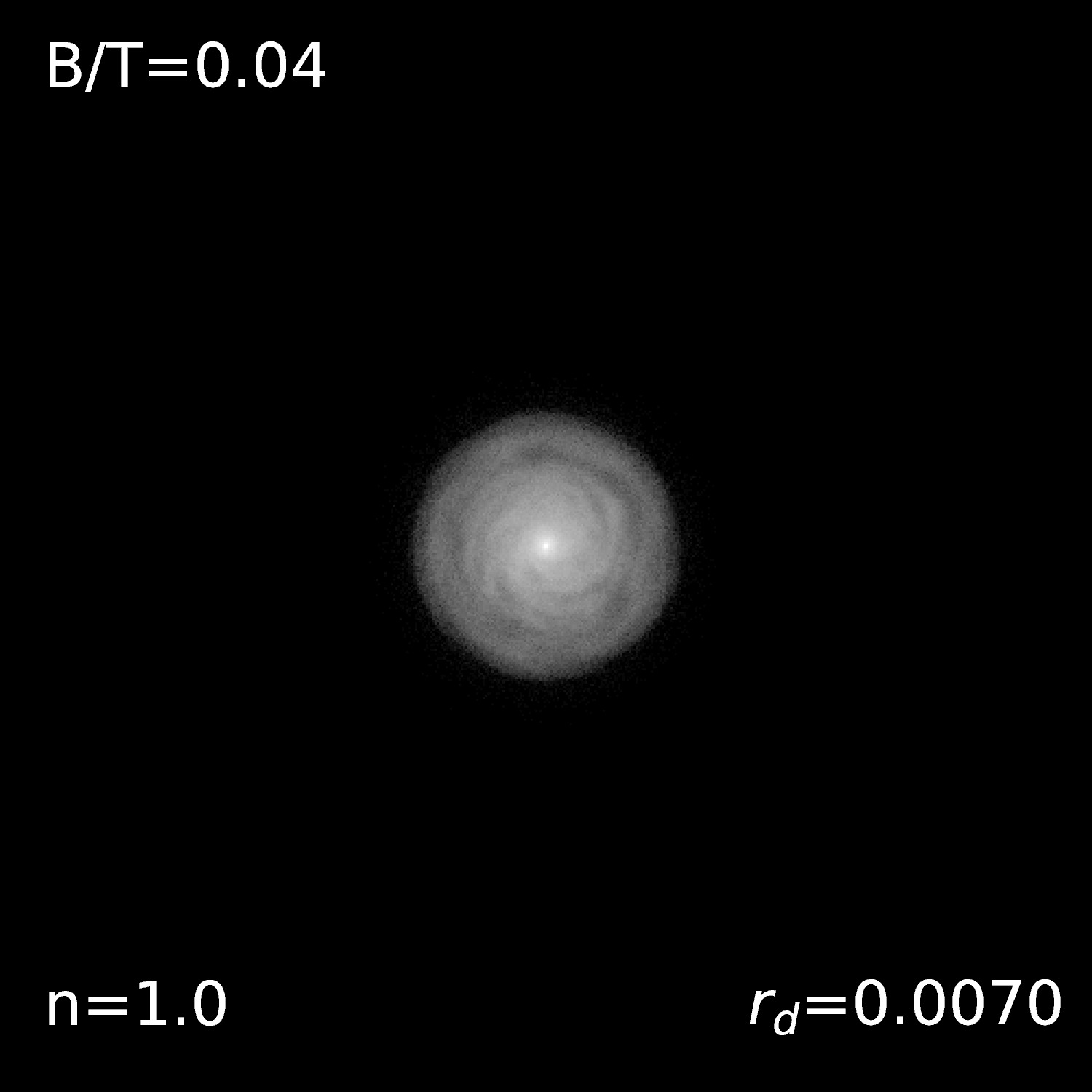}&&&\\
$m_{\rm d}=0.01$&
\includegraphics[width=1.122\hsize]{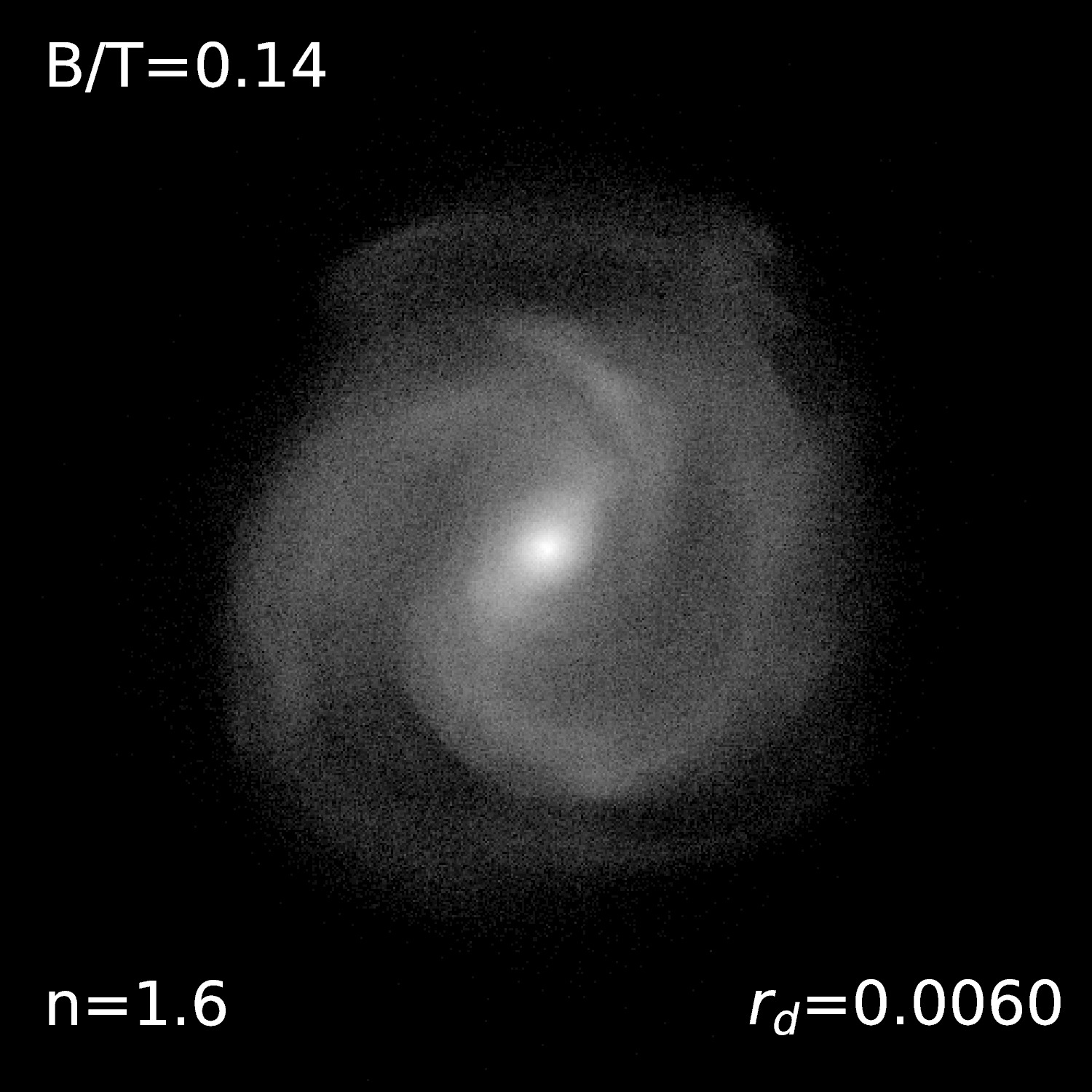}&
\includegraphics[width=1.122\hsize]{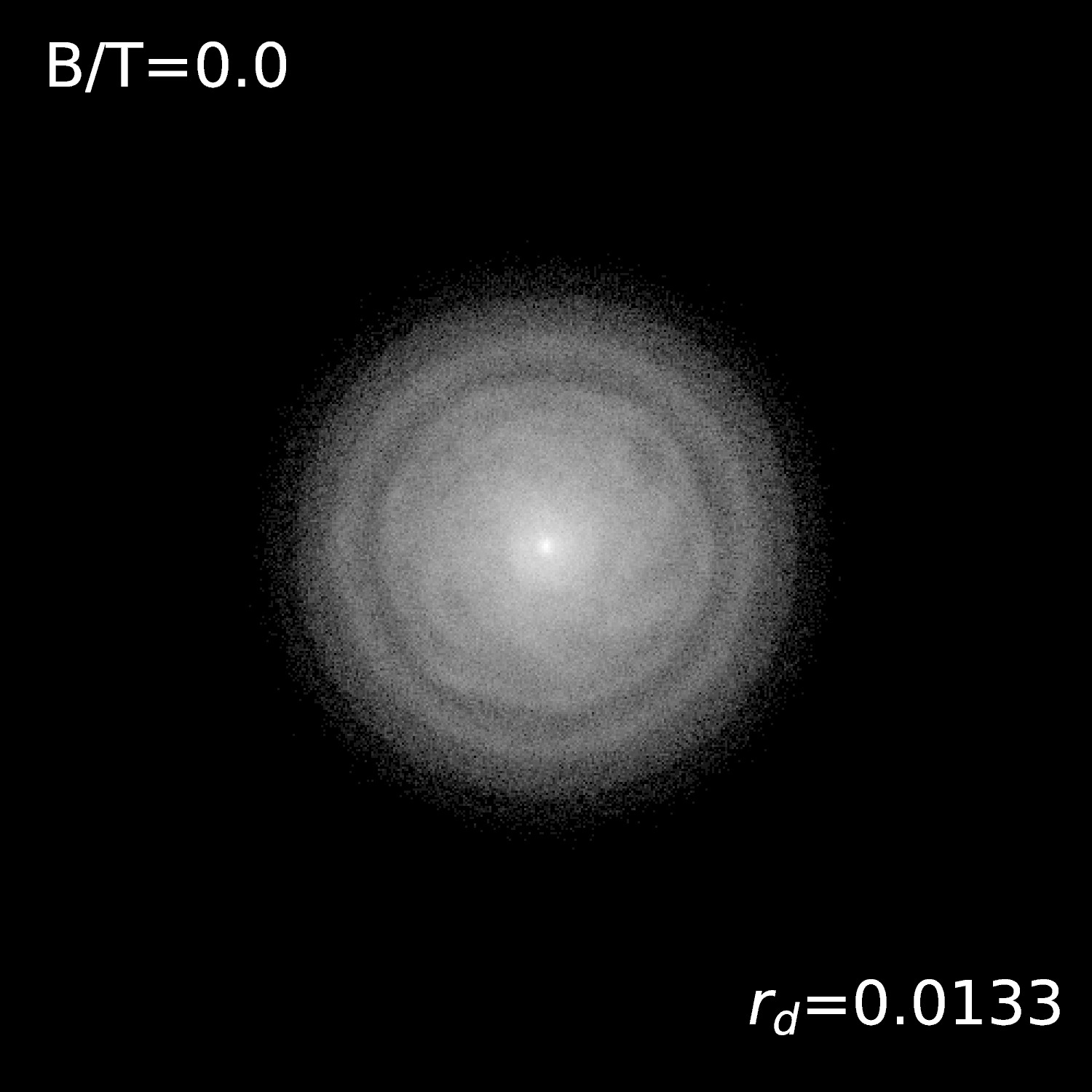}&&\\
$m_{\rm d}=0.02$& 
\includegraphics[width=1.122\hsize]{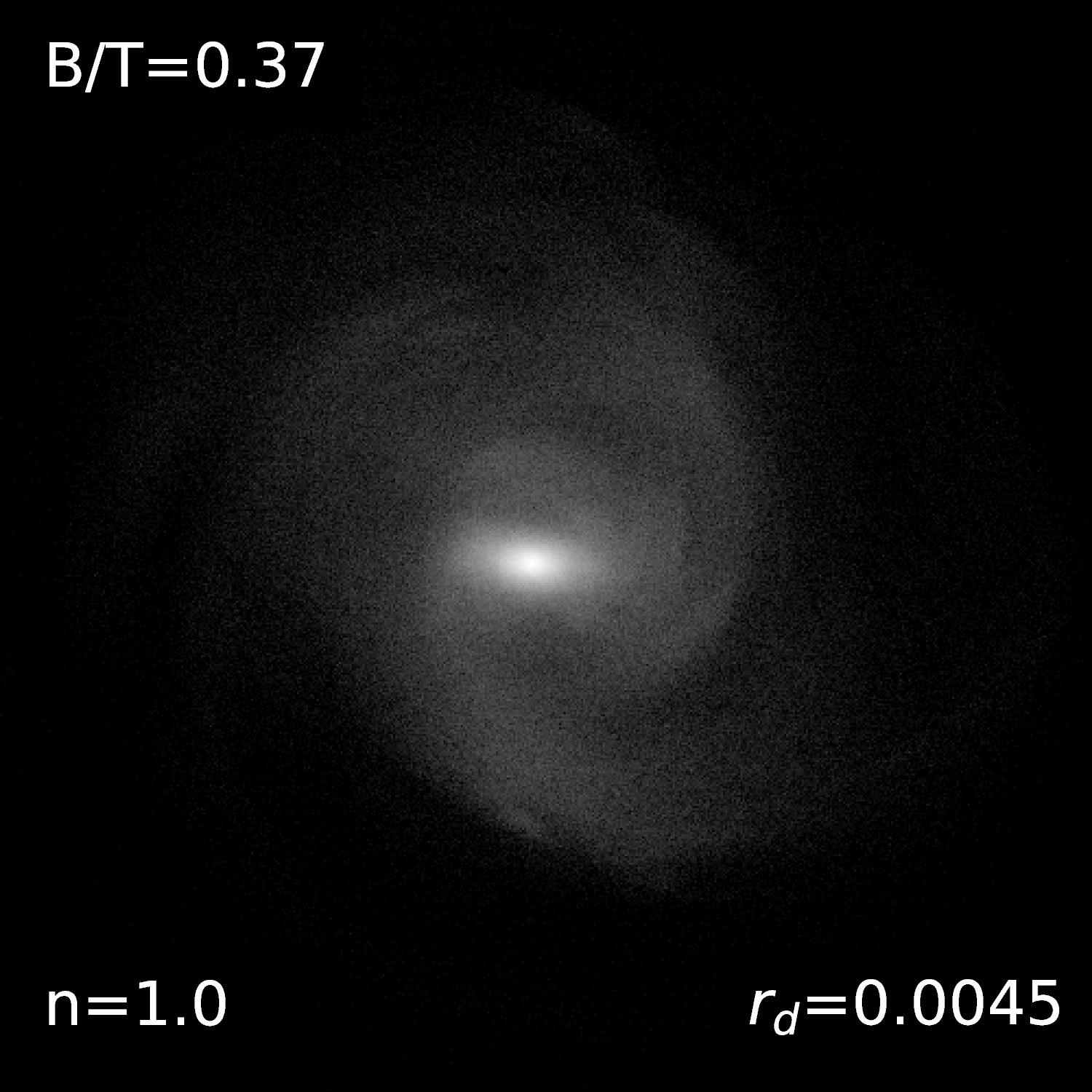}&
\includegraphics[width=1.122\hsize]{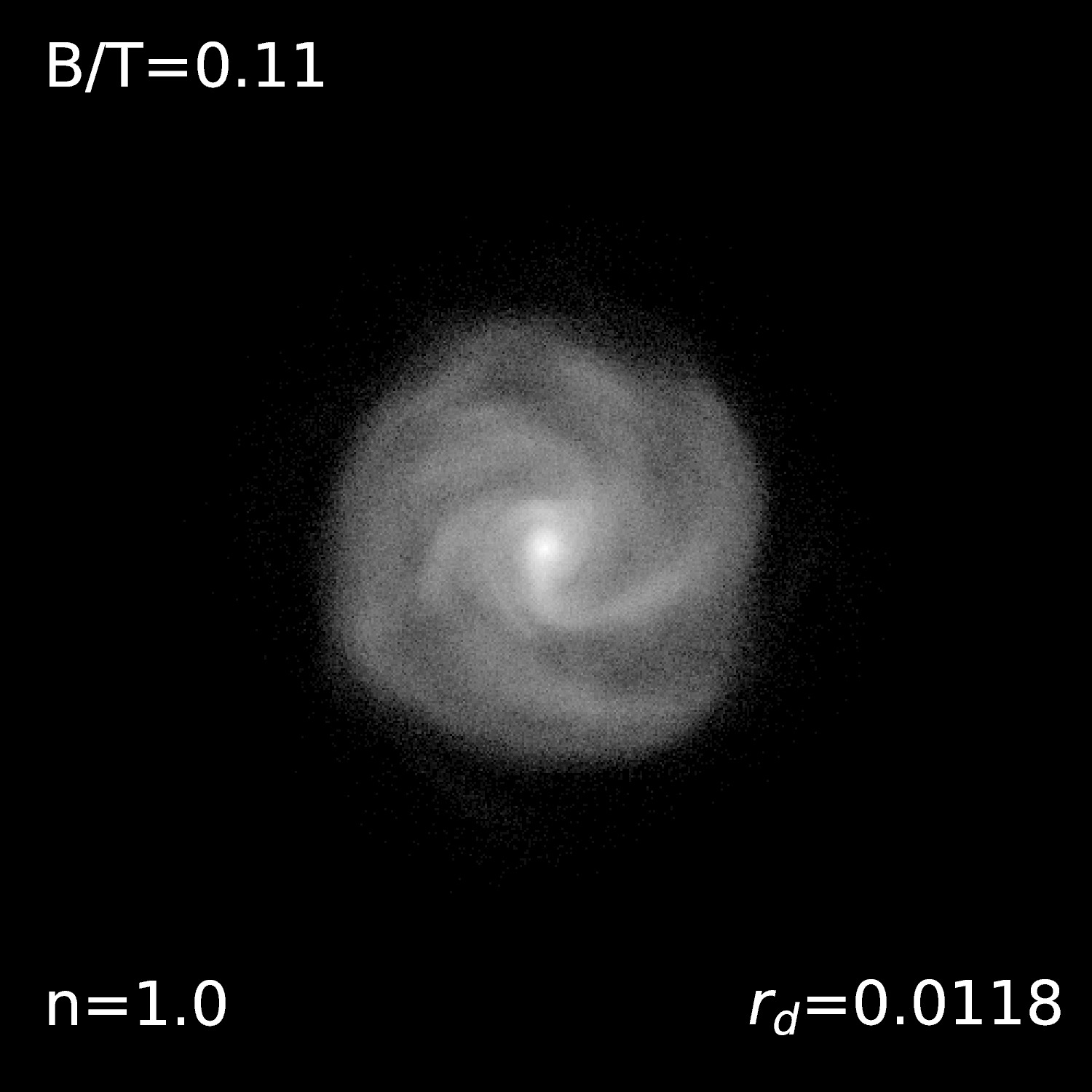}& 
\includegraphics[width=1.122\hsize]{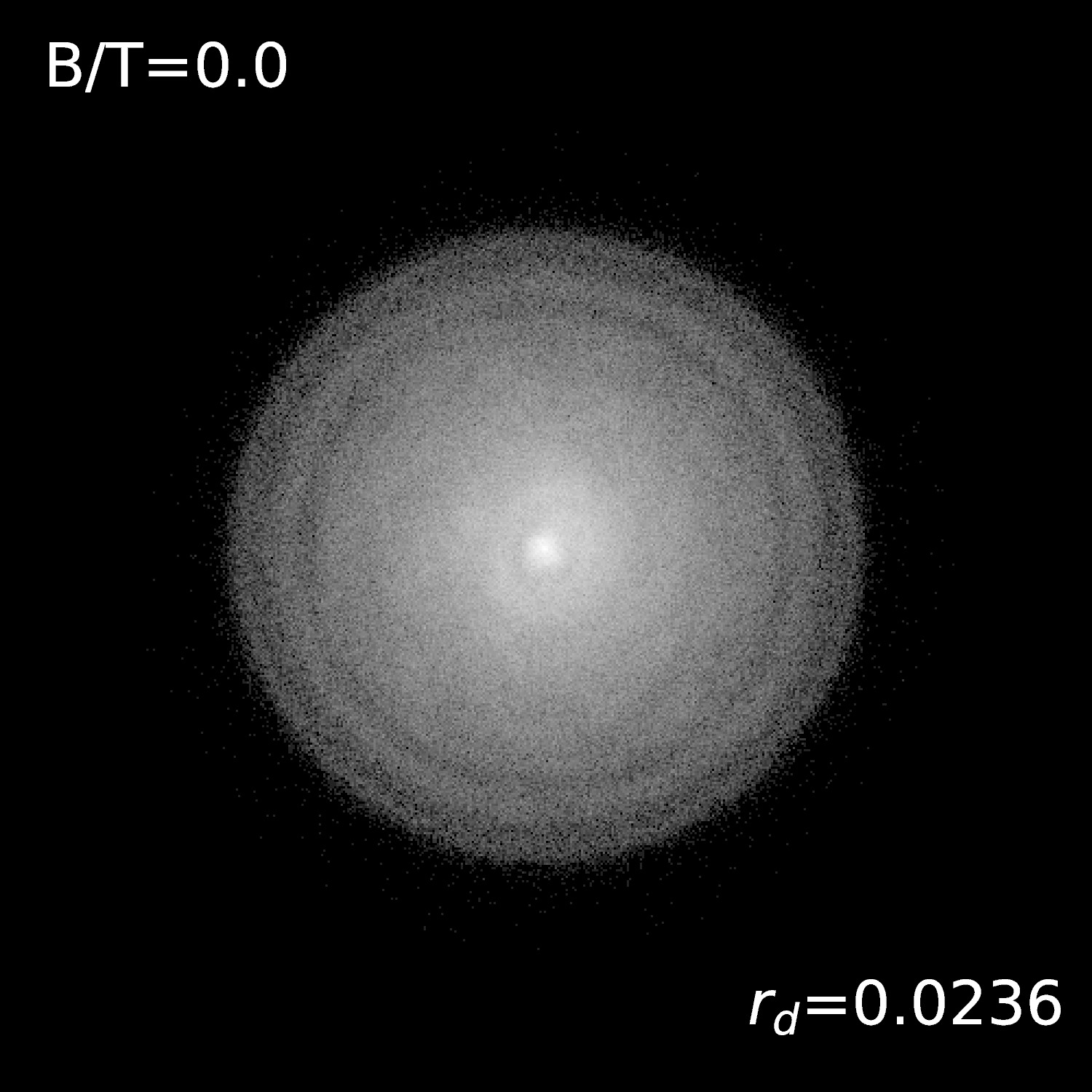}&\\ 
$m_{\rm d}=0.04$&
\includegraphics[width=1.122\hsize]{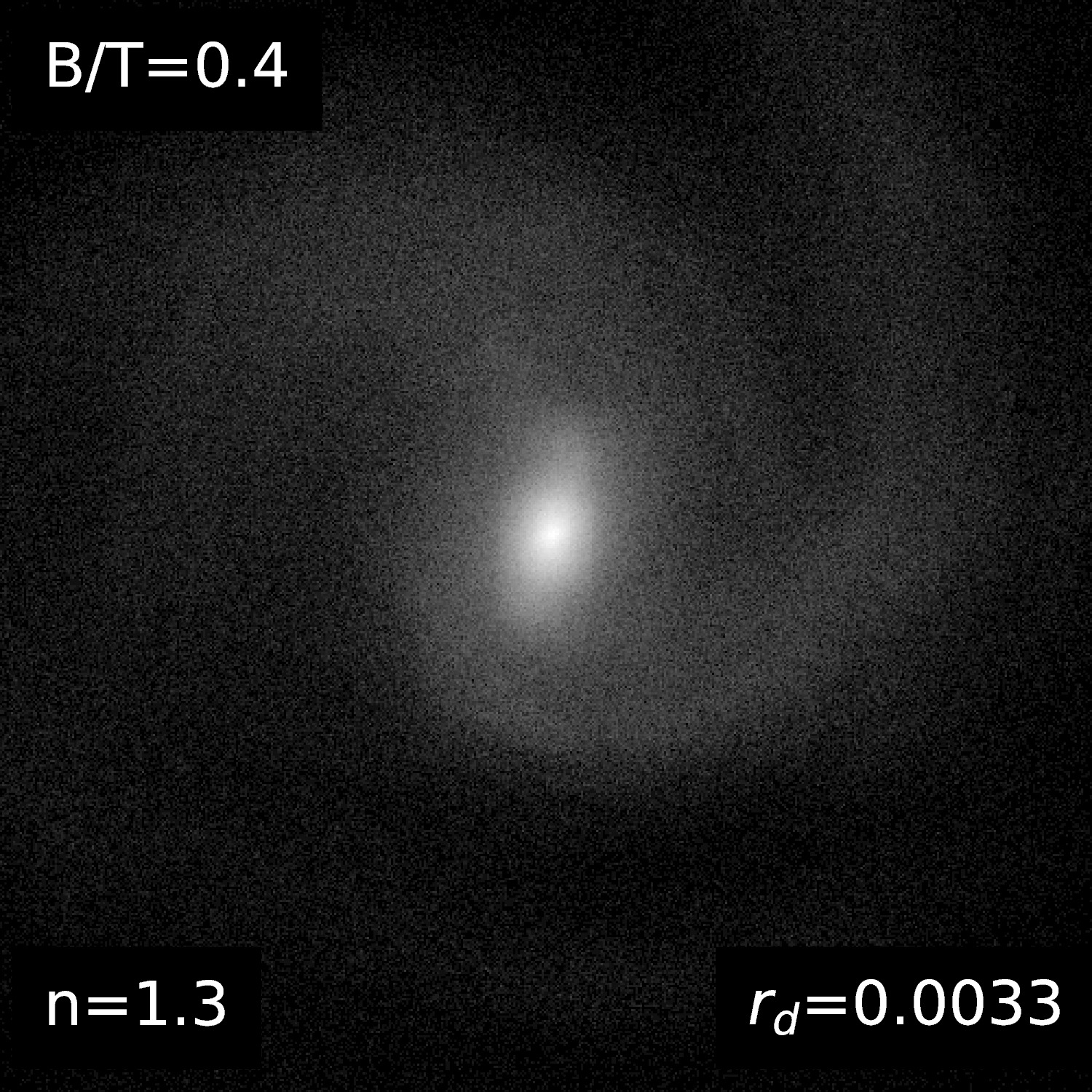}&
\includegraphics[width=1.122\hsize]{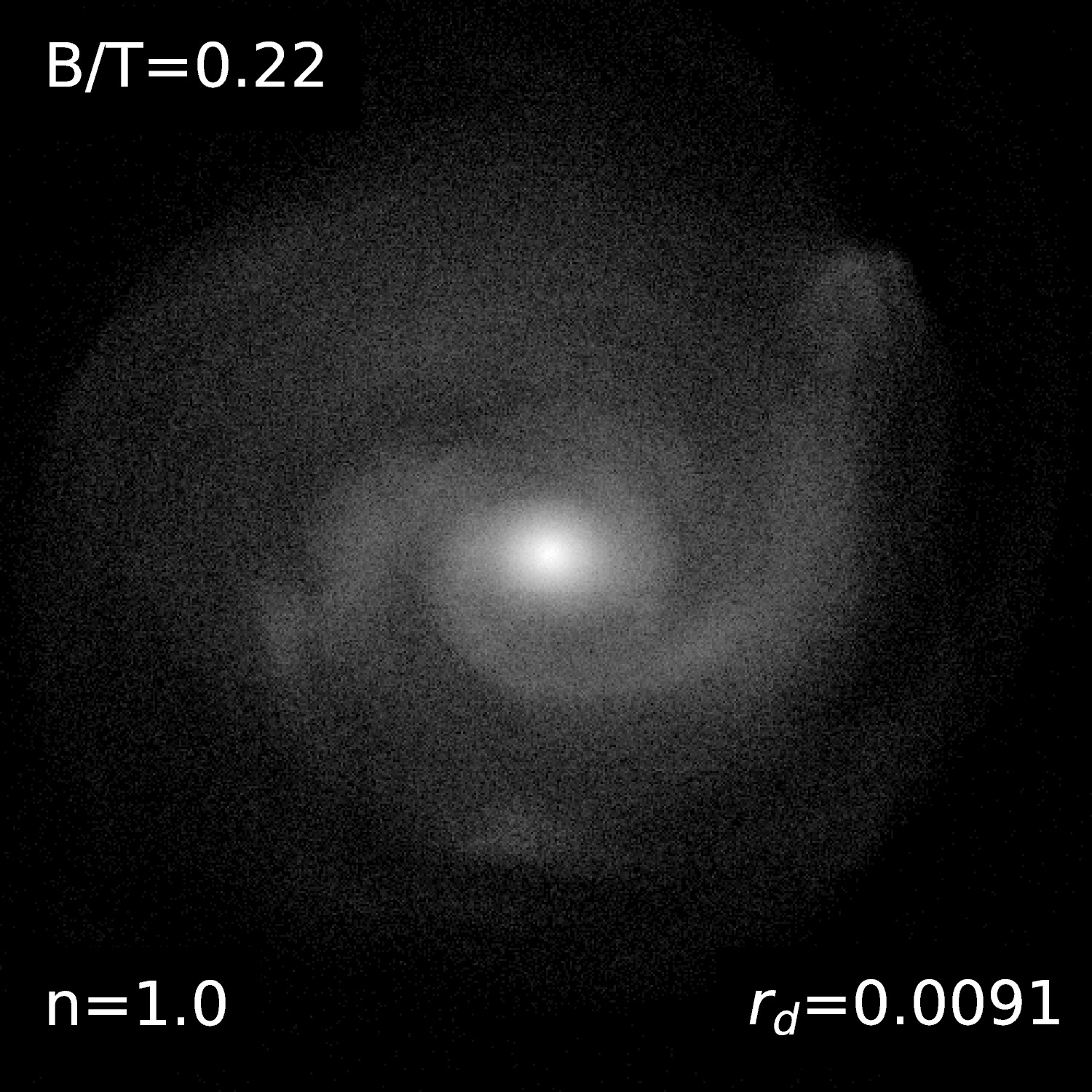}& 
\includegraphics[width=1.122\hsize]{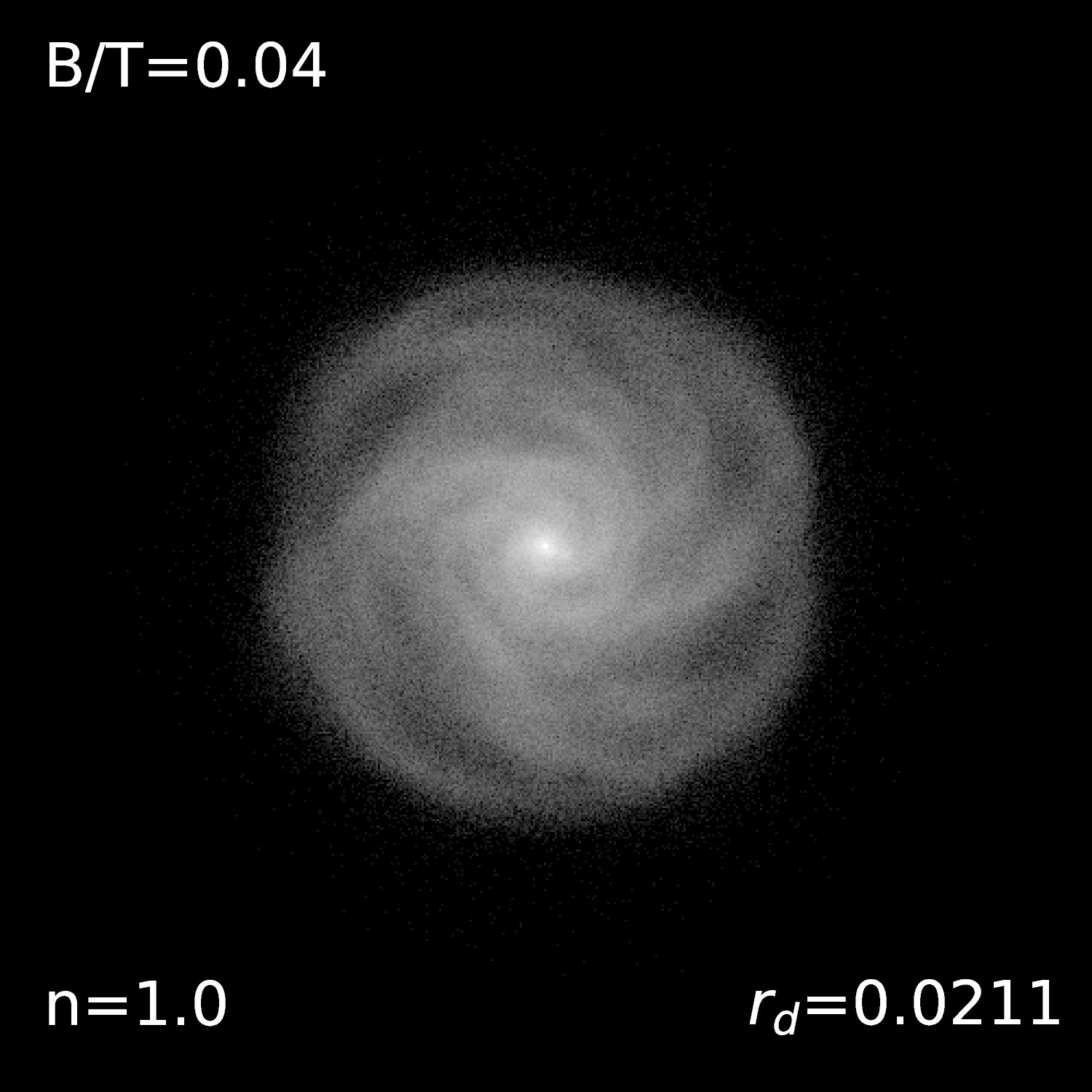}&
\includegraphics[width=1.122\hsize]{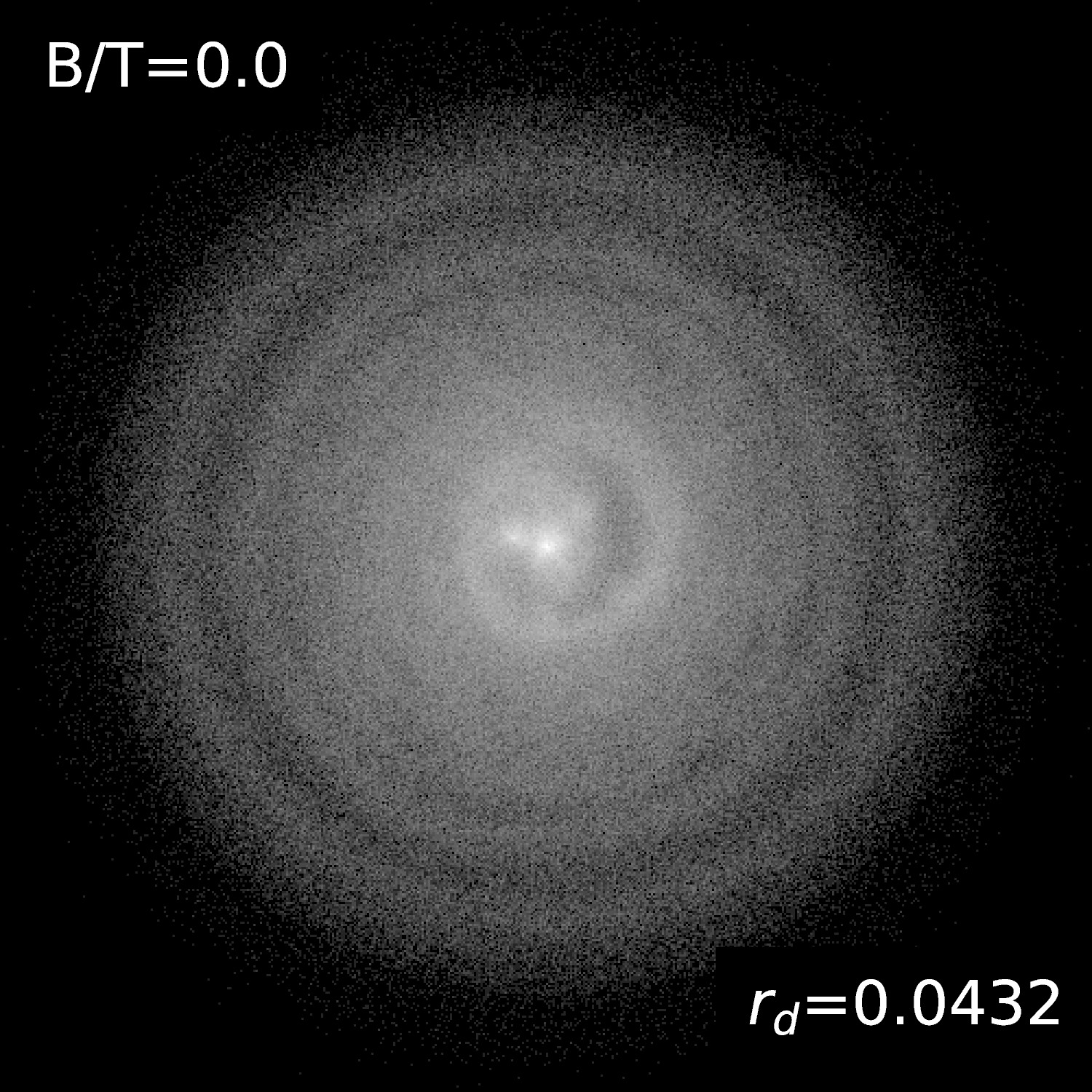}\\
\end{tabular} 
\end{center}
\caption{Same as Fig.~\ref{conc5} with $c=10$ instead of $c=5$. To avoid
  overcrowding the figure, we have not shown the simulations with
  $\lambda=0.018$ and $\lambda=0.35$.}
\label{conc10}
\end{figure*}

\begin{figure*}
\begin{center}
\begin{tabular}{ C D D D }
&$\lambda=0.011$&$\lambda=0.025$&$\lambda=0.05$\\
$m_{\rm d}=0.005$&
\includegraphics[width=1.122\hsize]{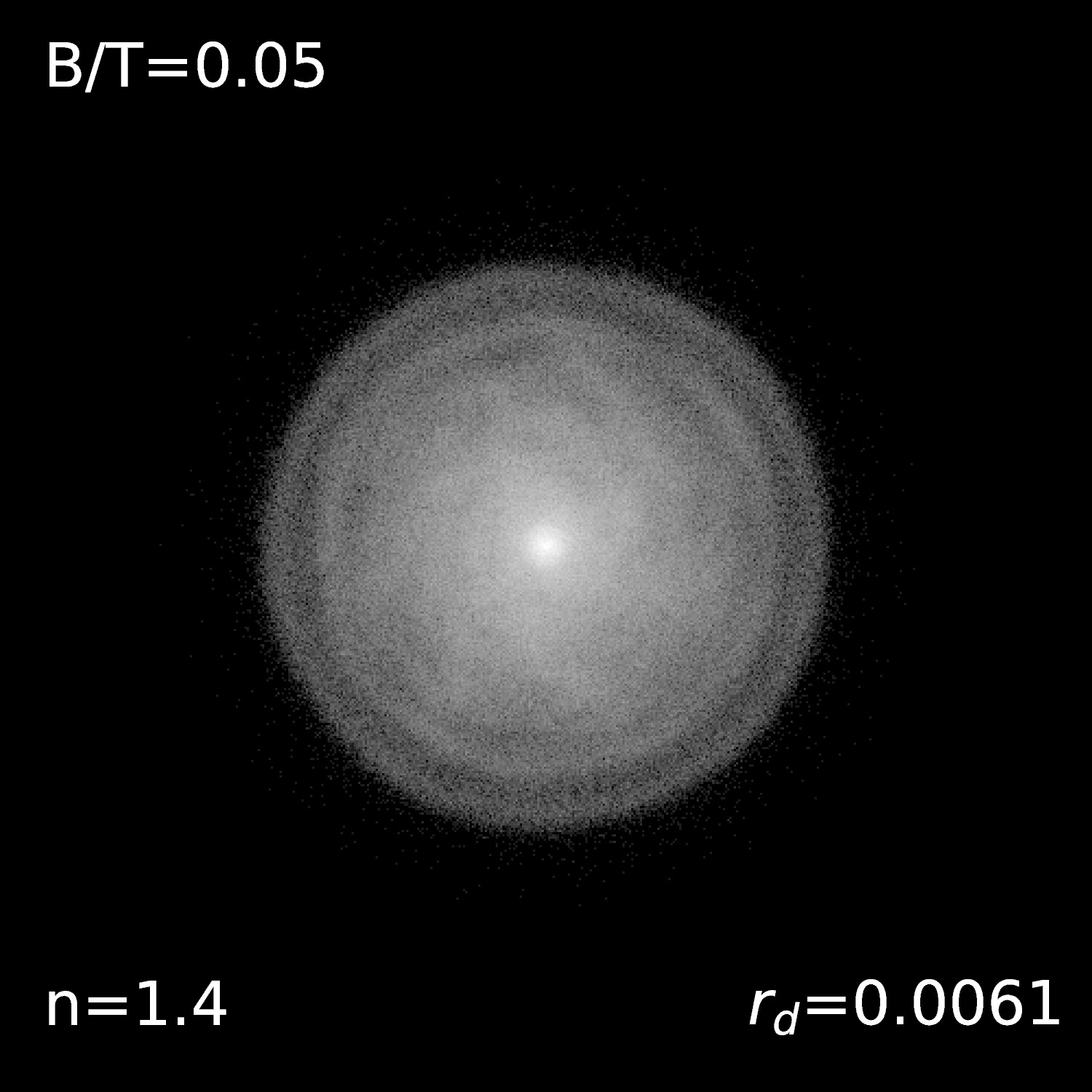}&&\\
$m_{\rm d}=0.01$&
\includegraphics[width=1.122\hsize]{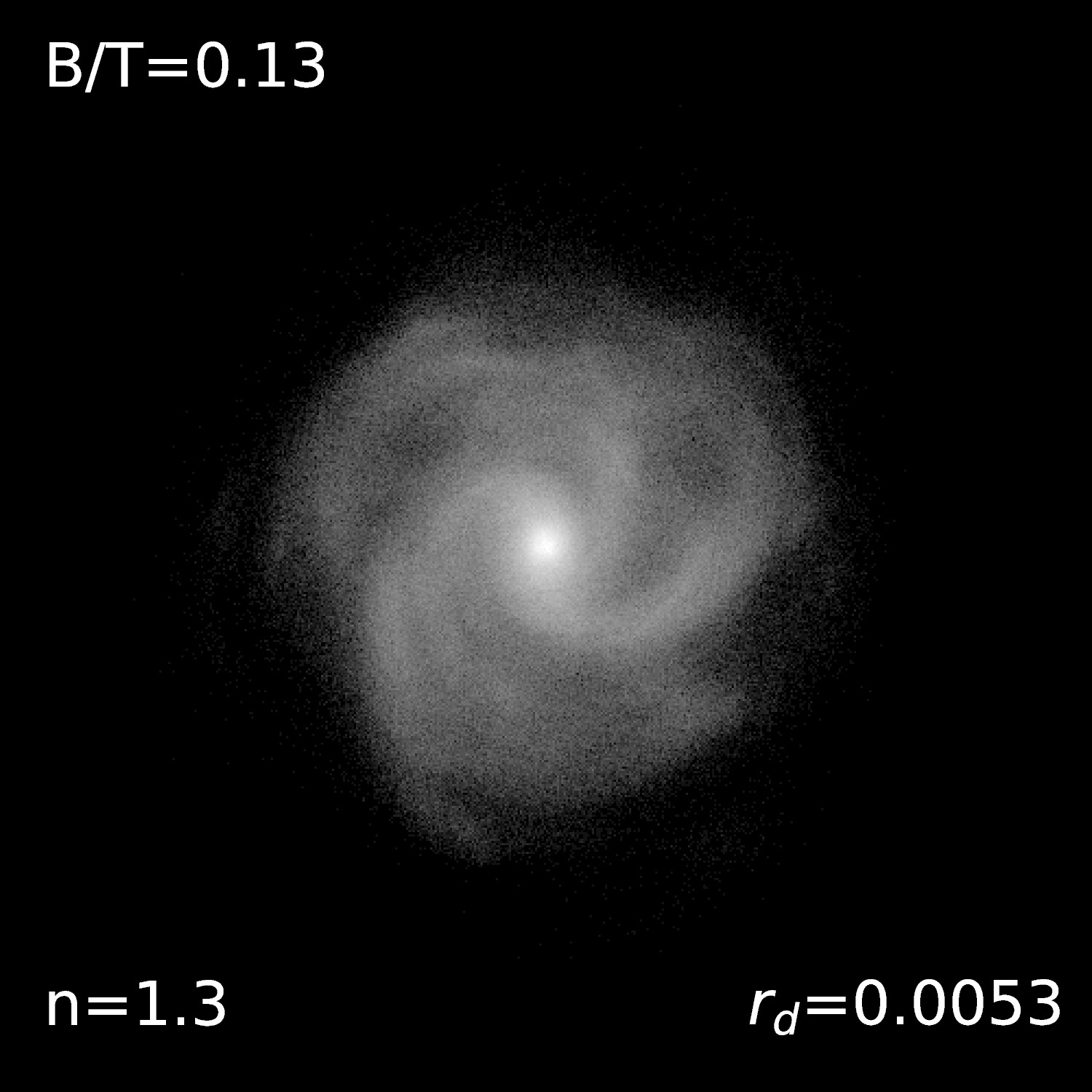}&&\\
$m_{\rm d}=0.02$& 
\includegraphics[width=1.122\hsize]{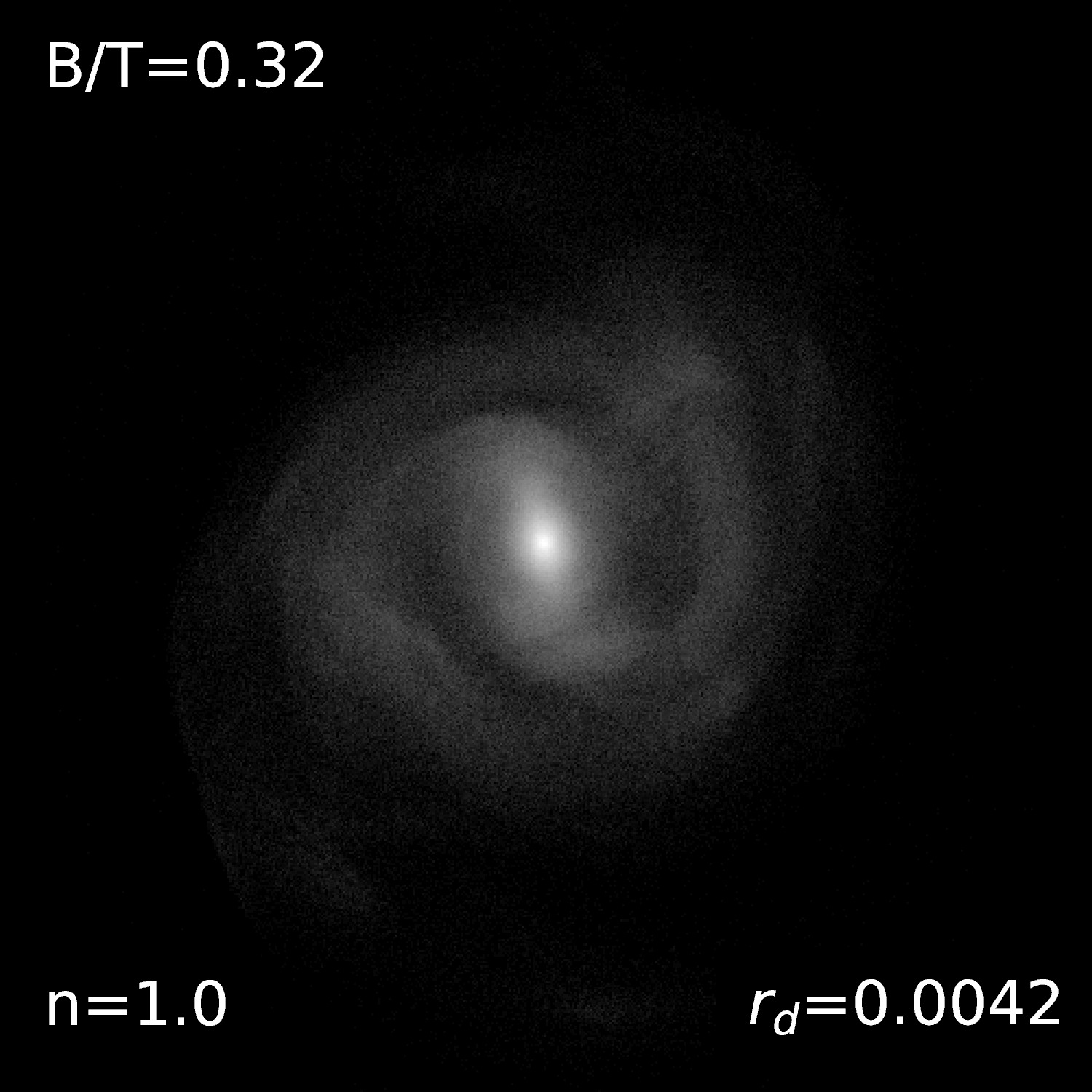}&
\includegraphics[width=1.122\hsize]{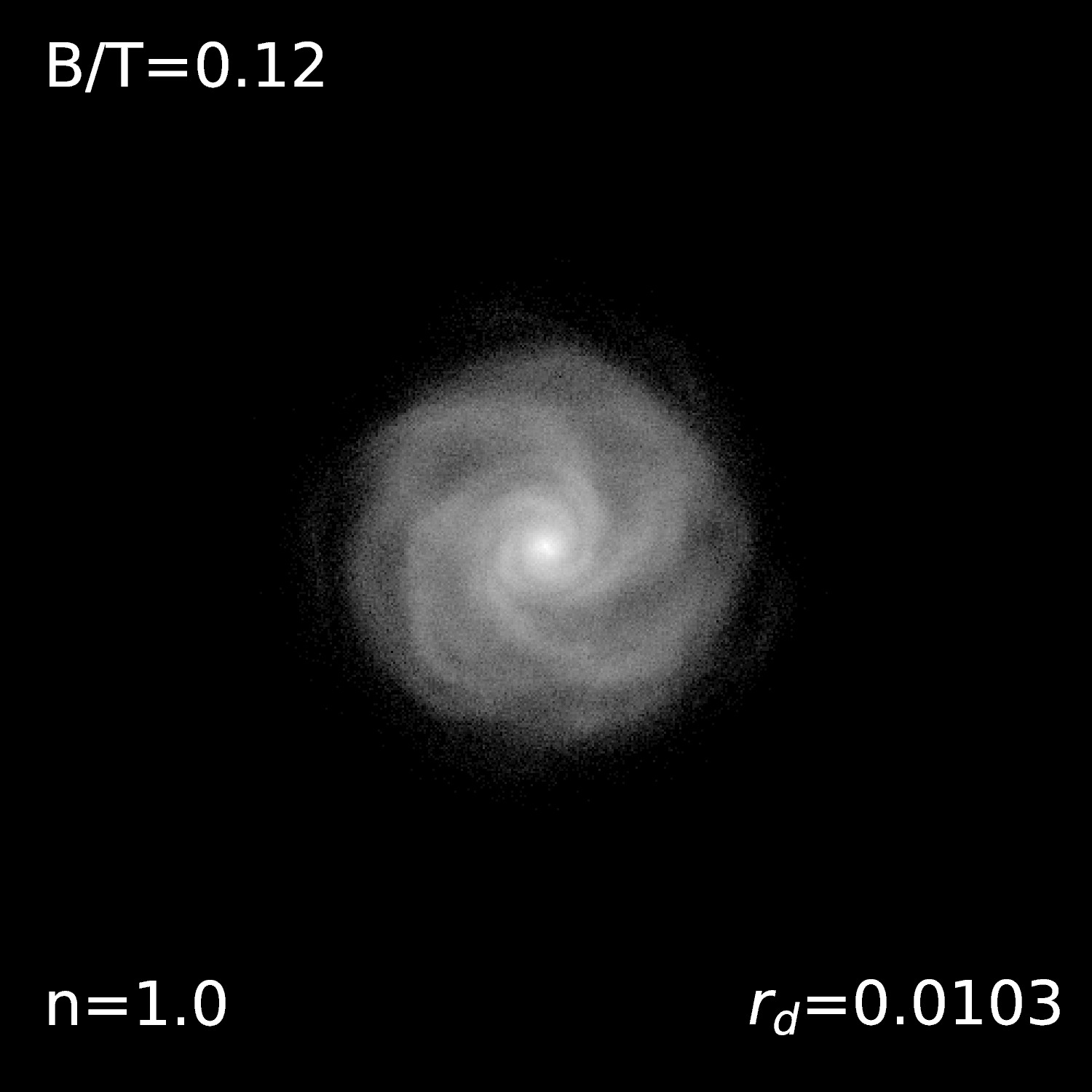}&\\ 
$m_{\rm d}=0.04$&
\includegraphics[width=1.122\hsize]{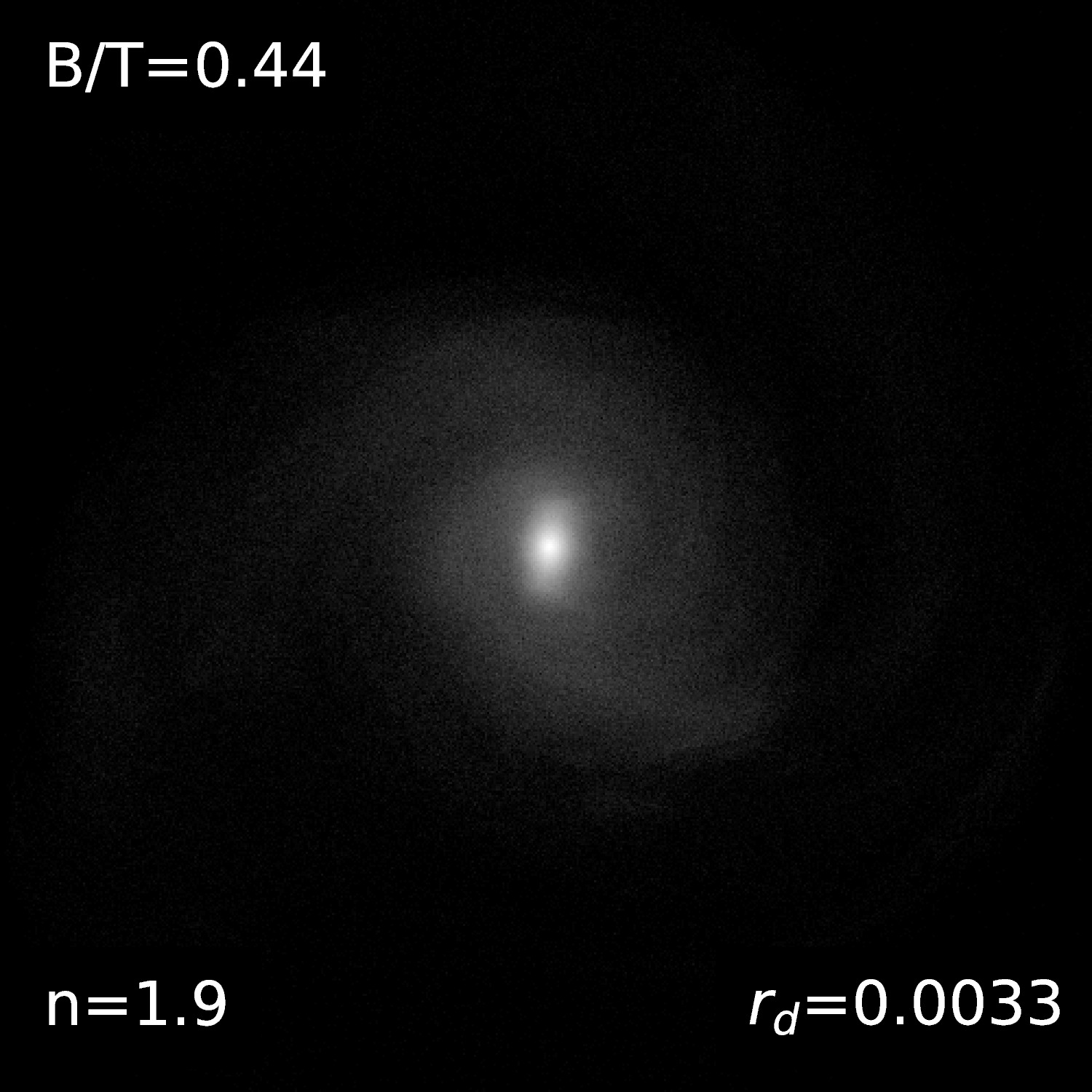}&
\includegraphics[width=1.122\hsize]{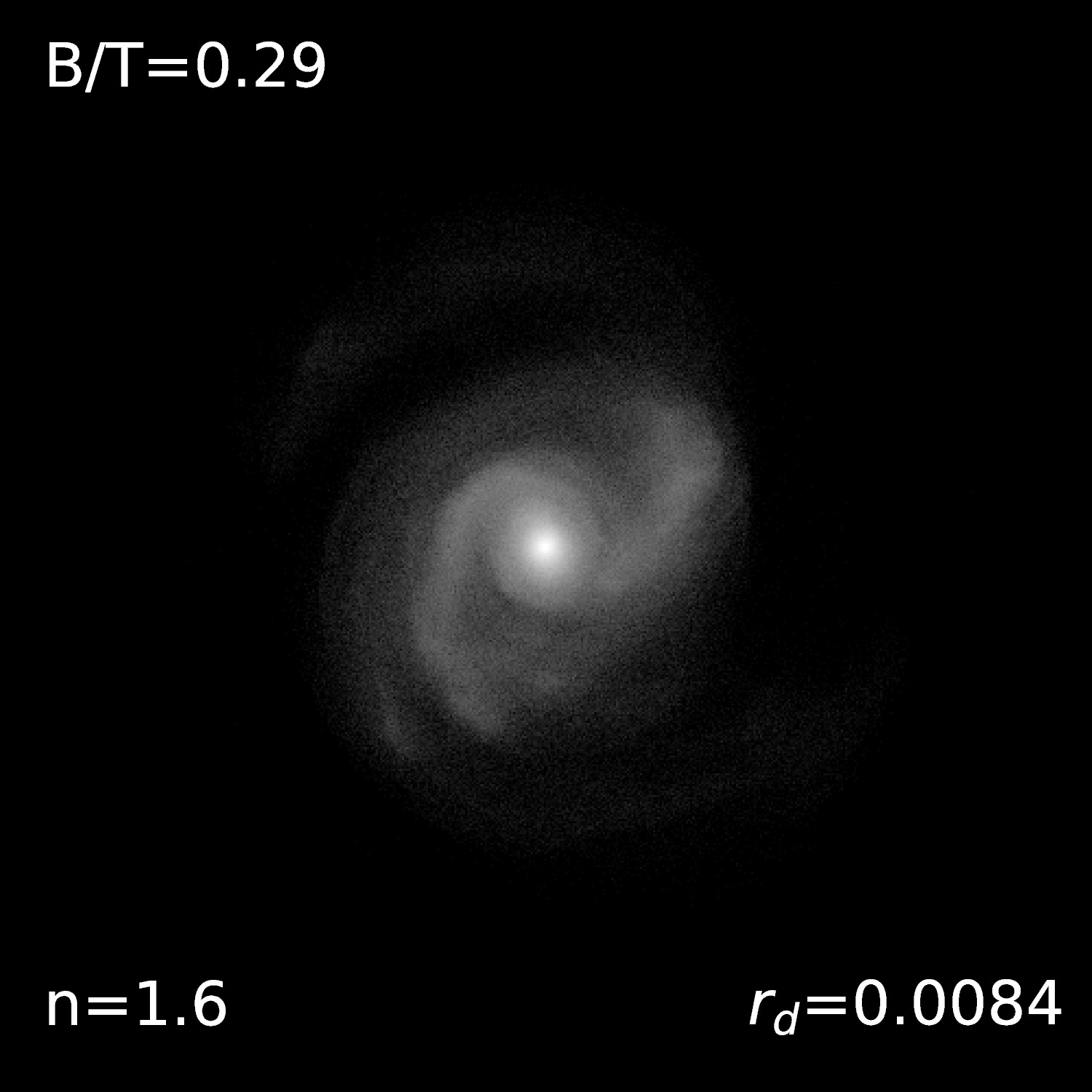}&
\includegraphics[width=1.122\hsize]{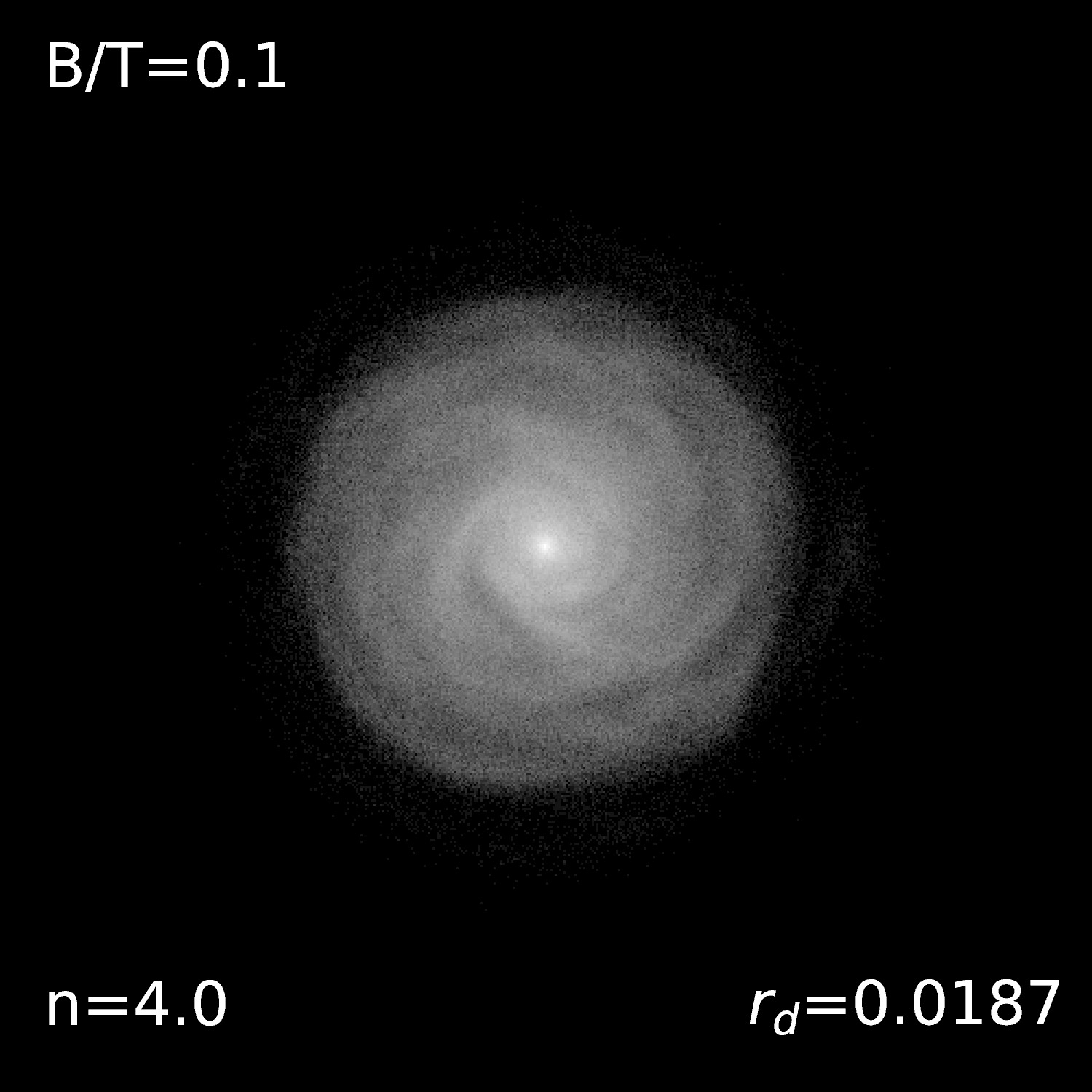}\\
\end{tabular}
\end{center}
\caption{Same as Fig.~\ref{conc5} with $c=15$ instead of $c=5$. 
The column $\lambda = 0.1$ is now missing because we know that
simulations with $\lambda=0.1$ will not form any (pseudo)bulge.}
\label{conc15}
\end{figure*}

The dimensionless radius of a thin exponential disc of dimensionless
mass $m_{\rm d}$ embedded in an NFW halo with concentration $c$, 
\begin{equation}
r_{\rm d}={\lambda\over\int_0^\infty e^{-x}v_{\rm c}x^2{\rm\,d}x},
\label{mmw}
\end{equation}
\citep{cattaneo_etal17} is completely determined by the disc's spin parameter
\begin{equation}
\lambda={J_{\rm d}\over M_{\rm d}R_{\rm vir}V_{\rm vir}},
\label{spinparameter}
\end{equation}
which is identical to the halo's spin parameter if we follow \citet*{mo_etal98} and we assume that the disc and the halo have the same specific angular momentum
(for $J_{\rm d}/M_{\rm d}=J_{\rm vir}/M_{\rm vir}$, Eq.~\ref{spinparameter} corresponds to the \citealp{bullock_etal01} definition of the spin parameter).

Eq.~(\ref{mmw}) implies that we can use the spin  distribution of DM haloes from cosmological N-body simulations to derive plausible values for $r_{\rm d}$.
This is true even if conservation of angular momentum is a crude approximation (\citealp{kimm_etal11,stewart_etal13}; Jiang et al. 2018) because 
the model by  \citet*{mo_etal98} reproduces the correct mass -- relation for disc galaxies (for example, \citealp{cattaneo_etal17}).

For a flat rotation curve with $v_{\rm c}=1$, Eq.~(\ref{mmw}) gives $r_{\rm d}=\lambda/2$. 
We do not make this approximation. We substitute Eq.~(\ref{vc}) into
Eq.~(\ref{mmw}) and solve Eq.~(\ref{mmw}) numerically {for each
  combination of $\lambda$, $m_{\rm d}$, and $c$ that we wish to consider.}
However, we find that the approximation  $r_{\rm d}\simeq \lambda/2$ is accurate at the $20\%$ level.

\begin{figure}
  \begin{center}
    \includegraphics[width=1.0\hsize]{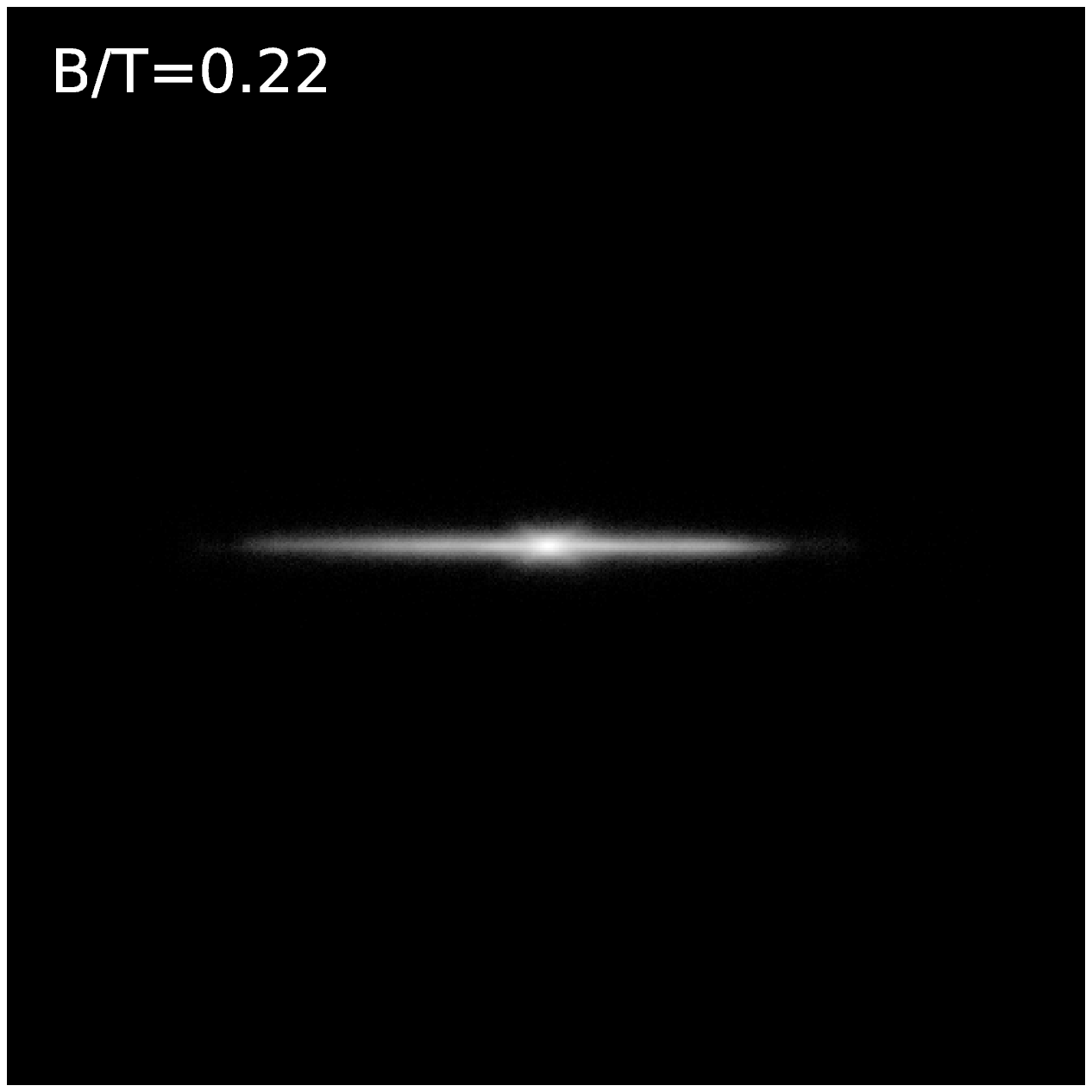}
    \end{center}
\caption{Galaxy in the simulation with {$c=10$}, $m_{\rm d}=0.04$, $\lambda=0.025$, viewed edge-on. An X-shaped pseudobulge is clearly visible. This galaxy has been chosen as an example and is by no means atypical. {The face-on image (Fig.~3) shows that this galaxy has a bar, but not a prominent one.}}
\label{EDGE_on}
\end{figure}

\citet{munoz_etal11} found that $\lambda$ follows a log-normal distribution with $\bar{\lambda}=0.044$ and $\sigma_{{\rm ln}\,\lambda}=0.57$.
\citet{burkert_etal16} found a similar distribution with $\bar{\lambda}=0.052$ and $\sigma_{{\rm ln}\,\lambda}=0.46$.
So did \citet{cattaneo_etal17} with $\bar{\lambda}=0.049$ and $\sigma_{{\rm ln}\,\lambda}=0.57$ (we use the same symbol $\lambda$ for the spin of the halo and the spin
of the disc because we have assumed they are identical).
From these figures, we conclude that $\lambda=0.05$ corresponds to a typical value and that $\sim 80\%$ of all haloes lie in the
interval $0.025<\lambda<0.1$.
Haloes with $\lambda<0.0125$ comprise less than $2\%$ of all systems.
Based on these considerations and the fact that $r_{\rm d}$ is the parameter to which our results are most sensitive,
we have explored six values of $\lambda$ (0.011, 0.018, 0.025, 0.035,
0.05, 0.1).

{We sample the parameter space by $\lambda$ rather than
  $r_{\rm d}$ because we started this  research to  find a prescription
  that may be useful to assign morphologies to galaxies in both semianalytic and halo-occupation-distribution models (for example,
  \citealp{behroozi_etal13,behroozi_etal19,moster_etal13,moster_etal18,tollet_etal17}).
If the objective is to find a recipe to populate haloes with galaxies,
then $\lambda$ and $c$ are the quantities that we
directly measure in N-body simulations, and we should like to find
$B/T$ as a function of them.
However, the halo spin has no direct effect on our simulations
because our haloes are spherical and static.
Hence, it is the dependence of $B/T$ on $r_{\rm d}$ that  we probe in reality.
}

In contrast to $\lambda$, which appears to be a true random quantity, $m_{\rm d}$ strongly correlates with $M_{\rm vir}$.
Considerable observational evidence shows that the stellar-to-halo mass ratio increases with $M_{\rm vir}$ over the range of halo masses where spiral morphologies are prevalent
\citep{papastergis_etal12,leauthaud_etal12,reyes_etal12,behroozi_etal13,moster_etal13,wojtak_mamon13,cattaneo_etal17,tollet_etal17}.
In a halo with $M_{\rm vir}=10^{12}\,M_\odot$, the typical stellar mass of the central galaxy is $M_\star = 4\times 10^{10}\,M_\odot$ ($m_{\rm d}=0.04$ if we assume that the
galaxy started as a pure disc).
In a halo with $M_{\rm vir}=10^{11}\,M_\odot$, the typical stellar mass is two orders of magnitudes lower and $m_{\rm d}$ is lower by about a factor of ten.
Hence, we explore four values of $m_{\rm d}$ (0.005, 0.01, 0.02, 0.04) that cover the typical range of $M_\star/M_{\rm vir}$ from dwarf to Milky-Way-type galaxies.

\begin{figure*}
\begin{center}$
\begin{array}{cc}
\includegraphics[width=0.48\hsize]{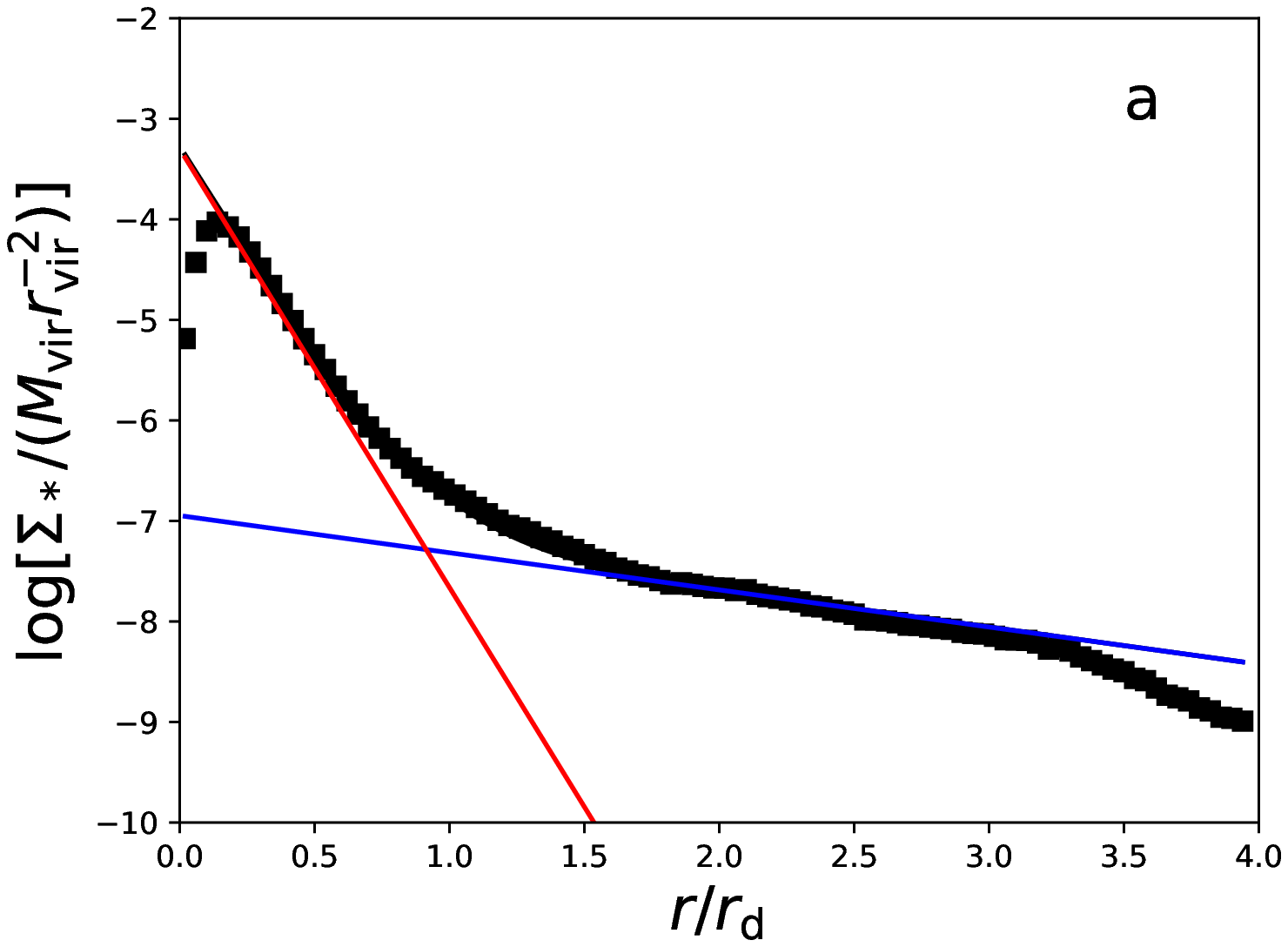}
\includegraphics[width=0.48\hsize]{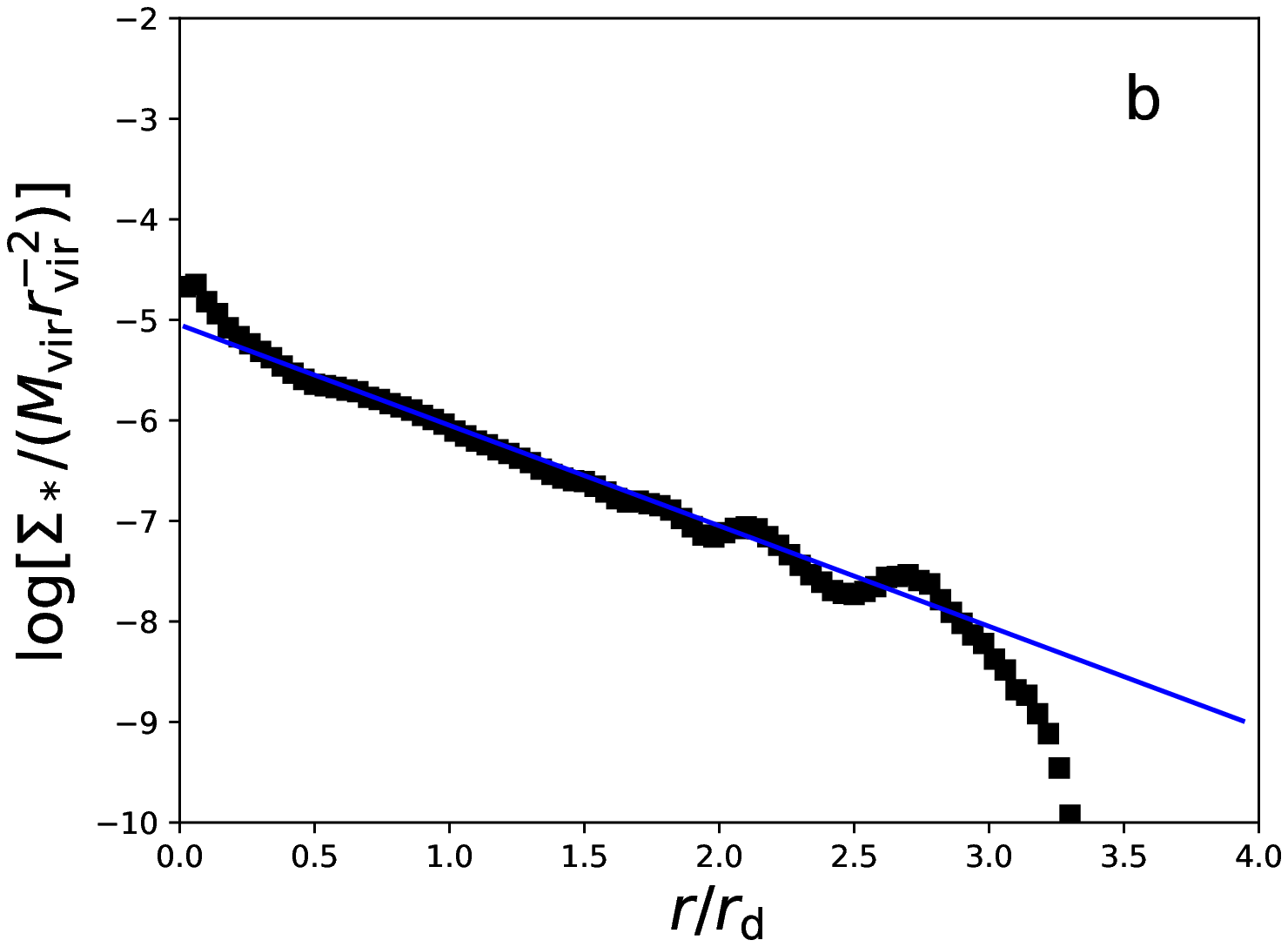} 
\end{array}$
\end{center}
\caption{Stellar surface-density for two of our simulated galaxies
  (black symbols). Galaxy $a$  has been fitted with the sum
of an exponential profile (blue line) and a S{\'e}rsic profile, with
$n=1$ in this particular case (red line). 
Its bulge-to-total mass ratio is $B/T=0.22$.
The black curve shows the
sum of the two components.
Galaxy $b$ is consistent with a single exponential
profile (blue line).
}
\label{serexp}
\end{figure*}

Concentration has a much weaker dependence on halo mass. \citet{dutton_maccio14} find $c=10.2M_{12}^{-0.097}$ with $M_{12}=M_{\rm vir}/10^{12}M_\odot$ at redshift zero (also see \citealp{munoz_etal11}). Thus, we expect the mean concentration to vary from $c=13$ for a dwarf galaxy to $c=10$ for a Milky-Way-type one.
The real concentration range is larger because the scatter is significant. We thus consider three concentration values (5, 10, 15) such that the interval $5<c<15$ contains $\sim 80\%$ 
of the haloes with $M_{\rm vir}>10^{11}\,M_\odot$.

Six values for $\lambda$, four values for $m_{\rm d}$ and three values for $c$ make $72$ combinations in total. We have
simulated only $34$ of these $72$ combinations  
because we have explored the cases $\lambda=0.018$ and $\lambda=0.035$ only for $c=10$ and because during our work it became obvious that certain sets of parameters would not form a bulge (Section~3).
Hence, it would have been pointless to study them.
 
 {In addition to these $34$ simulations, we have run another $14$ simulations to address specific issues, such as:
  1) relaxation effects due to the assumption of a razor-thin disc, 2) the assumption of a static halo, and 3) the impact of a small gas fraction.
 The parameter values for these simulations are listed in Tables~1, 2, and 3, respectively. The simulations run for this study are
 $34+14=48$ in total.}

\subsection{Refinement strategy}

We evolved our initial conditions with the AMR code {\sc ramses} \citep{teyssier02} until the galaxies converged to a stable configuration. This usually occurs within
$2\,$Gyr. 

As we use a Eulerian code, the most important numerical aspect of our work is the choice of the grid on which we integrate the equations of motion for the stellar fluid.
We centre our discs on a cubical grid with $128^3$ cells and side length $32\,r_{\rm d}$. Hence, the resolution on the scale of the entire computational volume is $l=0.25\,r_{\rm d}$.
The resolution is increased within six nested cylinders by a factor of two each time (Fig.~\ref{in_cond}).
The radii $r/r_{\rm d}\simeq 3.8,\,3.2,\,2.7,\,2.3,\,1.9$, and $1.7$ of the six cylinders enclose the isodensity contours that contain 80, 70, 60, 50, 40, and 30$\%$ of the disc mass
and correspond to  $l/r_{\rm d}=1/8,\,1/16,\,1/32,\,1/64,\,1/128$, and $1/256$, respectively.

Height equals radius in all cylinders except the innermost one,
{which} has a height-to-radius ratio of $1:5$ to ensure than the vertical structure of the thin gaseous disc is well resolved
even though the gas fraction in our simulations is so small that it has no dynamical effects.
{The second innermost cylinder (the one marked in yellow in
  Fig.~\ref{in_cond}) has a semi-height ($0.95\,r_{\rm d}$) large enough
to ensure that the disc remains well resolved when it buckles up, since
the scale-length (and therefore the scale-height) of a pseudobulge is usually several times smaller
than $r_{\rm d}$ \citep{gadotti09}.}

\section{Results}

{Some of our discs our stable. They do not show any bar or pseudobulge
even after $2\,$Gyr.
In unstable discs, a bar forms rapidly, but after $2\,$Gyr the growth
of $B/T$ becomes very slow and, in many cases, almost inexistent (see below for details of how we measure $B/T$).
Figs.~2 to 4 show {$27$ simulated galaxies from our main sample of $34$ galaxies in total (we have not shown the simulations
$\lambda=0.018$ or $\lambda=0.035$). The galaxies are shown face-on at the first time $t>2\,$Gyr when the growth of $B/T$ has stabilised.
In most galaxies, $B/T$ has converged at $t<2.5\,$Gyr.
}

Figs.~2, 3, 4 correspond to $c=5,\,10,\,15$, respectively.
They portray the variety of morphologies that we can reproduce, from Scd to SBa types, simply by changing the values of our parameters.
{Since our galaxies are isolated and the gas fraction is very small, bar instabilities are the only physical mechanisms through which we expect that a bulge should form in our simulations (see the discussion of this article). In agreement with this expectation, 10 out of 17 galaxies
with $B/T>0.1$ display morphologies that are clearly barred.
We have also looked at our galaxies edge-on. In most bulges, an X-shaped peanut is clearly visible (Fig.~\ref{EDGE_on}).

The galaxy with $m_{\rm d}=0.01$ and $\lambda=0.011$ on Fig.~2 is the most conspicuous example of a prominent bulge ($B/T=0.24$)
without any evidence for a bar. Such cases are rare. We have looked at this galaxy edge on. The bulge looks boxy, although the peanut shape is much less obvious than in Fig.~\ref{EDGE_on}.
}

The stellar surface-density profile $\Sigma_*(r)$ of each galaxy has been fitted
with the sum of an exponential and a \citet{sersic63} profile
(Fig.~\ref{serexp}a).
This decomposition has been used to compute a $B/T$ ratio for each galaxy and a S{\'e}rsic index $n$ for the pseudobulges of all galaxies with $B/T>0$.
{We assume $B/T=0$ when a visual inspection shows the
  surface-density profile is consistent with  a single exponential
  function (Fig.~\ref{serexp}b).}
  
  {The decrease in $\Sigma_*$ in the central region (Fig.~\ref{serexp}a) is an artefact due to how we set up the initial conditions for the velocity field. Its cause is the relaxation effect discussed in Section~2.1 after Eq.~(\ref{vrot}) coupled to the lower resolution in the central region. The radius $r\sim 0.1\,r_{\rm d}$ within which $\Sigma_*$ begins to decline is so small that it does not affect our results, except when 
$B/T$ is very low, in which case it limits our accuracy in fitting the surface-density profile of the bulge component.}

All galaxies with $B/T>0.1$ have $n\lsim 1.2$, as expected for
pseudobulges. 
In systems with $B/T\le 0.1$, the bulge is small and
the S{\'e}rsic index may not be robust, {also because of the effect mentioned above.
The galaxy with $m_{\rm d}=0.04$ and $\lambda=0.05$ on Fig.~4 exemplifies this situation.
Bulges with $n>2$ are usually classical bulges and are seldom encountered in galaxies with $B/T\lsim 0.1$ \citep{fisher_drory08}.
Hence, our simulations should not form any bulges with $n=4$, let alone one with $B/T=0.1$.
A visual inspection of the bulge/disc decomposition shows a poor fit to the stellar surface density in the central region. The S{\'e}rsic index $n=4$ measured for this galaxy cannot be considered meaningful.
}

For given $c$ and $\lambda$, higher values of $m_{\rm d}$ correspond
to earlier-type morphologies in the Hubble sequence and to higher
$B/T$ (Figs.~2 to 4).
We can also increase $B/T$ by reducing $\lambda$ at constant $c$ and $m_{\rm d}$.
At constant $\lambda$ and $m_{\rm d}$, $B/T$ usually decreases with $c$.
Hence, if a simulation does not form a pseudobulge, simulations with lower $m_{\rm d}$, higher $\lambda$ or higher $c$ will not form one either (supposing the other two parameters were kept fixed).
Thanks to this finding, which we have verified in a few cases, we could nearly half the simulations required for this work
(there is no need to simulate galaxies for which we know in advance that we shall find $B/T=0$).

The qualitative findings above have a simple interpretation, which will become apparent once we have examined the ELN criterion in greater detail.

For an exponential disc:
\begin{equation}
v_{\rm d}(2.2r_{\rm d}) = 0.62\sqrt{m_{\rm d}\over r_{\rm d}} = \sqrt{1.3{m_{\rm d}(2.2r_{\rm d})\over 2.2r_{\rm d}}},
\label{vd}
\end{equation}
where $m_{\rm d}(2.2r_{\rm d}) = 0.65m_{\rm d}$ is the disc mass within $2.2r_{\rm d}$ \citep{freeman70}.
The first equality in Eq.(\ref{vd}) implies that Eq.~(\ref{efstathiou}) can be rewritten as:
\begin{equation}
{v_{\rm d}(2.2r_{\rm d})\over v_{\rm c}(2.2r_{\rm d})}>{0.62\over\epsilon}.
\label{vdb}
\end{equation}
By introducing the new parameter $\alpha=0.31/\epsilon$,
Eq.~(\ref{vdb}) can be written in the alternative form:
\begin{equation}
{v_{\rm d}(2.2r_{\rm d})\over v_{\rm c}(2.2r_{\rm d})}>2\alpha,
\label{vdbis}
\end{equation}
where $\alpha = 0.28$ for $\epsilon = 1.1$ (ELN) and  $\alpha = 0.26$ for $\epsilon = 1.2$
\citep{christodoulou_etal95}\footnote{The factor of two in Eq.~(\ref{vdbis}) has been introduced so that our definition of $\alpha$ coincides with the one in \citet{christodoulou_etal95}.}.

The second equality in Eq.~(\ref{vd}) shows that Eq.~(\ref{vdbis}) is equivalent to a criterion on the disc fraction within $2.2r_{\rm d}$ because $v_{\rm c}^2(2.2r_{\rm d})/(2.2r_{\rm d})$ is 
the total gravitational acceleration at $r=2.2r_{\rm d}$ in dimensionless units and $v_{\rm d}^2(2.2r_{\rm d})/(2.2r_{\rm d})$ is the disc's contribution. Therefore,
\begin{equation}
f_{\rm d}(2.2r_{\rm d})=\left[{v_{\rm d}(2.2r_{\rm d})\over v_{\rm c}(2.2r_{\rm d})}\right]^2.
\label{fd}
\end{equation}
is the disc's fractional contribution to the gravitational acceleration at $r=2.2r_{\rm d}$ and is related to disc's mass fraction within $2.2r_{\rm d}$ by the equation:
\begin{equation}
f_{\rm d}(2.2r_{\rm d})={1.3 m_{\rm d}(2.2r_{\rm d})\over 1.3 m_{\rm d}(2.2r_{\rm d})+m_{\rm DM}(2.2r_{\rm d})}.
\label{fd2}
\end{equation}

Van den Bosch (1998) was the first to use the ELN criterion to predict
disc stability in a SAM, but he applied Eq.~(\ref{vdbis}) 
at $r=3r_{\rm d}$ rather than at $r=2.2r_{\rm d}$ for the reason that the rotation curve $v_{\rm c}(r)$
peaks at larger radii when the halo's contribution is considered. {
  We have verified that Eq.~(\ref{fd2}) keeps holding at $r=3.2r_{\rm
    d}$ (the factor $1.3$ does not change to the first decimal digit).}

Eqs.~(\ref{vd}) and (\ref{vdbis}) explain why
$B/T$ increases with $m_{\rm d}$ and why it decreases with $\lambda$
and $c$ (Figs.~\ref{conc5}, \ref{conc10}, and \ref{conc15}).
The higher $m_{\rm d}$, the higher the disc fraction within a given radius. The larger the disc, the more DM it will contain.
The more concentrated the halo is, the more DM there will be within the disc.
Our next step is to go beyond these semi-qualitative considerations and to explore the relation between $B/T$ and  $f_{\rm d}$ in quantitative detail.

\begin{figure*}
\begin{center}$
\begin{array}{cc}
\includegraphics[width=0.48\hsize]{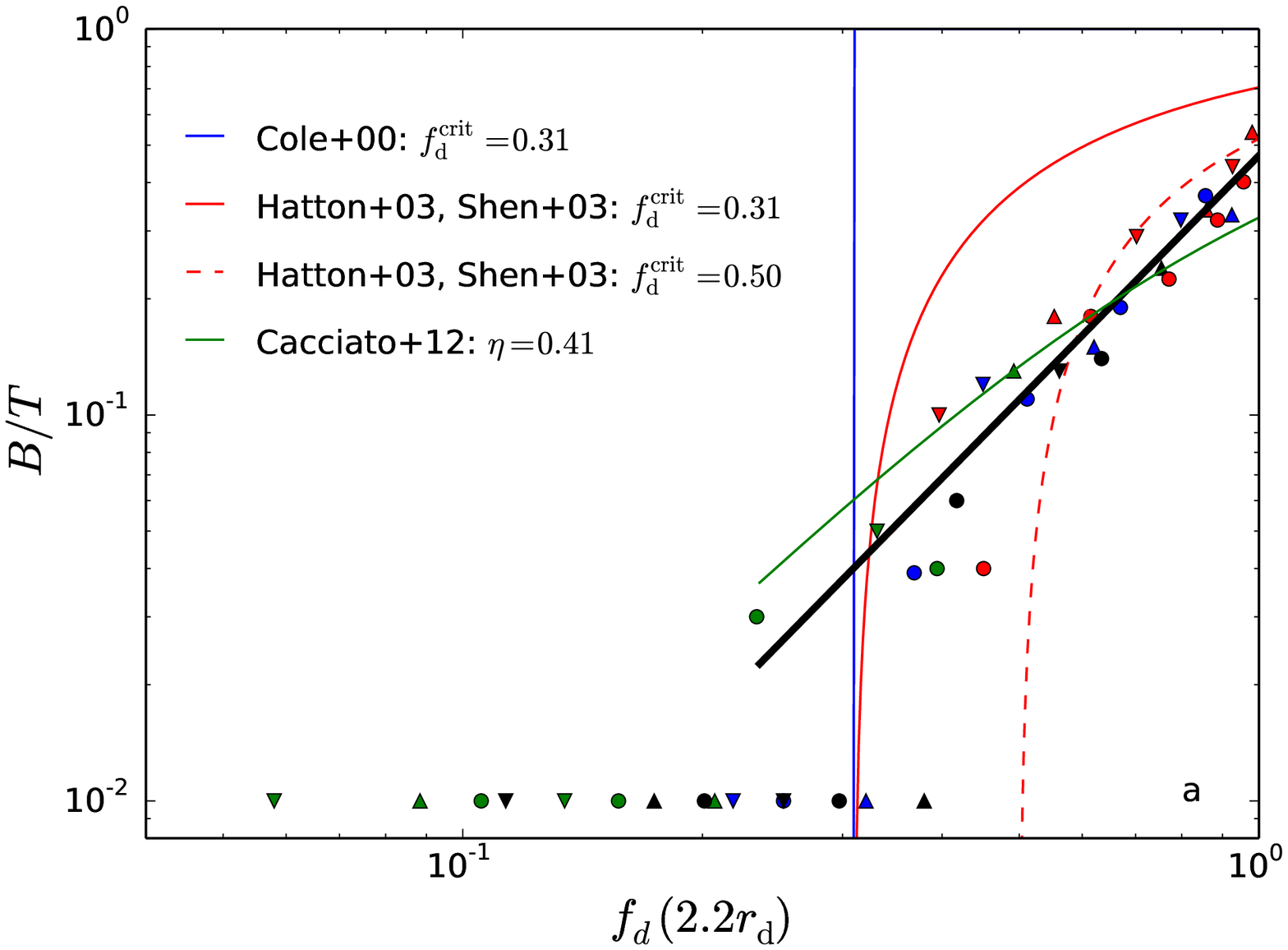} 
\includegraphics[width=0.48\hsize]{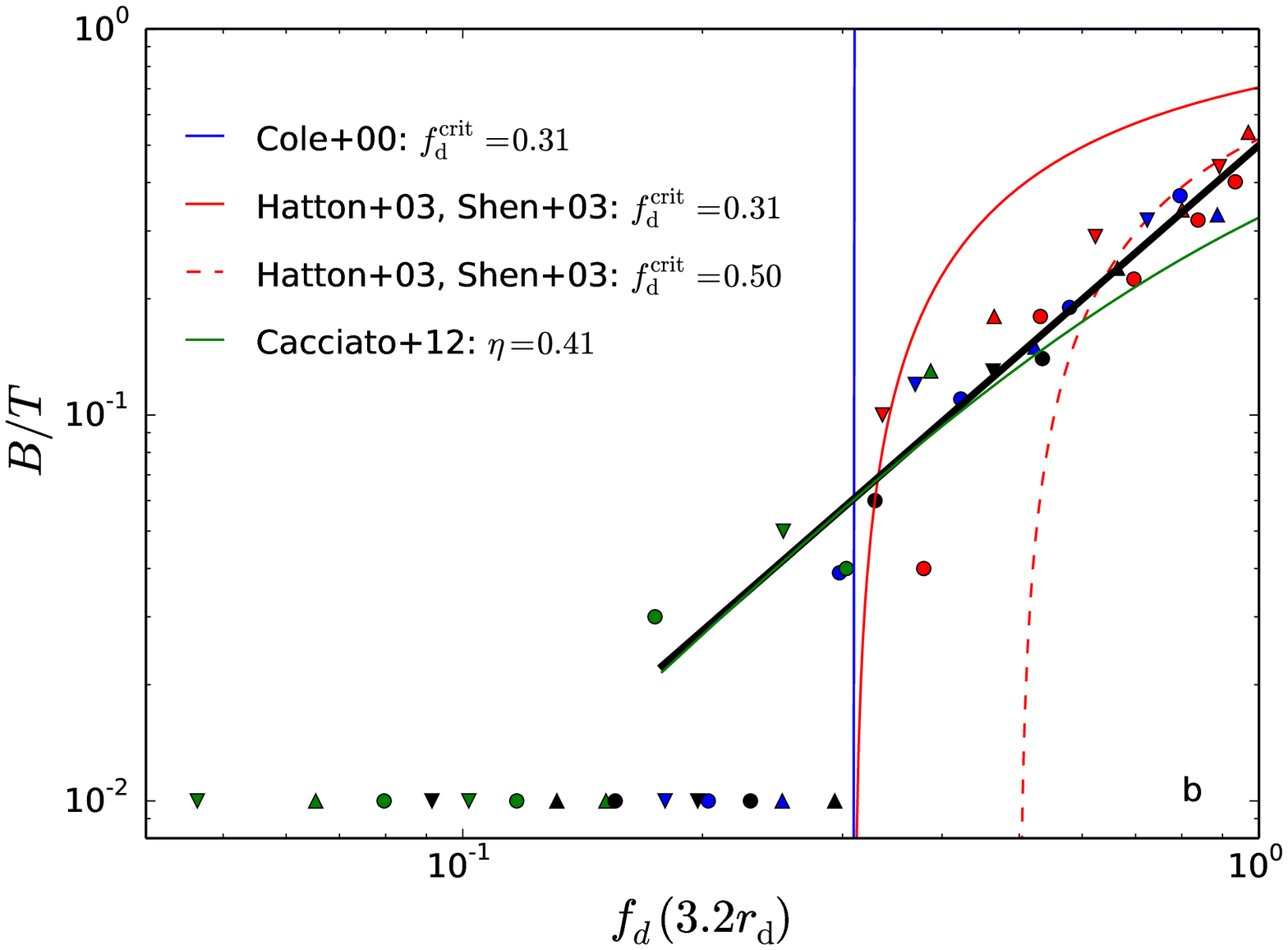} 
\end{array}$
\end{center}
\caption{Relation between bulge-to-total mass ratio $B/T$ and initial disc fraction $f_{\rm d}(r)$ for $r=2.2r_{\rm d}$ (left) and $r=3.2r_{\rm d}$ (right).
Triangles pointing up, circles, and triangles pointing down correspond to simulations with $c=5$, $c=10$, and $c=15$, respectively.
The colours correspond to different disc-to-virial mass ratios: $m_{\rm d}=0.005$ (green), $m_{\rm d}=0.01$ (black), $m_{\rm d}=0.02$ (blue), and $m_{\rm d}=0.04$ (red).
Symbols with $B/T=0.01$ correspond to bulgeless galaxies. They have been assigned $B/T=0.01$ merely to be able to show them on a logarithmic diagram.
The thick solid black lines are log-log linear least-squares fits to the symbols with $B/T>0.01$. 
They correspond to Eqs.~(\ref{BT22}) and~(\ref{BT32}).
The vertical blue lines show the critical $f_{\rm d}$ that corresponds to the
$\alpha=0.28$ (equivalent to assuming $\epsilon=1.1$ in Eq.~\ref{efstathiou})
{The coloured curves compare our results to previous models \citep{cole_etal00,hatton_etal03,shen_etal03,cacciato_etal12}}.
}
\label{ELN_fig}
\end{figure*} 

Fig.~\ref{ELN_fig} shows $B/T$ versus $f_{\rm d}$ at $r=2.2r_{\rm d}$ (where the rotation curve of an exponential disc peaks) and at the optical radius
$r_{\rm opt}=3.2r_{\rm d}$, although 
we have investigated the relation at other radii down to $r=0.5r_{\rm d}$.
The vertical blue lines correspond to:
\begin{equation}
f_{\rm d}^{\rm crit}=(2\alpha)^2\simeq 0.31
\end{equation}
for $\alpha\simeq 0.28$.

The qualitative picture is the same at $r=2.2r_{\rm d}$ and
$r=3.2r_{\rm d}$.
At $f_{\rm d}\ll f_{\rm d}^{\rm crit}$, all galaxies are pure discs (the galaxies with $B/T=0.01$ in Fig.~\ref{ELN_fig} are bulgeless galaxies; 
we have assigned them $B/T=0.01$ merely to be able to show them on a logarithmic plot).
At $f_{\rm d}\gg f_{\rm d}^{\rm crit}$, all galaxies develop a pseudobulge and display a tight correlation between $B/T$ and $f_{\rm d}$.
At intermediate $f_{\rm d}$, galaxies with $B/T=0$ and $B/T>0$ coexist. Therefore, the ELN criterion may fail 
to discriminate between stable and unstable discs \citep{athanassoula08,fujii_etal18}.
Nevertheless, the notion of a critical $f_{\rm d}$ that separates the two populations remains fundamentally valid.

If we restrict our attention to galaxies with $B/T>0$ and 
apply a linear least-squares fit to the relation between $\log(B/T)$
  and $\log f_{\rm d}$, we find:
\begin{equation}
{B\over T}=0.47f_{\rm d}^{2.1}(2.2r_{\rm d})
\label{BT22}
\end{equation}
and
\begin{equation}
{B\over T}=0.50f_{\rm d}^{1.8}(3.2r_{\rm d})
\label{BT32}
\end{equation}
(thick black solid lines in Figs.~\ref{ELN_fig}a and b, respectively).

Eq.~(\ref{BT22}) implies $B/T=0.04$ for $f_{\rm d}=f_{\rm d}^{\rm
  crit}=0.31$.
With Eq.~(\ref{BT32}), $B/T=0.04$ is attained for the slightly lower
value $f_{\rm d}=f_{\rm d}^{\rm
  crit}=0.25$.
If we use $B/T=0.04$ as the minimum bulge-to-total stellar mass ratio below which galaxies are effectively bulgeless,
then these figures imply that the critical $f_{\rm d}$ is lower at $3.2r_{\rm d}$ than it is at $2.2r_{\rm d}$.
This makes sense because $f_{\rm d}(r)$ decreases at large radii in both stable and unstable discs (the density of the baryons in the disc decreases faster than the density of the DM).
Hence, it is logical that the critical $f_{\rm d}$ that separates them should decrease with radius, too.

The standard deviation of $\log (B/T)$ from the linear least-squares
fit of $B/T$ versus $f_{\rm d}(r)$ (that is, the root-mean-square residuals)
decreases 
slowly but systematically from $0.12\,$dex at $r=0.5r_{\rm d}$ to $0.11\,$dex at $r=3.2r_{\rm d}$.
Hence, the ELN criterion is fairly insensitive to the radius at which
it is applied, but a criterion at $r=3.2r_{\rm d}$ appears to be preferable.

If the condition $f_{\rm d}^{\rm crit}\simeq 0.31$ is applied at $r=3.2r_{\rm d}$, then
all discs with $f_{\rm d}(3.2r_{\rm d})>f_{\rm d}^{\rm crit}$ are correctly identified as unstable (Fig.~\ref{ELN_fig}b).
For these galaxies, Eq.~(\ref{BT32}) gives $B/T$ with an accuracy of $\sim 30\%$.
In contrast, we find four galaxies with $f_{\rm d}(3.2r_{\rm d})<f_{\rm d}^{\rm crit}$ where the ELN criterion failed to predict instability.
However, all these galaxies have $B/T<0.06$.
To assume that galaxies with $B/T\lsim 0.06$ are bulgeless is not such a great approximation.
Even observers may not be able to detect bulges that small, particularly for galaxies at distances $\gsim 100\,$Mpc.

\section{Comparison with previous models}

  The blue curves and the red curves in Fig.~\ref{ELN_fig} compare  our results
  to the two most common assumptions encontered in SAMs.
The blue curves correspond to the model by  \citet{cole_etal00}:
as soon as a disc becomes unstable, it collapses into a bulge.
The difference from the black lines is plainly obvious.
The red solid curves correspond to the model by \citet{hatton_etal03}
and \citet{shen_etal03}: as soon as a disc becomes unstable, matter is
transferred from the disc to the bulge until the disc is marginally
stable.

We compute the analytic form of the red curves by assuming that initially all the stars are in the disc,
so that the stellar masses within $2.2\,r_{\rm d}$ and $3.2\,r_{\rm d}$ are $0.65\,m_*$ and $0.83\,m_*$, 
respectively.
Eq.~(\ref{fd2}) gives:
\begin{equation}
  {\beta m_*\over \beta m_*+m_{\rm DM}}=f_{\rm d},
  \label{eq1}
\end{equation}
with $\beta=1.3\times 0.65=0.85$ or $\beta = 1.3\times 0.83 = 1.1$,
depending on the radius at which the instability condition is considered
($m_{\rm DM}$ is the mass of the DM within this radius).
The disc becomes marginally stable when:
\begin{equation}
  {\beta(m_*-m_{\rm b})\over \beta(m_*-m_{\rm b)})+m_{\rm b}+m_{\rm
      DM}}=f_{\rm d}^{\rm crit},
   \label{eq2}
  \end{equation}
where $m_{\rm b}$ and $m_*-m_{\rm b}$ are the final bulge mass and the
final disc mass, respectively. 
We compute the final bulge fraction $B/T=m_{\rm b}/m_*$ by
solving Eqs.~(\ref{eq1}) and (\ref{eq2}), and find:
\begin{equation}
{B\over T}={1\over 1+(\beta^{-1}-1)f_{\rm d}^{\rm crit}}
\left(1-{f_{\rm d}^{\rm crit}\over f_{\rm d}}\right),
\label{hatton}
\end{equation}
where $f_{\rm d}^{\rm crit}\le f_{\rm d}\le 1$.
Eq.~(\ref{hatton}) shows that $B/T\rightarrow 0$ for $f_{\rm
  d}\rightarrow f_{\rm d}^{\rm crit}$ and that:
\begin{equation}
 {B\over T}\rightarrow{1-f_{\rm d}^{\rm crit}\over 1+(\beta^{-1}-1)f_{\rm d}^{\rm crit}}
  \end{equation}
  for $f_{\rm
  d}\rightarrow 1$. The limit for  $f_{\rm
  d}\rightarrow 1$ reduces to $B/T\rightarrow 0.71$ for $f_{\rm
  d}^{\rm crit}=0.31$ and $\beta=1.1$.

The red solid curves are in better agreement with the symbols
than the blue curves, but they systematically overestmate $B/T$ at high
masses and underestimate it at low masses.
The red dashed curves show that increasing $f_{\rm d}^{\rm crit}$ does not improve the overall agreement between
Eq.~(\ref{hatton}) and the simulations because it reduces the
discrepancy with the symbols at high $f_{\rm d}$, but it makes it worse at
low $f_{\rm d}$.

{We have also compared our results to a third model based on the conservation of energy
\citep{cacciato_etal12}. The specific energy of the stars in a cold disc is:
\begin{equation}
e = {1\over 2}v_{\rm c}^2+\Phi,
\label{total_e}
\end{equation}
where $\Phi$ is the gravitational potential.
When stars migrate from the disc to the bulge, they move to a lower energy level. The gravitational energy released by this process increases the stellar velocity dispersion $\sigma$. \citet{cacciato_etal12} used the conservation of energy to study the evolution of two-component discs (gas plus stars) in a cosmological context.
Here, we consider a simpler version of their model, in which the disc is isolated (there is no accretion) and purely stellar (there is no gas and thus no star formation). The internal energy of the stars, ${3\over 2}M_*\sigma^2$, equals the energy 
released by the formation of a bulge, $M_{\rm b}\Delta e$.

To compute $\Delta e$, one should know the initial and final radii of the stars that migrated to bulge, but one can expect
$\Delta e\sim V_{\rm c}^2$ within a factor of order unity \citep{cacciato_etal12} and compute $B/T=M_{\rm b}/M_*$ by solving the equation:
\begin{equation}
M_{\rm b}V_{\rm c}^2\simeq {3\over 2}M_*\sigma^2.
\label{en_eq}
\end{equation} 

The calculation of $\sigma$ is based on the  assumption that $\sigma$ self-regulates so that the 
\citet{toomre64} stability parameter is always close to critical value $Q_{\rm c}$ for local stability. This assumption gives:
\begin{equation}
{\sigma\Omega\over{\rm G}\Sigma}=Q_{\rm c},
\label{metastability}
\end{equation}
with $Q_{\rm c}=1.68$ for a razor-thin sheet of stars with a Maxwellian velocity distribution (see \citealp{binney_tremaine08}).

Let expand Eq.~(\ref{metastability}) by using $\Omega = V_{\rm c}/R$:
\begin{equation}
\sigma={{\rm G}\Sigma R\over V_{\rm c}}Q_{\rm c}={{{\rm G}\Sigma R^2\over R^2} \over {V_{\rm c}^2\over R}}Q_{\rm c}V_{\rm c}.\label{metastability2}
\end{equation}
$V_{\rm c}^2(R)/R$ is the total gravitational acceleration at radius $R$. The numerator measures the disc's contribution 
after the formation of a bulge (the mass of the disc within $R$ scales as $\Sigma R^2$). This contribution can be decomposed into 
the product of two terms:
the stellar contribution relative to the total stars plus DM (the $f_{\rm d}$ factor, since all the stars are in the disc at $t=0$) and the stellar fraction in the disc after a bulge has formed, $M_{\rm d}/M_*$. 
The product is not exact because matter changes geometry when it migrates from the disc to the bulge, but neglecting that has a small effect.
Eq.~(\ref{metastability2}) can thus be rewritten:
\begin{equation}
\sigma\simeq \eta f_{\rm d}(R){M_{\rm d}\over M_*}Q_{\rm c}V_{\rm c},
\label{metastability3}
\end{equation}
where $\eta$ is a fudge factor (the disc mass within $R$ scales as, but is not equal to, $\Sigma R^2$).

By substituting  Eq.~(\ref{metastability3}) into Eq.~(\ref{en_eq}) and using $M_{\rm d}=M_*-M_{\rm b}$, we find:
\begin{equation}
{M_{\rm b}\over M_*} \simeq {3\over 2}\eta^2Q_{\rm c}^2f_{\rm d}^2\left(1-{M_{\rm b}\over M_*}\right)^2,
\label{an_mo}
\end{equation}
where all uncertainties, and notably the one on $\Delta e/V_{\rm c}^2$, are reabsorbed in the value of $\eta$.

It is straightforward to solve the second-degree Eq.~(\ref{an_mo}) and compute $B/T$ as a function of $f_{\rm d}$.
The results are shown by the green curves on Fig.~\ref{ELN_fig}. The  best agreement with our simulations is for $\eta=0.41$.

The model by \citet{cacciato_etal12} predicts a shallower slope for $B/T$ versus $f_{\rm d}$ than what we find in our simulations, especially at high
$f_{\rm d}$ and when it is applied at $r=2.2\,r_{\rm d}$. However, it is the model that works best.}

\section{Comparison with observations}

Our simulations predict that $B/T$ is related to $V_{\rm d}/V_{\rm c}$.
\citet{persic_salucci95} and \citet{persic_etal96} used 967 galaxies with $I$-band photometry and {\sc H}$\alpha$ data from \citet{mathewson_etal92} to conduct one of the most careful and systematic study of the rotation curves of galaxies.
The main result from \citet{persic_etal96} was that $V_{\rm d}/V_{\rm c}$ increases with $I$-band luminosity
from $V_{\rm d}(R_{\rm opt})/V_{\rm c}(R_{\rm opt})=0.4$ at  $M_I=-18.5$ to $V_{\rm d}(R_{\rm opt})/V_{\rm c}(R_{\rm opt})=0.96$ at $M_I=-23.2$, where $R_{\rm opt}=3.2R_{\rm d}$
(the brightest discs are close to being maximal).
This finding is just another way to say that the stellar-to-total mass ratio increases with stellar mass (for example, \citealp{dekel_silk86,papastergis_etal12,behroozi_etal13,moster_etal13}). 
We can use it to transform Eq.~(\ref{BT32}) into a prediction for how $B/T$ grows with $I$-band luminosity
(Fig.~\ref{BT_vs_MI}, black curve). In this section,  we use three data sets to compare this prediction with observations.

\begin{figure}
\includegraphics[width=1.1\hsize]{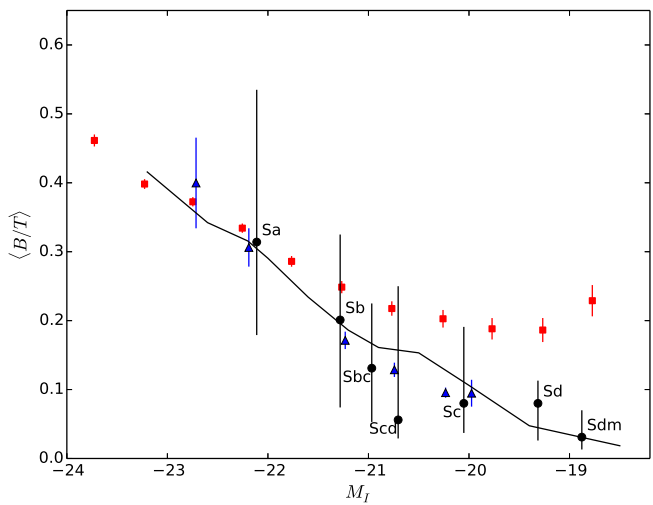} 
\caption{$\langle B/T\rangle$ versus $I$-band magnitude $M_I$. The black curve shows our prediction.
The black circles are from \citet{mathewson_etal92} when we assign to each Hubble type the $\langle B/T\rangle$
from Graham \& Worley (2008; the uncertainty of estimating $B/T$
determines the size of the error bars).
The red squares are from \citet{simard_etal11} after removing all galaxies with $B/T>0.7$.
{The up-pointing filled blue triangles and the down-pointing empty blue triangles are from \citet[S$^4$G]{salo_etal15} when we assign
the bar to the bulge and the disc, respectively.}
The error bars on the red squares and the blue filled triangles show the
standard error on the mean. {We have not shown the error bars on the empty triangles so that they do not overlap with those on the filled
triangles, but their width is comparable.}
}
\label{BT_vs_MI}
\end{figure}

 The catalogue in \citet{mathewson_etal92} contains only 1355 galaxies and no quantitative morphologies, but it has the advantage of being the parent sample in \citet{persic_salucci95}:
using the morphology -- luminosity relation from the same data set
that we have used to pass from $V_{\rm d}(R_{\rm opt})/V_{\rm
  c}(R_{\rm opt})$ to $M_I$ ensures that our analysis is internally
consistent.

Hubble-type information can obviate the lack of quantitative morphologies.
In Fig.~\ref{BT_vs_MI}, the black circles with error bars show the mean $B/T$ from  \citet{graham_worley08} and the mean $I$-band magnitude $M_I$ in  the catalogue from  \citet{mathewson_etal92}
for each Hubble type.
Our predictions are in good agreement with the black circles within the observational uncertainties.
The analysis by Hubble type has two limitations, however.
First, Hubble type is not a quantitative measurement of $B/T$.
Second, we have not considered the frequency of different Hubble types in \citet{mathewson_etal92}'s catalogue (Table~4).

\citet{simard_etal11}  fitted the $r$-band surface-brightness profiles 
of $1.12$ million galaxies from the SDSS
with a S{\'e}rsic  plus
exponential model. Their $B/T$ values are therefore directly comparable to
ours (to the extent that $r$-band luminosity traces stellar mass).

{To pass from $r$ to $I$-band, we
  use Lupton's formula\footnote{http://www.sdss3.org/dr10/algorithms/sdssUBVRITransform.php
  \#Lupton2005}:
\begin{equation}
  M_I = M_i - 0.3780(M_i - M_z)  -0.3974.
\end{equation}
We retrieve $M_i$ and $M_z$ for galaxies in the SDSS Data Release 7 (DR7) by using the selection query:}
\begin{verbatim}
SELECT
  objid, ra, dec, run, camcol, rerun, 
  petromag_i, petromag_z, 
  extinction_i, extinction_z
FROM
  PhotoPrimary 
WHERE 
  flags 
  AND (dbo.fPhotoFlags('SATURATED')+
  dbo.fPhotoFlags('DEBLENDED_AS_PSF'))= 0 
  AND (petroMag_r- extinction_r) 
  BETWEEN 14.0 and 18.0 
  AND Type=3
\end{verbatim}
{and by cross-correlating the results with the catalogue from \citet{simard_etal11}.

\begin{table}
\begin{center}
\caption{Morphological composition of the catalogue in \citet{mathewson_etal92}. Hubble types $T$ are from \citet{devaucouleurs_etal76}.}
\begin{tabular}{ lrr}
\hline
\hline 
Class         & $T$ &  $N_{\rm gal}$        \\
 \hline
Sa  & 1 & 1\\
Sab& 2 & 0\\
Sb & 3 & 362\\
Sbc& 4 & 124\\
Sc  & 5 & 114\\
Scd& 6 & 558\\
Sd  & 7 & 34\\
Sdm& 8& 66\\
Sm & 9 & 0\\
Im & 10 & 1\\
\hline
\hline
\end{tabular}
\end{center}
\label{model_parameters}
\end{table}

 This sample selection  includes elliptical galaxies, while we are interested in
the $B/T$ -- $M_I$ relation for spiral galaxies. We circumvent
this problem by removing all {\it bona fide} ellipticals with
$B/T>0.8$ and $n>2$.
This selection is based on the distribution of galaxies
  on a $B/T$ versus $n$ diagram (Fig.~14 of \citealp{simard_etal11}), which
shows two main populations: a spiral/S0 population with $0<B/T<0.7$
and $0<n<7$ (the mean $B/T$ grows with $n$), and an elliptical
population with $B/T>0.8$ and $n>2$.
The diagram also shows a third population with $B/T>0.7$ and $n<1.5$
(most of these galaxies have $B/T\sim 1$). Visual inspection (for example, Fig.~\ref{simard})
suggests that the bulges of these galaxies are misclassified discs (discs for which 
a S{\'e}rsic model with $n<1$
worked better than an exponential fit).
We therefore assign $B/T=0$ to galaxies for which
\citet{simard_etal11} find $B/T>0.7$ and $n<1.5$.

The red squares in Fig.~\ref{BT_vs_MI} show
the $B/T$ -- $M_I$ relation for the subsample from
\citet{simard_etal11} 
selected according to the above criteria.
The error bars are small
  because they show the error on the mean:
  \begin{equation}
  {\sigma_{B/T}\over\sqrt{N-1}}=\sqrt{\Sigma_{i=1}^N[(B/T)_i-\overline{B/T}]^2\over
      N(N-1)},
  \end{equation}
  which is much smaller than the standard deviation $\sigma_{B/T}$
  ($N$ is the number of galaxies per magnitude bin).}

Our predictions are in good agreement with  the red squares at $M_I\lsim -20.5$. 
At $M_I> -20.5$,  the $B/T$ measured by \citet{simard_etal11} are larger than those measured by \citet{graham_worley08} for galaxies with Hubble type
$T\ge 5$, which comprise the dominant population in this luminosity range.
One explanation for this discrepancy is the challenge of fitting correctly a
faint bulge
component that contributes to $\lsim 10\%$ of the total light when the
uncertainty on $B/T$ can be as large as $20\%$
(see Figs.~18 and 19 of  \citealp{simard_etal11}).
{Another explanation is that bulge/disc decompositions based on a
  single photometric band are not reliable (observations discussed in
  the next section suggest that this may be the case).}

\citet{salo_etal15} used the Spitzer Survey of Stellar Structure in Galaxies (S$^4$G) to study the morphologies of $2352$ with distances $<40\,$Mpc.
The S$^4$G's excellent image quality allowed \citet{salo_etal15} to
fit  three components: {an exponential disc, a {S{\'e}rsic bulge, and a bar modelled with a
    Ferrer profile.
The drawback is a smaller sample and thus poorer statistics.

{In Fig.~8, the filled blue triangles with error bars show $B/T$ in the S$^4$G when the bar is assigned to the bulge.
The down-pointing open triangles show the difference when the bulge is identified with the S{\'e}rsic component only.}

We note that \citet{salo_etal15} published stellar masses.
  We converted them to baryonic masses using the relation between
  gas fraction and stellar mass from
  \citet{boselli_etal14} and we used the relation between
  baryonic mass  and $M_I$ relation for the galaxies of
  \citet{persic_salucci95} to obtain a final result in $I$-band
  magnitudes (the baryonic masses for the galaxies of \citealp{persic_salucci95} 
  were measured dynamically by performing a galaxy/halo
  decomposition, they were not inferred by fitting a stellar
  population synthesis model).

The higher-quality measurements of $B/T$ from \citet{salo_etal15} are in agreement
with those by \citet{simard_etal11} at $M_I<-22$ and confirm our
suspicion that the latter are
probably overestimated  at low luminosities. 
{Our agreement with the filled blue triangles is very good over the
 entire luminosity range $-23<M_I<-20$, especially given the simplifying assumptions we have made (isolated system, axisymmetric initial conditions, purely stellar disc).
 The fact that the agreement with the filled triangles is better than the agreement with the open ones suggests that our two-component fit effectively assigns most of the bar mass to the bulge component.
 }

{We have also checked if our predictions are consistent with the data for the Milky Way.
\citet{bland_ortwin16} studied the disc's contribution $f_{\rm d}$  to the gravitational acceleration at $2.2\,R_{\rm d}$ and found
$f_{\rm d}=0.42$--$0.74$. The main uncertainty is value of $R_{\rm d}$. We predict $B/T=0.09$--$0.16$ for $f_{\rm d}=0.53$ 
($R_{\rm d}=2.6\,$kpc; \citealp{robin_etal03,juric_etal08}) and  $B/T=0.19$--$0.33$ for $f_{\rm d}=0.74$ 
($R_{\rm d}=2.15\,$kpc; \citealp{bissantz_gerhard02,bovy_rix13}).
Observationally, $B/T$ ranges between $0.19$ and $0.32$ depending on whether the long bar is assigned to the disc or the bulge \citep{bland_ortwin16}.
Hence, our predictions are consistent with the data for the Milky Way within the observational uncertainties.

}

\begin{figure}
\begin{center}
\includegraphics[width=1.0\hsize]{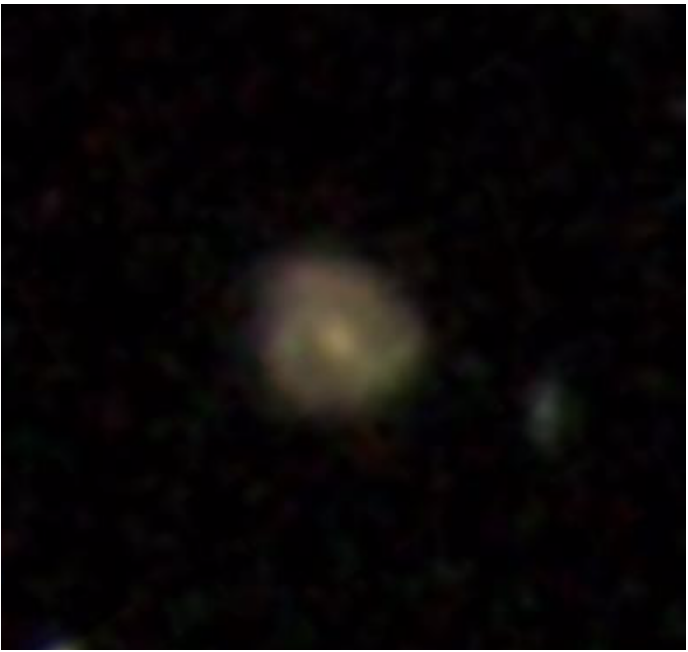} 
\caption{Composite image of SDSS J141007.94-002348.4 from skyserver.sdss.org. SDSS J141007.94-002348.4 is a spiral galaxy with $M_I=-21.8$.
 \citet{simard_etal11} assign to it $B/T=0.93$ but a bulge S{\'e}rsic
 index  of only $n=0.5$.}
\label{simard}
\end{center}
\end{figure}
\section{Application to SAMs}

\begin{figure*}
\begin{center}$
\begin{array}{cc}
\includegraphics[width=0.5\hsize]{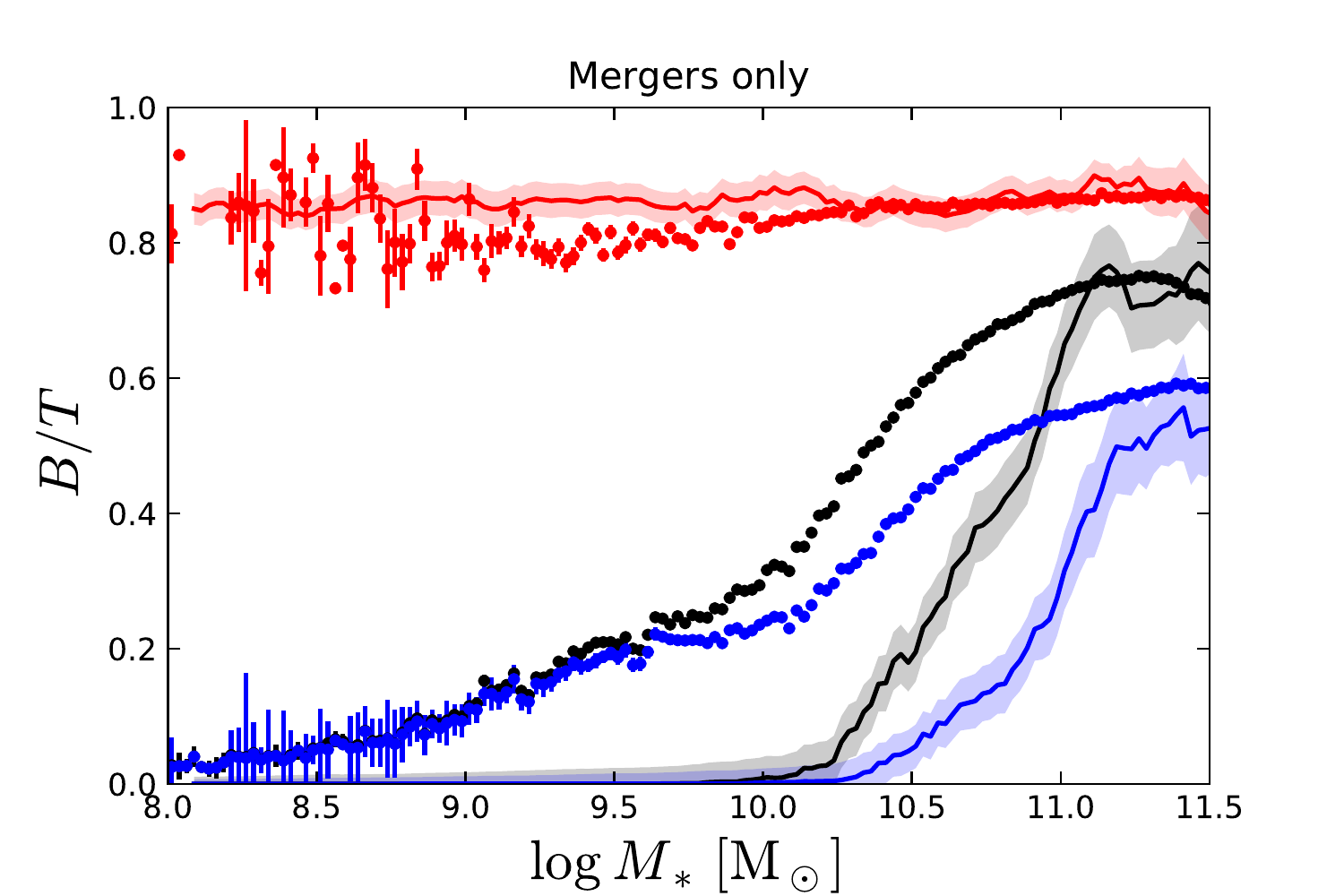} 
\includegraphics[width=0.5\hsize]{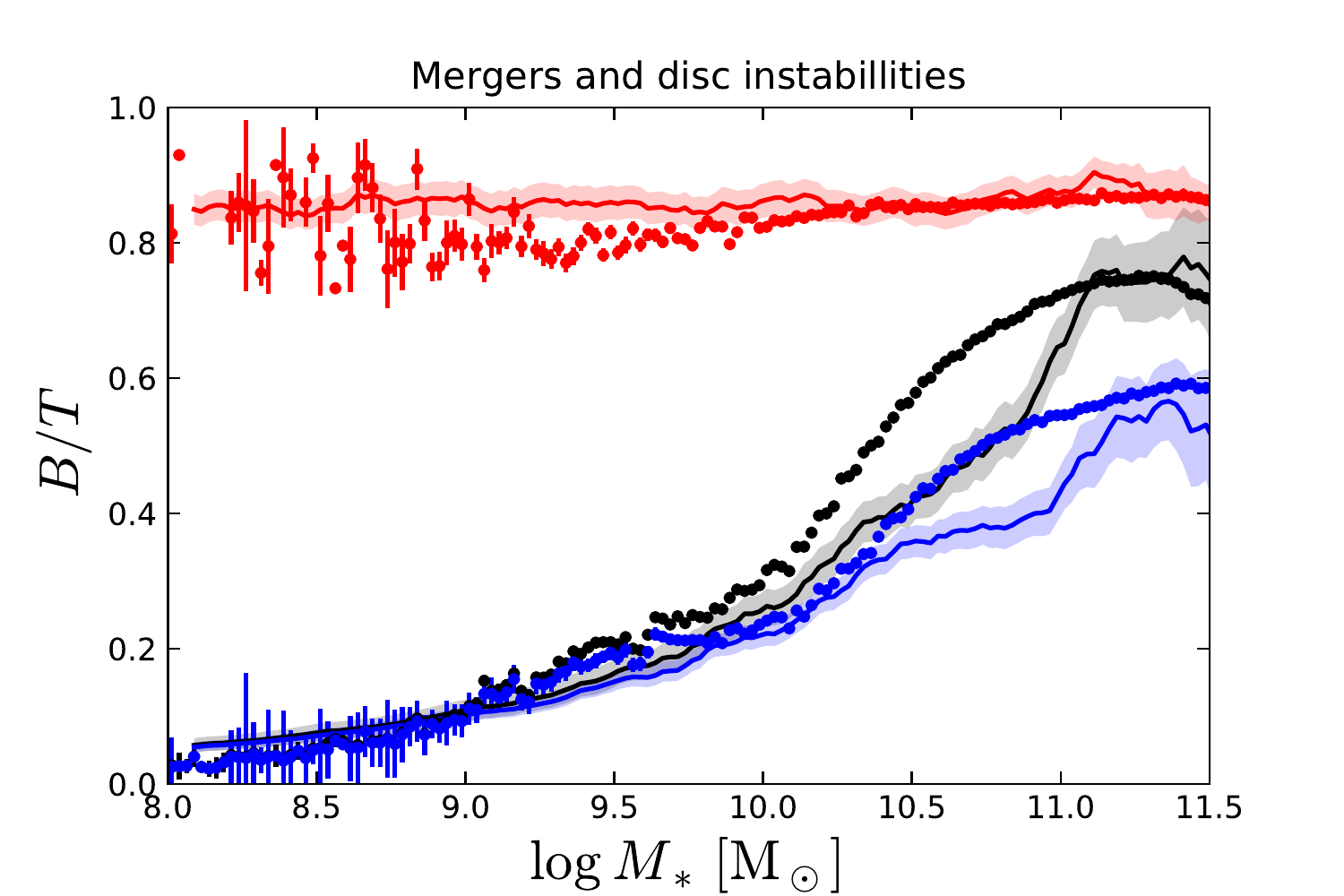} 
\end{array}$
\end{center}
\caption{$B/T$ versus $M_*$ for all galaxies (black), spiral galaxies
  ($B/T<0.7$, blue), and elliptical galaxies ($B/T>0.7$, red). The
  curves are medians in bins of stellar mass. They refer to GalICS~2.0 without (left) and with (right) disc
  instabilities.
  The data points are the observations by
  \citet{mendel_etal14}. The width of the shaded areas around the
  curves shows the standard error on the median.}
\label{GalICS}
\end{figure*}

Our key result is a model for $B/T$ as a function of $f_{\rm d}$, the disc's
contribution  to the total gravitational acceleration. To
use this model predictively, we must know how $f_{\rm d}$ depends on
observable quantities such as luminosity or stellar mass. In
Section~4, we followed a semi-empirical approach. The conversion from
$M_I$ to $f_{\rm d}$ was based on observational data. Here,
Eq.~(\ref{BT32}) is incorporared into the {\sc GalICS~2.0} SAM
\citep{cattaneo_etal17,koutsouridou_cattaneo19}.
Hence, the values of $f_{\rm d}$ used to compute $B/T$ are those
predicted by our SAM.

The main difference with the semi-empirical approach is that, in our SAM,
mergers cannot be switched off.
Hence, we must consider two mechanisms for the formation of bulges:
mergers and disc instabilities.
{\sc GalICS~2.0} accounts for these two mechanisms by separating galaxies into four components: the disc, the pseudobulge, the classical
  bulge, and the central cusp. The cusp is composed of baryons that
  fell to the centre at the last major merger. It is absent in core
  ellipticals, which were formed in dissipationless mergers
  \citep{kormendy_etal09}.
The cusp component is irrelevant for this article.
Relevant is the distinction  between classical bulges formed through
mergers and pseudobulges formed through disc instabilities
({\sc GalICS~2.0} is the only SAM to operate such a distinction to the
best of our knowledge).

In {\sc GalICS~2.0}, the distinction between major and minor mergers
is based on the mass ratio ${\cal M}_1/{\cal M}_2$, where ${\cal M}_1$
and ${\cal M}_2$ are the total masses of the merging galaxies within
their respective baryonic half-mass radii (${\cal M}_1>{\cal M}_2$).
\citet{cattaneo_etal17} assumed a sharp transition at a critical mass ratio
${\cal M}_1/{\cal M}_2=\epsilon_{\rm m}=4$.
In major mergers (${\cal M}_1/{\cal M}_2<\epsilon_{\rm m}$), all stars go  into a classical bulge and  all gas
goes into the central cusp
(\citealp{toomre_toomre72,barnes92,negroponte_white83,barnes_hernquist96,springel_etal05};
but also see \citealp{hopkins_etal09} and \citealp{hammer_etal18}).
In minor mergers (${\cal M}_1/{\cal M}_2>\epsilon_{\rm m}$), the merging
galaxies are added component by component.

Our assumptions for minor mergers are the same as in \citet{cattaneo_etal17}.
Based on N-body simulations by \citet{eliche_etal12}, we assume thay
the scale-lengths of the disc and the bulge of the larger galaxy are
not affected by minor-merging events.
The difference is in major mergers. The model by
\citet{cattaneo_etal17} had the shortcoming of predicting too many pure
ellipticals.  The median $B/T$ for early-type
galaxies with $B/T>0.7$ was close to unity in the model while it is $B/T\sim
0.85$ in the
observations (the red symbols with error bars in Fig.~\ref{GalICS},
data from \citealp{mendel_etal14}).
We obviate to this shortcoming by assuming that
the mass fraction $f_{\rm tr}$ transferrred to the
bulge (or the cusp in the case of gas) is not always unity but varies gradually from
$f_{\rm tr}=0$ for ${\cal M}_1/{\cal M}_2=\epsilon_{\rm m}$ to $f_{\rm
  tr}=0$ for ${\cal M}_1/{\cal M}_2=1$.
We assume:
\begin{equation}
 f_{\rm tr}={\epsilon_{\rm m}-{\cal M}_1/{\cal M}_2\over\epsilon_{\rm
     m}-1},
 \label{ftr}
\end{equation}
where $\epsilon_{\rm m}=3$. Using $\epsilon_{\rm m}=4$ would increase
$B/T$ at $10^{10.5}\,M_\odot<M_*<10^{11}\,M_\odot$ but also at
$M_*<10^{11}\,M_\odot$,
where there is no need for such an increase (Fig.~\ref{GalICS}).

The reader should keep in mind that the choice of a linear
relation in Eq.~(\ref{ftr}) is motivated by its simplicity rather than physical
considerations.
The study of the dependence of $f_{\rm tr}$ on ${\cal M}_1/{\cal M}_2$
would be a major project in its own right and
would require an extensive campaign of numerical simulations.
The GalMer database of major and minor mergers experiments
\citep{chilingarian_etal10} has been the most significant effort in
this direction but does not answer our question because the questions
that  were important for those who have analysed the GalMer database (for example,
\citealp{eliche_etal18} for a recent study) are not those that are
important for our SAM.

Besides the model for major mergers, the only other differences with
the version of  {\sc GalICS~2.0} described in
\citet{koutsouridou_cattaneo19} are the model
for disc instabilities and a detail in the calculation of disc sizes.
Disc radii are important because compact discs are less stable
(Section~3). In {\sc GalICS~2.0}, $r_{\rm d}$ is
determined by the spin parameter $\lambda$ of the DM halo
(Eq.~\ref{mmw}), which we measure from the N-body simulation
used to construct the merger trees.
The gaseous disc and the stellar disc are assumed to have the same
exponential scale-length.
If $\lambda$ fluctuates from one
timestep to the next, so does $r_{\rm d}$.
The effect is small for central galaxies, but it can be significant for 
satellites and
is particularly strong in the central regions of groups and clusters, where the
halo finder\footnote{{\sc GalICS~2.0} uses a halo finder that is called {\sc HaloMaker}
  \citep{tweed_etal09} and is based on {\sc AdaptaHOP}
  \citep{aubert_etal04}.} may incorrectly assign to the host system
particles that in reality belong to a subhalo.
\citet{koutsouridou_cattaneo19} showed that
this numerical effect causes the discs of satellite galaxies to shrink at pericentric passages.
Discs regain their previous sizes
after moving away from their orbital pericentres, but the temporary
contraction affects their stability and drives the formation of spurious bulges.
In this article, we prevent this phenomenon by requiring that the
sizes of discs cannot decrease.

Our model for disc instability is based on Eq.~(\ref{BT32}); $f_{\rm
  d}$ is the contribution of the disc to the total gravitational
acceleration at $r=3.2r_{\rm d}$. The other two contributions are
those of the bulge and the halo. Let $M_{\rm d}$, $M_{\rm cb}$, and
$M_{\rm bar}$ be the masses of the disc, the classical bulge inclusive
of the cusp, and the
bar inclusive of the pseudobulge, respectively, where $M_{\rm bar}=0$ initially. If the $B/T$
computed with Eq.~(\ref{BT32}) is $B/T>M_{\rm cb}/(M_{\rm cb}+M_{\rm
  bar})$,
then a bar will form until:
\begin{equation}
  {B\over T}={M_{\rm cb}+M_{\rm bar}\over M_{\rm cb}+M_{\rm
      bar}+M_{\rm d}}.
  \label{BTsam}
  \end{equation}

To understand how $f_{\rm d}$ is computed in the case of a galaxy that
already has a bar, one should consider that, in our SAM, the bar is
the disc's inner part.
The formation of a bar changes the azimuthal distribution of stars in
the inner disc and causes the inner disc to buckle, but does not change
the total contribution of the disc-plus-bar system
to the circular velocity at $r=3.2r_{\rm d}$. We therefore compute
$f_{\rm d}$ as if the mass in the bar belonged to the disc (in the
beginning this mass will be zero because all galaxies start as pure
discs).
$M_{\rm bar}$ does not change the $B/T$ computed with
Eq.~(\ref{BT32}) but affects the right-hand side of Eq.~(\ref{BTsam}).
If the right-hand side is larger than the left-hand side, the disc is
stable and there is no need for any further increases in $M_{\rm
  bar}$.
A classical bulge from a previous major merger lowers $f_{\rm d}$,
makes the disc more stable and can
prevent the formation of a bar.

The black, blue, and red curves in Fig.~\ref{GalICS} show $B/T$ versus stellar mass $M_*$
in {\sc GalICS~2.0} for all galaxies, galaxies with $B/T<0.7$, and
galaxies with $B/T>0.7$, respectively.
 The shaded area around each curve shows the standard error on the median.
The left panel refers to a model with mergers only. The right panel
shows a model that contains both mergers and disc instabilities.

We now focus on the blue curves ($B/T<0.7$) and 
compare them to the SDSS data by \citet[blue symbols with error
bars]{mendel_etal14}.
The model without disc instabilities is in reasonably good agreement with the
data points for $M_*>10^{11}\,M_\odot$ but fails completely at lower masses, where mergers make a negligible contribution
to the mass growth of galaxies (\citealp{cattaneo_etal11}; also see \citealp{cattaneo_etal17}).

Including disc instabilities improves the SAM considerably. The agreement with the observations by \citet[data points with error
bars]{mendel_etal14} is reasonably good at all masses but especially
at $M_*<10^{10.5}\,M_\odot$. The transition from pseudobulges formed
at low masses to classical bulges formed by mergers at high masses is
not only a theoretical predictions. The data themselves show that the
$B/T$ -- $M_*$ relation changes slope at $M_*\sim 10^{10}\,M_\odot$

We remark that a model could reproduce the
mean $B/T$ galaxies with $B/T<0.7$ (blue symbols) and with $B/T>0.7$
(red symbols)
separately and still not reproduce the mean $B/T$ of the total galaxy
population (black symbols) if the
fraction of galaxies with $B/T>0.7$ is not reproduced correctly.
The reasonably good agreement between the black curve and the black
symbols suggests that the fraction of galaxies with $B/T>0.7$ is
reproduced reasonably well at both low  ($M_*\lsim 10^{10}\,M_\odot$)
and high  ($M_*\gsim 10^{11}\,M_\odot$) masses. It is only in the
intermediate mass-range $M_*\sim 10^{10.5}\,M_\odot$ that mergers
seem to struggle to form enough elliptical and S0 galaxies compared to
what the observations suggest they should be doing.

We note that the catalogue by Mendel et al. (2014; 660000 galaxies) is
based on the one by \citet{simard_etal11} used to produce the
red squares in Fig.~\ref{BT_vs_MI}.
The difference is the use of additional photometric bands, which
allows a more accurate fitting of the spectral energy distributions and thus
more accurate mass estimates, as well as a more detailed analysis of
the effects of dust absorption. Comparing the red squares in
Fig.~\ref{BT_vs_MI} with the blue circles in Fig.~\ref{GalICS} shows
that the bulge-to-total mass ratios of faint galaxies in Fig.~\ref{BT_vs_MI} are almost
certainly overestimated and highlights the danger of $B/T$
decompositions based on one photometric band only.

\section{Discussion and conclusion}

Our simulations of the stability of a thin disc embedded in a static spherical halo improve the classical work by ELN 
because they are three-dimensional and have higher resolution, but the underlying assumptions are very similar.
The only substantial difference is the density distribution of the DM.
Hence, it is unsurprising and reassuring that our results and those by ELN are in substantial agreement.

However, not only do our simulations confirm that all discs with $v_{\rm d}/v_{\rm c}>0.6$ are unstable unless the threshold is exceeded by a 
narrow margin.
They also take the work by ELN one step forwards and show that $v_{\rm d}/v_{\rm c}$ can be used to predict a galaxy's bulge-to-total mass ratio $B/T$.
This is the main difference between our work and that of ELN. They used $f_{\rm d}=(v_{\rm d}/v_{\rm c})^2$ to decide whether an instability would develop.
We use the same parameter to compute $B/T$, so that our answer is more than yes or no. A pseudobulge may form, but its mass may be very small.
Therefore, it is completely unrealistic to assume as \citet{cole_etal00}
that the entire disc collapses into a bulge whenever ELN's stability criterion is not satisfied.

\citet{cattaneo_etal17} computed the sizes and masses of pseudobulges by solving the equation:
\begin{equation}
{v_{\rm d}(r)\over v_{\rm c}(r)}=2\alpha,
\label{C17}
\end{equation}
where $\alpha$ was treated as an input parameter of the SAM. Discs were bulgeless when the equation
admitted no solution for $r$. This approach has no physical justification because the ELN criterion is a global one and the stability threshold
depends on the radius at which it is applied.
A disc with $f_{\rm d}(2.2r_{\rm d})< 2\alpha$ is likely to be globally stable and should not form any pseudobulge, statistically at least.
However, it is possible that $f_{\rm d}(r)=2\alpha$ for $r<2.2r_{\rm d}$, in which case the SAM of \citet{cattaneo_etal17}
would attribute to it a pseudobulge of radius $r$.
\citet{cattaneo_etal17} compensated this overpropensity of their model to form pseudobulges by using $\alpha=0.45$ instead of $\alpha=0.28$
(it is more difficult to form pseudobulges when the instability threshold is set to a higher level) 
and none of their key results (stellar mass function of galaxies, disc sizes, and the Tully-Fisher relation) 
are sensitive to their model for disc instabilities. Still, Eq.~(\ref{C17}) should not be regarded as a physically correct model for the formation of pseudobulges.
In contrast, using Eq.~(\ref{BT32}) would represent a major step forwards with respect to the prescriptions currently used in
SAMs.

{Having summarised the main results of our simulations, we now discuss a number of caveats. The first is the cell size.}
We have run some of our simulations at lower resolution and we did not see substantial differences that would alter our conclusions.

{Another potential issue is the local stability of our initial conditions. The local stability of a stellar thin disc is determined by the \citet{toomre64} parameter:
\begin{equation}
Q={\sigma\kappa\over 3.36{\rm G}\Sigma},
\label{ToomreQ}
\end{equation}
where $\sigma$ is the radial one-dimensional velocity dispersion, $\Sigma$ is the stellar surface density and 
\begin{equation}
\kappa=\sqrt{R{{\rm d}\over{\rm d}R}\Omega^2+4\Omega^2}
\end{equation}
is the epicyclic frequency. The disc is stable for $Q>1$. In our razor-thin discs:
\begin{equation}
{\sigma^2\over{\rm G}\Sigma}=\pi {h\over r_{\rm d}}R_{\rm d}\simeq 0.138 R_{\rm d}
\label{sigma2GSigma}
\end{equation}
(Section~2.1). By substituting the $\sigma$ from Eq.~(\ref{sigma2GSigma}) and $\Omega=V_{\rm c}/R$ with $V_{\rm c}$ from Eq.~(\ref{vc}) into Eq.~(\ref{ToomreQ}), we find that our discs have $Q=0.3$--$0.4$ over most of their surface and that $Q>1$ only at the centre.
Hence, our initial conditions  do not satisfy Toomre's stability criterion. When $Q<1$, spiral shocks heat the disc and contribute to its global stability, but a disc with $Q<1$ can also develop clumps \citep{hockney_hohl69}, which can sink and merge to the centre because
of dynamical friction. Relaxation effects will also lead to a redistribution of mass that could increase the central density.

To explore how this may affect our results, we have run six simulations in which we have let our initial conditions relax before we let them evolve. The relaxation procedure is as follows.
We compute the initial gravitational potential and let the disc evolve in this fixed potential for one orbital time at the disc's half-mass radius.
We recompute the gravitational potential and perform a second iteration, then a third one.
After three orbital times we let the disc's potential become live and we run the simulations as usual.
Table~1 lists the simulations that we have rerun in this manner and the final difference in $B/T$ with respect to the values presented in Section~3.

The difference for relaxed and unrelaxed initial conditions is fairly random: $\Delta(B/T) = ({B/T})_{\rm rel}-({B/T})_{\rm unrel}$ is sometimes negative and sometimes positive.
Statistically, there is a systematic trend towards lower $B/T$ when relaxed initial conditions are used, but the difference is small: $-6\%$ on average. No observer can reliably distinguish between
$B/T=0.2$ and $B/T=0.18$, and certainly no model can claim to predict $B/T$ with such accuracy. Hence, the effect of using unrelaxed, locally unstable initial conditions has no bearing on our conclusions 
within the precision that we aim to achieve
(we remind the reader that the fit in Eq.~\ref{BT32} is accurate within to $30\%$).

{This conclusion appears at odds with recent simulations of the dynamical evolution of cold stellar discs. 
\citet{saha_cortesi18} studied how such discs evolve for different stellar velocity dispersions when all other parameters are kept the same.
Discs with higher $\sigma$ and $Q\sim 1$ formed a bar. They did not fragment into clumps. Discs with very low $\sigma$ and $Q<0.5$ fragmented in less than $1\,$Gyr. The stellar clumps migrated to the centre via dynamical friction and there coalesced into a bulge that reached $B/T\simeq 0.35$ at the end of the simulations.
We shall argue that the simulations by \citet{saha_cortesi18} are more physical than ours, but also that their finding does not invalidate our results.

The key point is how dynamical friction works. Stellar clumps sink to the centre because they move with respect to the DM and dynamical friction transfers their angular momentum to the halo. We assume a static halo. Hence, this process cannot occur in our simulations. The disc can fragment into clumps, but the clumps do not migrate to the centre. One can see this as a limitation of our simulations. If some stellar discs really started with $Q\ll 1$, their evolution into lenticular galaxies through the process described by \citet{saha_cortesi18} will not captured by our analysis. One can also see it as  a justification for our assumption of a static spherical halo. We find that discs that start with $Q\sim 0.5$ reach $Q\sim 0.8$--$1.1$ in just one rotation time (measured at the half-mass radius). By assuming a static halo, we avoid that our final results are too contaminated by relaxation effects due to the Toomre instability. The latter view makes more sense if we consider that our initial conditions are artificial and that stellar discs did not start with $Q\ll 1$.

To check that this explanation is correct, we have rerun four simulations with a live halo (Table~2). Orbital decay and coalescence of clumps through dynamical friction increase $B/T$ by 12$\%$ on average. The effect is strongest in galaxies where clumps are 
clearly visible (those with $m_{\rm d}=0.04$ and $\lambda=0.1$ in Figs.~2 and~3).

An effect of 10--20$\%$ (Table~2) is ultimately small. \citet{saha_cortesi18} did not report the final $B/T$ for the simulations with $Q>0.5$.
Hence, we cannot make a quantitative comparison with their results. However, a glance at their images shows that the simulations with $Q>0.5$ display a significant bulge component, too. Hence, there is no reason to suppose that our final $B/T$ are inconsistent with theirs.

That is not to say that the final visual morphologies look identical in the simulations with $Q<0.5$ and the simulation with $Q\sim 1$ (Fig.~2 of 
\citealp{saha_cortesi18}). The simulations with $Q<0.5$ form S0 galaxies with a prominent bulge but no bar or spiral structure. In the simulation with $Q\sim 1$, the final galaxy is clearly barred. This is a case where a two-component bulge/disc fit and a fit with an additional bar component may give substantially different $B/T$ and show a much stronger dependence on $Q$. 
We do not perform such an analysis, which would be more accurate in reality, because it would then be difficult to compare our results to SDSS studies based on a two-component analysis (for example, \citealp{mendel_etal14}).

Finally, to understand the extent to which $2\%$ gas could affect our
  results, we have run four simulations in which the gas fraction is
  exactly zero (Table~3) and we have analysed them at $t=1$,1.5, 2, 2.5
  and $3\,$Gyr. The mean variation in $B/T$ without gas was $(-2.7\pm
  1.8)\%$  with minimum and maximum variations with respect to the
  case with gas of $-9.3\%$ and $+3.2\%$, respectively. 
  These variations show that gas tends to concentrate in the
  inner regions and thus to increase $B/T$, but the effect is
  negligible for the $2\%$ gas fraction assumed in this article,
  especially since
the scatter in $B/T$ for a given $f_{\rm d}(3.2r_{\rm d})$ is much
larger ($\sim 30\%$).

}

Based on the numerical tests above}, we are confident that Eq.~(\ref{BT32}) provides the correct answer to our problem within the approximations that we have made.
The adequacy of these approximations is another problem, the one that we now discuss.

Already ten years ago, \citet{athanassoula08} had run more realistic simulations with a live halo 
and she found that supposedly stable discs (according to the ELN criterion) could still form a pseudobulge by transferring angular momentum to the DM through resonances. 
Conversely, supposedly unstable discs could still be stabilised by random motions in the disc or the halo or by a central DM concentration (the effects of the latter have been considered in this article).
She concluded that the ELN criterion is too simplistic to describe a complex process such as the formation of pseudobulges.

\citet{fujii_etal18} ran simulations with a live halo at extremely high resolution and they confirmed cases (6 out of 18) where the ELN criterion failed to predict the formation of a bar,
although, in all those cases, $V_{\rm c}(2.2R_{\rm d})/\sqrt{{\rm G}M_{\rm d}/R_{\rm d}}$
exceeded the threshold value $\epsilon=1.1$ by only $30\%$ on average.
Their result is not inconsistent with ours (we, too, find a pseudobulge in a galaxy with $v_{\rm d}/v_{\rm c}<0.5$ at $r=2.2r_{\rm d}$, which should not
have one according to the ELN criterion) and it does not surprise us given the additional possibilities (such as resonances) that come into play when a live
halo is considered. {Our simulations with a live halo (Table~2) confirms that it can make a difference, but also that the bulge-to-total mass ratios measured with a two-component fit are not too affected by it}.

The comparison with \citet{fujii_etal18} is interesting also in relation to two specific claims they make.
Fujii et al. claim that all discs will develop a bar if one waits long enough and that what $f_{\rm d}(2.2r_{\rm d})$ really measures is the timescale $t_{\rm b}$
for bar formation. The discs with $f_{\rm d}(2.2r_{\rm d})<f_{\rm d}^{\rm crit}\simeq 0.31$ are simply those with $t_{\rm b}>10\,$Gyr.
They also claim that the shear rate:
\begin{equation}
\Gamma(r) = -{{\rm d\,ln}\Omega\over{\rm d\,ln}r}=1 -{{\rm d\,ln}v_{\rm c}\over{\rm d\,ln}r}, 
\label{Gamma}
\end{equation} 
is the physical quantity that is most relevant to determining $B/T$ ($\Omega=v_{\rm c}/r$; also see \citealp{donghia15}), 
and demonstrate a tight correlation between $B/T$ and $\Gamma(2.2r_{\rm d})$.

\begin{figure}
\includegraphics[width=1.05\hsize]{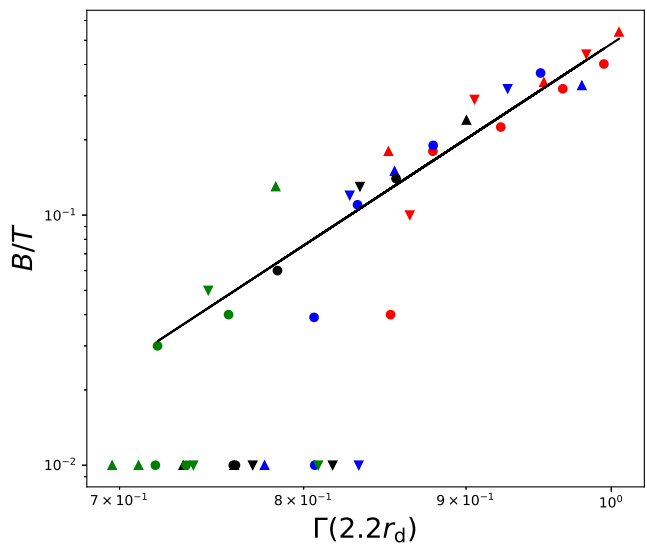} 
\caption{$B/T$ versus the exponent $\Gamma$ of $\Omega(r)$ at $r=2.2r_{\rm d}$.
Symbols of the same shape and colour correspond to the same galaxies as in Fig.~\ref{ELN_fig}.
The black line is a log-log linear least-square fit to the symbols with $B/T>0.01$.}
\label{fig_gamma}
\end{figure}

We cannot comment on the long-term evolution of our galaxies because we have not run any simulations for more than $3\,$Gyr
and since our resolution is lower than that of \citet{fujii_etal18} anyway (the originality of our study is rather in our systematic exploration of the parameter space).
However, {
  we have verifed that our discs evolved very little in the
last gigayear before we stopped following them, in agreement with
\citet{debattista_etal17} and \citet{martinez_etal17}, who find that
the bar strength grows slowly 
after the initial rapid growth and the formation of a pseudobulge in
the first $2\,$Gyr. 
Furthermore,}
any result for isolated discs is purely academical when used to discuss the evolution of galaxies on timescales comparable to the Hubble time\footnote{Even 
in the absence of mergers, discs still accrete gas from their environment.
Slow accretion lowers the $B/T$ ratio and could compensate the growth of pseudobulges in galaxies with $t_{\rm b}\gsim 10\,$Gyr.
Rapid gas accretion, whereby the gas fraction becomes significant, is beyond the scope of this article, which is on stellar discs.}.

In relation to the second claim, our simulations confirm that $B/T$ and $\Gamma(2.2r_{\rm d})$ are tightly related
 (Fig.~\ref{fig_gamma}).
However, we do not consider that $\Gamma$ adds substantial information with respect to $f_{\rm d}$ and the reason is simple.
 $\Gamma(r)$ is directly linked to the form of the rotation curve
($\Gamma<1$ for a rising rotation curve;  $\Gamma>1$ for a declining one), which is the sum in quadrature of $v_{\rm d}$ and $v_{\rm h}$.
Substituting Eq.~(\ref{vc}) into Eq.~(\ref{Gamma}) gives:
\begin{equation}
\Gamma(r)=1-\beta_{\rm d}f_{\rm d}-\beta_{\rm h}(1-f_{\rm d}),
\label{Gammabis}
\end{equation}
where:
\begin{equation}
\beta_{\rm d}={{\rm d\,ln}v_{\rm d}\over{\rm d\,ln}r}
\end{equation}
and:
\begin{equation}
\beta_{\rm h}={{\rm d\,ln}v_{\rm h}\over{\rm d\,ln}r}={1\over 2}\left({{\rm d\,ln}m_{\rm DM}\over{\rm d\,ln}r}-1\right).
\end{equation}
At $r=2.2r_{\rm d}$, $v_{\rm d}$ has a maximum, so that $\beta_{\rm d}(2.2r_{\rm d})=0$; $\beta_{\rm h}(2.2r_{\rm d})$ depends on $cr_{\rm d}$, but,
for reasonable values of $c$ and $r_{\rm d}$, its value is always $\beta_{\rm h}(2.2r_{\rm d})\simeq 0.4$ ($\pm 0.1$ at most).
In contrast, $f_{\rm d}$ can vary by an order of magnitude from one galaxy to another.
Hence, it is $f_{\rm d}$ that determines $\Gamma$, as we intended to demonstrate.

Studies such as those by \citet{athanassoula08} and \citet{fujii_etal18} show that the formation of a bar is sensitive to the instability threshold $\alpha$ and that $\alpha$
is not universal but depends on parameters that our crude assumption of a static spherical halo cannot capture.
However, predicting the stability of individual galaxies with the highest possible accuracy is not the goal of our article.
Our goal is to develop a model that may be able to make statistical predictions for the most likely $B/T$ ratios of galaxies 
given the halo masses, spins, concentrations, and merging histories measured in cosmological N-body simulations.

Considering the quantitative dependence of $B/T$ on $v_{\rm d}/v_{\rm c}$ helps to overcome the problem highlighted by the above studies.
We may erroneously predict the formation of a pseudobulge in a galaxy with $v_{\rm d}/v_{\rm c}>\alpha$ that should not have formed one in reality.
However, if $v_{\rm d}/v_{\rm c}$ is small, $B/T$ will be small, and galaxies with $B/T<0.1$ are, for many practical purposes, indistinguishable from galaxies with $B/T=0$,
particularly when one considers the difficulty of obtaining accurate quantitative morphologies from observations at high redshift.
In the same way, our criterion may fail to predict the  formation of a pseudobulge in a galaxy with $v_{\rm d}/v_{\rm c}<\alpha$ that should develop one.
However, the bulges that we miss are usually small.
The clearest example is galaxy md0.3mb1 of \citet{fujii_etal18}. For that galaxy, the left-hand side of Eq.~(\ref{efstathiou}) is equal to $1.87$,
so the galaxy should be stable. The simulations by  \citet{fujii_etal18} show that it is not.
However, visual inspection of the morphology after 5$\,$Gyr does not show any massive bulge.

The assumptions under which we have derived Eq.~(\ref{BT32})
are consistent with the spirit of the
semi-analytic approximation, which is to separate the evolution of the DM from that of the baryons within haloes.
Cosmological hydrodynamic simulations have shown that baryonic physics can affect the structural properties of DM haloes 
\citep{pontzen_governato12,teyssier_etal13,dicintio_etal14,tollet_etal16}
and cause them to deviate from the NFW profile
that SAMs assume, when they do not make the even cruder approximation of a singular isothermal sphere.
Using Eq.~(\ref{BT32}) to assign morphologies to galaxies that have not experienced any mergers is no greater approximation than to model the DM with an NFW profile or to assume
that major mergers transform discs into spheroids instantaneously.
In fact, \citet{robertson_etal06}, \citet{hopkins_etal09}, and \citet{hammer_etal09} have shown examples of discs that survived gas-rich major mergers.
Yet, a lot has been learned about the Universe from this simple picture.

Finally,  despite their simplifying assumptions, our simulations reproduce the magnitude -- morphology relation observed in the catalogues from \citet{mathewson_etal92},
\citet{simard_etal11}, and \citet{salo_etal15}.
SAMs based on our findings are in very good agreement with the $B/T$
-- $M_*$ relation \citep{mendel_etal14}, especially at $M_*\lsim
10^{10}\,M_\odot$, where most bulges are pseudobulges formed by disc
instabilities {in both our SAM and observations.
Our results support models in which pseudobulges grow gradually
  through non-axisymmetric secular disc instabilities as opposed to
  violent instabilities where discs collapse rapidly into bulges
  ({\`a} la \citealp{cole_etal00}).}

We conclude with a note on our future plans.
`Pseudobulge formation through bar instabilities' might have been a more precise title for our current article, but we have chosen `Bulge formation through disc instabilities' because
we are planning a second article, in which explore how our results can
be extended to gas-rich systems.
We have seen that a $2\%$ gas fraction increases the final $B/T$ by
$\sim 3\%$ on average (Section~4). For high gas fractions,
the instability may be not only quantitatively but also qualitatively different.
Already \citet{noguchi_shlosman94} found that unstable gas-rich discs do not develop a bar. They fragment into clumps that merge into a central bulge.
Bulges formed through this process may be closer to classical bulges than to peanut-shaped pseudobulges \citep{bournaud_etal07}.
Hence our choice of a title that is broad enough to encompass our future projects.

\section*{Acknowledgements}

This work constitutes the eight-week undergraduate thesis of TD.
Our simulations were run on the OCCIGEN supercomputer of the Centre Informatique National de l'Enseignement Sup{\'e}rieur and on the
HPC resources of MesoPSL financed by the R{\'e}gion Ile de France and the project Equip@Meso (reference
ANR-10-EQPX-29-01) of the Agence Nationale pour la Recherche.

We also thank the anonymous referee and the editor F.~Combes for many remarks that have strengthened our manuscript.

\bibliographystyle{aa}

\bibliography{ref_av}
\label{lastpage}
\end{document}